\documentstyle[11pt,amssymb]{article}
\newlength{\dinwidth}
\newlength{\dinmargin}
\newcommand{\xsection}[1]{\section{#1}\setcounter{equation}{0}}

\newcommand{\beq}{\begin{equation}}
\newcommand{\eeq}{\end{equation}}
\newcommand{\beqa}{\begin{eqnarray}}
\newcommand{\eeqa}{\end{eqnarray}}
\newcommand{\al}{\alpha}
\newcommand{\be}{\beta}
\newcommand{\de}{\delta}
\newcommand{\De}{\Delta}
\newcommand{\ga}{\gamma}
\newcommand{\Ga}{\Gamma}
\newcommand{\ka}{\kappa}
\newcommand{\la}{\lambda}
\newcommand{\La}{\Lambda}
\newcommand{\om}{\omega}
\newcommand{\Om}{\Omega}
\newcommand{\si}{\sigma}
\newcommand{\Si}{\Sigma}
\newcommand{\ze}{\zeta}
\newcommand{\th}{\theta}
\newcommand{\Th}{\Theta}
\newcommand{\ve}{\varepsilon}
\newcommand{\vp}{\varphi}
\newcommand{\cD}{{\cal D}}
\newcommand{\cE}{{\cal E}}
\newcommand{\cF}{{\cal F}}
\newcommand{\cL}{{\cal L}}
\newcommand{\cN}{{\cal N}}
\newcommand{\rz}{{\mathbb R}}
\newcommand{\nz}{{\mathbb N}}
\newcommand{\gz}{{\mathbb Z}}
\newcommand{\kz}{{\mathbb C}}
\newcommand{\qz}{{\mathbb Q}}
\newcommand{\hz}{{I\!\! H}}
\newcommand{\ez}{{I\!\! E}}
\newcommand{\unmat}{1 \;\!\!\!\! 1}
\newcommand{\rto}{\rightarrow}
\begin{document}
\thispagestyle{empty}
\noindent ULM-TP/97-2\\
\noindent January 1997\\
\vspace*{5mm}
\begin{center}
{\LARGE\bf Semiclassical Trace Formulae and}\\
\vspace*{5mm}
{\LARGE\bf Eigenvalue Statistics in Quantum Chaos\footnote{
Lectures held at the 3rd International Summer School/Conference
{\it Let's face chaos through nonlinear dynamics} at the University of
Maribor, Slovenia, 24 June -- 5 July 1996}} \\
\vspace*{3cm}
{\large\bf Jens Bolte}\footnote{Electronic address: {\tt 
bol@physik.uni-ulm.de}}\\ \vspace*{5mm}
Abteilung Theoretische Physik, Universit\"at Ulm\\ 
Albert-Einstein-Allee 11, D-89069 Ulm\\ 
Federal Republic of Germany 
\end{center}

\vfill
\begin{abstract}
A detailed discussion of semiclassical trace formulae is presented
and it is demonstrated how a regularized trace formula can be derived 
while dealing only with finite and convergent expressions. Furthermore,
several applications of trace formula techniques to quantum chaos
are reviewed. Then local spectral statistics, measuring correlations
among finitely many eigenvalues, are reviewed and a detailed
semiclassical analysis of the number variance is given. Thereafter
the transition to global spectral statistics, taking correlations among
infinitely many quantum energies into account, is discussed. It is
emphasized that the resulting limit distributions depend on the
way one passes to the global scale. A conjecture on the distribution
of the fluctuations of the spectral staircase is explained in this 
general context and evidence supporting the conjecture is discussed.  
\end{abstract}
\newpage
\setcounter{page}{1}
\section*{Introduction}
\label{intro}
In a dynamical system deterministic chaos manifests itself in
an effective unpredictability of the dynamics for large times.
Different degrees of a chaotic behaviour can be observed by
identifying the properties of ergodicity, mixing, positive 
dynamical entropies, etc.\ \cite{CFS}. All of these concepts 
require to investigate the limit $t\rto\infty$ of the dynamics. 
In the following we will restrict attention to the --important-- 
cases of autonomous, bound Hamiltonian dynamics. Nevertheless, 
one can still find examples for all types of different chaotic 
behaviour. However, upon quantizing such classical dynamical 
systems chaos in the above sense disappears. In a loose manner 
one could explain this observation by the regularizing effect 
the quantum mechanical uncertainty principle has on an irregular 
classical dynamics. More precisely, the discrete spectrum of the 
quantum Hamiltonian $\hat H$ enforces the time evolution in 
quantum mechanics to be almost periodic. In contrast, an almost 
periodic classical time evolution is only observed for integrable 
systems. If one prepares a quantum system at time $t_0=0$ in a 
state $\psi_0$, the solution $\psi(t)$ of the Schr\"odinger 
equation at time $t>0$ reads
\beq
\label{qtimeevolve}
\psi(t)=\sum_n c_n\,\vp_n\,e^{-\frac{i}{\hbar}E_n t}\ ,
\eeq
where $\vp_n$ and $E_n$ denote the eigenvectors and eigenvalues
of the quantum Hamiltonian, respectively: $\hat H\vp_n=E_n\vp_n$.
As $t\rto\infty$ the state described by (\ref{qtimeevolve})
fluctuates in a possibly very wild manner, but cannot approach
zero. More importantly, the same observation holds for the 
correlation function
\beq
\label{timecorr}
\left< \psi_0,\psi(t)\right> =\sum_n |c_n|^2\,e^{-\frac{i}{\hbar}
E_n t}\ ,
\eeq
which has the same structure as (\ref{qtimeevolve}). Thus no
mixing can occur, nor can any of the other features conventionally
characterising chaos be observed. However, in the classical limit, 
which is formally obtained by setting Planck's constant $\hbar=0$,
one recovers integrable as well as chaotic dynamics and all possible
situations in between. The precise behaviour of the classical system
is dictated by the Hamilton function, which generates the classical
time evolution. The latter can produce correlation functions that
range from quasi-periodic ones to exponentially decaying ones. The 
reason for this seemingly paradoxical finding is that the two limits 
involved, $\hbar\rto 0$ and $t\rto\infty$, do not commute.

The field of quantum chaos \cite{Gutz,LesHouches} now is concerned 
with a search for fingerprints the chaotic behaviour of its classical 
limit leaves on a quantum system. Since due to (\ref{qtimeevolve}) 
the spectrum and the eigenfunctions of the quantum Hamiltonian
completely determine the time evolution of a quantum system, it
seems natural to investigate statistical properties of eigenvalues 
and eigenfunctions. The goal of the following lectures now is to 
explain in some detail an approach to eigenvalue statistics
employing trace formulae. In the first part we introduce semiclassical
trace formulae and discuss some of their applications, including 
a discussion of the use of zeta functions. The second part then 
is devoted to an application of trace formulae to eigenvalue
statistics. A certain aspect of the latter, namely the approach
to the global distribution of eigenvalues, will be the central
theme of the third part.  

\xsection{Introduction to Trace Formulae and Zeta Functions}
\label{sec1}
The basic idea behind semiclassical trace formulae is to take the
trace of the quantum mechanical time evolution operator, thereby
loosing all information on eigenfunctions, and to exploit the
Fourier duality between the time dependent and the energy dependent
picture. In order to deal with well behaved quantities, in the 
course of the required manipulations one has to employ a certain 
regularization procedure. Finally then, one can express certain
sums over the spectrum of a quantum Hamiltonian by sums over
periodic orbits of the corresponding classical dynamics. In
certain cases one is furthermore able to determine explicitly 
the leading order contributions as $\hbar\rto 0$.

To get an idea of the kind of quantum systems to which one can
apply the following procedure we now list three of the more
prominent types of examples:
\begin{enumerate}
\item Schr\"odinger operators of the form $\hat H=-\frac{\hbar^2}
{2m}\De+V(x)$, where here $\De$ denotes the Laplacian for $\rz^d$ 
and $V(x)$ is a suitable potential. If $V(x)\rto\infty$ for $|x|
\rto\infty$ sufficiently fast, it is ensured that $\hat H$ has
a discrete spectrum. The corresponding classical dynamics
on the phase space $\rz^d\times\rz^d$ is generated by the 
Hamilton function $H(p,x)=\frac{p^2}{2m}+V(x)$.
\item Quantum billiards: Let $D\subset\rz^d$ be a compact domain
with a sufficiently well behaved boundary $\partial D$. Then
$\hat H=-\frac{\hbar^2}{2m}\De$, where now $\De$ denotes either 
the Dirichlet- or the Neumann-Laplacian for $D$. The corresponding
classical dynamics on the phase space $D\times S^{d-1}$ is that
of a free motion inside $D$ with elastic reflections from
$\partial D$.
\item Let $M$ be a compact Riemannian manifold of dimension $d$.
Then $\hat H=-\frac{\hbar^2}{2m}\De$, where now $\De$ is the
Laplace-Beltrami operator for the Riemannian metric on $M$.
The classical phase space is provided by the unit cotangent 
bundle over $M$, and the classical dynamics is that of the 
geodesic flow, i.e., the free motion along geodesics.
\end{enumerate}
When it comes to explicit formulae, we will in the following 
primarily adhere to case 1. However, possibly after some more or 
less obvious modifications, the results carry over to the other
cases as well.
 
In all of the above cases the spectrum of $\hat H$ is discrete
and bounded from below. One can therefore add a suitable
constant to $\hat H$ in order to render the quantum Hamiltonian
non-negative. Thus
\beq
\label{spectrum}
0\leq E_1\leq E_2\leq E_3\leq \dots
\eeq
provides the list of quantum energies in which we count each eigenvalue
with its respective multiplicity.

\subsection{Semiclassical Trace Formulae}
\label{sec1.1}
The crudest semiclassical approximation to quantum mechanics
can be summarized in the following rule: Each eigenstate of the
quantum Hamiltonian corresponds to a cell of volume $(2\pi\hbar
)^d$ in the classical phase space. Brought in a more mathematical 
form this statement yields the leading semiclassical asymptotics
for the spectral staircase,
\beq
\label{stairsemicl}
N(E):=\#\left\{n;\ E_n\leq E\right\}\sim\frac{1}{(2\pi\hbar)^d}
\int\int\Th\left(E-H(p,x)\right)\,dp\,dx\ ,\ \ \ \ 
\hbar\rto 0\ .
\eeq
For the density of states this means
\beq
\label{densesemicl}
d(E)=\frac{d}{dE}N(E)=\sum_n\de\left(E-E_n\right)\sim\frac{1}
{(2\pi\hbar)^d}\int\int\de\left(E-H(p,x)\right)\,dp\,dx
\ ,\ \ \ \ \hbar\rto 0\ .
\eeq
The ultimate goal of a semiclassical trace formula now is to
provide correction terms to the asymptotics on the r.h.s.\ of
(\ref{stairsemicl}) and (\ref{densesemicl}). But since the
spectral density $d(E)$ is a highly singular object, with
distributional singularities at the eigenvalues $E_n$, one
has to employ some smoothing procedure. 

To start with let us consider the time evolution operator
$\hat U(t)=e^{-\frac{i}{\hbar}\hat H t}$, so that
\beq
\label{timeevolve}
\psi(t,x)=\left(\hat U(t)\,\psi_0\right)(x)=\int K(x,y;t)\,
\psi_0(y)\,dy
\eeq
is the solution of the Schr\"odinger equation with initial
condition $\psi(0,x)=\psi_0(x)$. The distributional kernel
$K(x,y;t)$ then obeys the Schr\"odinger equation
\beq
\label{SchroedingerK}
\left( i\hbar \frac{\partial}{\partial t}-\hat H_x\right)\,
K(x,y;t)=i\hbar\,\de (x-y)\,\de(t)\ ,
\eeq
with initial condition
\beq
\label{ini}
\lim_{t\rto 0}\ K(x,y;t)=\de (x-y)\ .
\eeq
It can for $t\geq 0$ be expanded in an orthonormal basis of 
eigenfunctions $\{\vp_n(x)\}$ of $\hat H$,
\beq
\label{kernelonb}
K(x,y;t)=\sum_n\vp_n(x)\,\overline{\vp_n(y)}\,e^{-\frac{i}{\hbar}
E_n t}\ ,
\eeq
so that the formal trace of the time evolution operator is given
by $\mbox{Tr}\,\hat U(t)=\sum e^{-\frac{i}{\hbar}E_n t}$. This
object does in general not exist as a smooth function of $t$
because it has singularities, the most obvious one being at
$t=0$. In order to regularize the trace consider a smooth
function $\rho (t)$ with a compact support and define the bounded
operator
\beq
\label{regUoft}
\hat U [\rho]:=\int_{-\infty}^{+\infty}\rho (t)\,\hat U(t)\,dt\ ,
\eeq
which has a finite trace, since
\beq
\label{regtrace}
\mbox{Tr}\,\hat U[\rho]=\int dx\int_{-\infty}^{+\infty}dt\ 
\rho(t)\,K(x,x;t)=\sum_n\int_{-\infty}^{+\infty}\rho(t)\,
e^{-\frac{i}{\hbar}E_n t}\,dt=\sum_n\hat\rho\left(-\frac{E_n}
{\hbar}\right)\ .
\eeq
Here $\hat\rho(E)$ denotes the Fourier transform of the function
$\rho(t)$. Since $\rho(t)$ was required to be smooth and compactly 
supported, its Fourier transform is a Schwartz-class test function, 
i.e., $\hat\rho(E)$ and all of its derivatives decrease faster
than any power. Once then the eigenvalues $E_n$ behave as $E_n\sim 
const.\,n^\al$, $n\rto\infty$, for some constant $\al$, the sum on the
r.h.s.\ of (\ref{regtrace}) is finite. Therefore, the map $\rho
\mapsto\mbox{Tr}\,\hat U [\rho]$ is a tempered distribution. 

We now choose the test function $\rho(t)\,e^{\frac{i}{\hbar}Et}$
and hence observe
\beq
\label{regtrace1}
\mbox{Tr}\,\hat U\left[\rho (t)\,e^{\frac{i}{\hbar}Et}\right]=
\int dx\int_{-\infty}^{+\infty}dt\ \rho (t)\,e^{\frac{i}{\hbar}Et}
\,K(x,x;t)=\sum_n\hat\rho\left(\frac{E-E_n}{\hbar}\right)\ .
\eeq
For convenience we rename the test functions in such a way that
$\vp(E):=\hat\rho(-E)$, and thus $\hat\vp(t)=2\pi\rho(t)$. Then
(\ref{regtrace1}) reads
\beq
\label{regtrace2}
\sum_n\vp\left(\frac{E_n-E}{\hbar}\right)=\frac{1}{2\pi}\int dx
\int_{-\infty}^{+\infty}dt\ \hat\vp (t)\,e^{\frac{i}{\hbar}Et}    
\,K(x,x;t)\ .
\eeq
Here $\vp(E)$ and $\hat\vp (t)$ are required to be smooth functions,
and $\hat\vp(t)$ shall have a compact support. The strategy now is
to find a semiclassical approximation for the r.h.s., which then
yields a semiclassical approximation for the spectral density
$d(E)$. To see this one notices that
\beq
\label{hbarspecd}
\hbar\,d(E+\hbar\ve)=\sum_n\hbar\,\de\left(E+\hbar\ve-E_n\right)
=\sum_n\de\left(\ve-\frac{E_n -E}{\hbar}\right)\ ,
\eeq
so that upon evaluating (\ref{hbarspecd}) on the test function
$\vp(\ve)$ one obtains 
\beq
\label{tracephid}
\left<\hbar\,d(E+\hbar\ve),\vp(\ve)\right> =\sum_n\vp\left(
\frac{E_n-E}{\hbar}\right)\ .
\eeq

Some of the general features of a semiclassical approximation 
for the r.h.s.\ of (\ref{regtrace2}) can be revealed once one
chooses an ansatz for the kernel of the time evolution operator.
Let us therefore assume that
\beq
\label{Kansatz}
K(x,y;t)=\frac{1}{(2\pi\hbar)^d}\int e^{\frac{i}{\hbar}\phi(x,y,t;q)}
\,a_\hbar(x,y,t;q)\,dq\ ,
\eeq
where $q\in\rz^d$ is some auxiliary variable. In order to achieve
the initial condition (\ref{ini}) for $K(x,y;t)$ at $t=0$
one demands that
\beq
\label{ansatzini}
\phi(x,y,0;q)=(x-y)q\ \ \ \ \mbox{and}\ \ \ \ a_\hbar (x,y,0;q)
=1\ .
\eeq
Inserting the ansatz (\ref{Kansatz}) into the Schr\"odinger
equation (\ref{SchroedingerK}) for $t>0$ then yields an
equation for the phase $\phi$ and the amplitude $a_\hbar$.
Assuming that the amplitude allows for a formal expansion in
powers of $\hbar$,
\beq
\label{ampexpand}
a_\hbar (x,y,t;q)=\sum_{k\geq 0}(i\hbar)^k\,a_k(x,y,t;q)\ ,
\eeq
one compares like powers of $\hbar$. As a result, one obtains to 
lowest order in $\hbar$ the Hamilton-Jacobi equation 
\cite{Arnold,Robert}
\beq
\label{HJEPhi}
\frac{\partial\phi(x,y,t;q)}{\partial t}+H\left(\nabla_x\phi(x,y,
t;q),x\right)=0
\eeq
for the phase. The higher orders yield transport equations for
the $k$-th order coefficients $a_k$. In principle, starting with
the Hamilton-Jacobi equation (\ref{HJEPhi}) one could now 
successively solve these equations order by order and thus 
construct the time evolution kernel (\ref{Kansatz}) to any
desired precision in $\hbar$. Our ambition in what follows, 
however, will be restricted to obtain the leading order semiclassical
asymptotics.  

A solution $\phi(x,y,t;q)$ of the Hamilton-Jacobi equation 
(\ref{HJEPhi}) with initial condition (\ref{ansatzini}) turns out 
to be closely related to a generating function for a canonical 
transformation in phase space that describes the classical dynamics 
backwards in time. One introduces $S(x,q,t):=\phi(x,y,t;q)+yq$,
which obviously solves the Hamilton-Jacobi equation (\ref{HJEPhi})
with initial condition $S(x,q,0)=xq$. Then, if $(q,y)$ is an initial 
condition for the solution of the classical equations of motion that 
reaches the phase space point $(p,x)$ at time $t$ with energy $E$, 
the function $S(x,q,t)$ generates the canonical transformation $(p,x)
\mapsto (q,y)$. The formalism of canonical transformations and 
generating functions \cite{Arnold,Robert} hence provides one with 
necessary conditions to be imposed on the phase appearing in the 
ansatz (\ref{Kansatz}),
\beqa
\label{neccond}
\nabla_x\phi(x,y,t;q)=p\ \ \ \ &\mbox{and}&\ \ \ \ \nabla_y\phi
(x,y,t;q)=-q\ ,\nonumber \\
\frac{\partial}{\partial t}\phi(x,y,t;q)=-E\ \ \ \ &\mbox{and}&
\ \ \ \ \nabla_q\phi(x,y,t;q)=0\ ,
\eeqa
whenever $(q,y)$ is the initial condition for a classical 
trajectory with energy $E$ in phase space that passes through 
$(p,x)$ at time $t$.

We are now in a position to insert the ansatz (\ref{Kansatz})
into (\ref{regtrace2}),
\beq
\label{tracephiansatz}
\sum_n\vp\left(\frac{E_n-E}{\hbar}\right)=\frac{1}{2\pi(2\pi\hbar)^d}
\int dx\int dt\int dq\ \hat\vp (t)\,e^{\frac{i}{\hbar}[\phi(x,x,t;q)+
Et]}\,\sum_{k\geq 0}(i\hbar)^k\,a_k (x,x,t;q)\ .
\eeq
Inspecting the r.h.s.\ of (\ref{tracephiansatz}) one realizes
that an application of the method of stationary phase 
\cite{Robert,Sjoe} to the multiple integral allows to obtain the 
asymptotic behaviour of this expression as $\hbar\rto 0$. To this 
end one has to identify the points $(x,t,q)$ at which the total 
phase $\phi(x,x,t;q)+Et$ becomes stationary, i.e., where
\beqa
\label{statcond}
0 &=& \left[\nabla_x\phi(x,y,t;q)+\nabla_y\phi(x,y,t;q)
      \right]_{y=x}\ , \nonumber \\
0 &=& \frac{\partial}{\partial t}\phi(x,x,t;q)+E\ , \\
0 &=& \nabla_q\phi(x,x,t;q)\ .\nonumber
\eeqa
A comparison with (\ref{neccond}) now immediately reveals that
(\ref{statcond}) picks out those triples $(x,t,q)$ such that
$(q,x)$ is a point in phase space that is connected to itself by
a solution of the classical equations of motion with energy $E$
and initial condition $(q,y)=(q,x)$ at $t_0=0$, and $(p,x)=(q,x)$
at $t$. Hence, exactly when $t$ is a period of some classical 
periodic orbit with energy $E$, and $(q,x)$ is a point in phase 
space on such an orbit, then $(x,t,q)$ is a stationary point for 
the total phase in (\ref{tracephiansatz}). On the contrary, if $t$ 
does not correspond to any period of a periodic orbit, no stationary
point in the integral over $(x,q)$ occurs. If one therefore chooses 
the test function $\hat\vp(t)$ in such a way that it vanishes on 
all periods of classical periodic orbits with energy $E$, the 
method of stationary phase \cite{Robert,Sjoe} yields the estimate 
$O(\hbar^\infty)$, as $\hbar\rto 0$ in (\ref{tracephiansatz}).

Apart from terms of $O(\hbar^\infty)$, hence all relevant
contributions to (\ref{tracephiansatz}) come from periodic orbits
of the classical dynamics. In this context a distinguished role
is played by the time $t=0$, since then the whole hypersurface
$\Om_E:=\{(p,x);\ H(p,x)=E\}$ of energy $E$ in phase space is
one huge manifold of stationary points. The leading order as
$\hbar\rto 0$ of its contribution to (\ref{tracephiansatz}) can, 
however, be explicitly calculated. To this end let $\chi(t)$ be 
a smooth function with $\chi(t)=1$ in a small neighbourhood of 
$t=0$, which vanishes outside a somewhat larger neighbourhood of 
$t=0$. It shall in particular vanish on all periods of non-trivial 
classical periodic orbits with energy $E$. The trivial identity $1=
\chi(t)+[1-\chi(t)]$ will then be introduced under the multiple
integral in (\ref{tracephiansatz}) such that the contribution
\beq
\label{partun1}
\frac{1}{2\pi(2\pi\hbar)^d}\int dx\int dt\int dq\ \chi(t)\,\hat\vp 
(t)\,e^{\frac{i}{\hbar}[\phi(x,x,t;q)+Et]}\,\sum_{k\geq 0}(i\hbar)^k
\,a_k (x,x,t;q)
\eeq
of the stationary points $(x,t=0,q)$ is separated from the further 
stationary points with $t>0$. If one now introduces polar coordinates
for $q$, $q=\la\om$, $\la=|q|$, $|\om|=1$, one can perform the 
integrals over the variables $t$ and $\la$ by the method of stationary
phase. As a result, (\ref{partun1}) yields 
\beq
\label{leaddense}
\frac{\mbox{vol}\,(\Om_E)}{(2\pi\hbar)^{d-1}}\,\frac{\hat\vp(0)}
{2\pi}\,\left\{1+O(\hbar)\right\}\ ,\ \ \ \ \hbar\rto 0\ ,
\eeq
see \cite[ch.12]{Sjoe} for a closely related problem. If one exploits 
the relation (\ref{tracephid}), the expression (\ref{leaddense}) indeed 
reproduces for the spectral density $d(E)$ the leading asymptotic term 
as given in (\ref{densesemicl}).

Although the leading asymptotic behaviour has already been determined,
the real challenge is to calculate corrections to this. One
therefore has to consider the remaining part
\beq
\label{partun2}
\frac{1}{2\pi}\int dx\int dt\ [1-\chi(t)]\,\hat\vp (t)\,
e^{\frac{i}{\hbar}Et}\,K(x,x;t)\ ,
\eeq
of (\ref{regtrace2}), which due to the above considerations is 
modulo terms of $O(\hbar^\infty)$ completely dominated by the
contributions of non-trivial classical periodic orbits.
Since in (\ref{partun2}) the test function $\hat\vp(t)$ is
required to be of compact support the time integral indeed
extends only over a finite interval. It therefore suffices to
know $K(x,x;t)$ for a certain bounded range of $t$-values.
This observation turns out to be essential for the further
manipulations performed in (\ref{partun2}) in order to obtain
a semiclassical approximation. Namely, the first step in this
direction consists of obtaining the leading semiclassical
asymptotics for the time evolution kernel. In this context one
naturally considers $\hbar\rto 0$ for fixed $t$. An application
of the resulting asymptotic expression in (\ref{partun2}), 
however, requires to use the latter on the whole range of 
integration. If it were not for the test function $\hat\vp(t)$,
this would involve $t\rto\infty$. But the two limits $\hbar\rto 0$
with $t$ fixed, and $t\rto\infty$ with $\hbar$ fixed, do not 
commute, as already mentioned in the Introduction. In mathematical
terms this means that the semiclassical estimate is not uniform 
in $t$. This remark should stress the indispensible role played 
by the regularization procedure applied in the course of the 
present discussion.  

At least for small $t$, the calculation of the leading order 
semiclassical asymptotics for the time evolution kernel goes back 
to Pauli \cite{Pauli}. It reads
\beq
\label{Paulipropagator}
K(x,y;t)=\frac{1}{(2\pi i\hbar)^{\frac{d}{2}}}\,\sum_{\ga_{x,y}}
\left|\det\left( -\frac{\partial^2 R_{\ga_{x,y}}}{\partial x_k
\partial y_l}\right)\right|^{\frac{1}{2}}\,e^{\frac{i}{\hbar}
R_{\ga_{x,y}}(x,y;t)-i\frac{\pi}{2}\nu_{\ga_{x,y}}}\left\{ 
1+O(\hbar)\right\}\ ,
\eeq
where the sum extends over all solutions $\ga_{x,y}$ of the
classical equations of motion with boundary conditions $\ga_{x,y}
(0)=y$ and $\ga_{x,y}(t)=x$. Furthermore, the quantity $R_{\ga_{x,y}}
(x,y;t)$ is defined as the integral of the Lagrangian $L(x,\dot{x})
=\frac{m}{2}\dot{x}^2-V(x)$ along $\ga_{x,y}$,
\beq
\label{Rdef}
R_{\ga_{x,y}}(x,y;t):=\int_0^t L(\ga_{x,y}(t'),\dot{\ga}_{x,y}(t'))
\,dt'\ .
\eeq
Finally, $\nu_{\ga_{x,y}}$ denotes the number of points on the
trajectory $\ga_{x,y}$ which are conjugate to the initial point $y$. 
For small $t$, $\nu_{\ga_{x,y}}$ vanishes, and this is the case covered
by Pauli. The modification required when passing through conjugate
points as $t$ increases was later provided by Gutzwiller \cite{Gutz1}. 

Inserting the semiclassical approximation (\ref{Paulipropagator})
for the time evolution kernel into (\ref{partun2}) yields the
expression
\beqa
\label{tracepo}
&&\frac{1}{2\pi\,(2\pi i\hbar)^{\frac{d}{2}}}\int dx\int dt\ \left[
1-\chi(t)\right]\,\hat\vp(t)\,\sum_{\ga_{x,x}}\left|\det\left( 
-\frac{\partial^2 R_{\ga_{x,y}}}{\partial x_k\partial y_l}\right)_{
y=x}\right|^{\frac{1}{2}}\cdot \\
&&\hspace*{4cm}\cdot e^{\frac{i}{\hbar}[R_{\ga_{x,x}}(x,x;t)+Et]
-i\frac{\pi}{2}\nu_{\ga_{x,x}}}\left\{ 1+O(\hbar)\right\}\ .
\nonumber
\eeqa
Since in general an explicit evaluation of (\ref{tracepo}) is
out of reach one employs the method of stationary phase to obtain
the leading semiclassical asymptotics. In order to determine the
stationary points of the phase $R_{\ga_{x,x}}(x,x;t)+Et$ one
has to solve
\beqa
\label{staphase} 
0&=&\frac{\partial}{\partial t}\,\left[R_{\ga_{x,y}}(x,y;t)+Et
    \right]_{y=x}\ ,\nonumber\\
0&=&\left.\nabla_x R_{\ga_{x,y}}(x,y;t)\right|_{y=x}+\left.\nabla_y 
    R_{\ga_{x,y}}(x,y;t)\right|_{y=x}\ .
\eeqa
From classical mechanics \cite{Arnold} one recalls the identities
\beq
\label{Rident}
\nabla_x R_{\ga_{x,y}}(x,y;t)=p\ ,\ \ \ \ \nabla_y R_{\ga_{x,y}}
(x,y;t)=-q\ ,\ \ \ \ \frac{\partial}{\partial t}R_{\ga_{x,y}}(x,y;t)
=E_{\ga_{x,y}}\ ,
\eeq
where $q$ and $p$ denote the momenta along $\ga_{x,y}$ at the 
instants $t_0=0$ and $t$, respectively. $E_{\ga_{x,y}}$ then is 
the energy of this trajectory. Thus, (\ref{staphase}) picks out
those closed classical trajectories $\ga_{x,x}$ that share 
identical initial and final momenta, $q=p$; hence these are
periodic orbits with energy $E$. That way one recovers the previous
observation, see (\ref{tracephiansatz})--(\ref{statcond}), that
all relevant contributions to (\ref{regtrace1}) are due to classical
periodic orbits of energy $E$. However, one now is in a position
to calculate the leading semiclassical contribution to
(\ref{regtrace2}) explicitly. 

As a first step, consider the contribution of non-trivial periodic
orbits to the regularized Green function,
\beq
\label{regGreen}
G_{po}^\vp (x,y;E):=\frac{1}{2\pi}\int_{-\infty}^{+\infty}\left[
1-\chi(t)\right]\,\hat\vp(t)\,K(x,y;t)\,e^{\frac{i}{\hbar}Et}\,dt\ .
\eeq
In (\ref{tracepo}) one therefore disregards the integration over $x$
and moreover reintroduces $y\neq x$. A well known calculation 
of the integral over $t$ with the help of the method of stationary
phase, see for example \cite{Gutz1,Gutz}, then yields 
\beq
\label{Greensemicl}
G^\vp_{po}(x,y;E)=\sum_{\ga_{x,y}}A_{\ga_{x,y}}^\vp(x,y;E)\ 
e^{\frac{i}{\hbar}S_{\ga_{x,y}}(x,y;E)-i\frac{\pi}{2}\mu_{
\ga_{x,y}}}\,\left\{1+O(\hbar)\right\}\ .
\eeq
Here 
\beq
\label{Greenamplitude}
A_{\ga_{x,y}}^\vp(x,y;E):=\frac{\hat\vp\left(T_{\ga_{x,y}}\right)}
{2\pi(2\pi i\hbar)^{\frac{d-1}{2}}}\ \sqrt{D_{\ga_{x,y}}(x,y;E)}
\eeq
is an amplitude factor attached to each classical trajectory $\ga_{
x,y}$ connecting $y$ and $x$ with fixed energy $E$. The time
$T_{\ga_{x,y}}$ needed for this may depend on the particular trajectory.
Moreover,
\beqa
\label{Dgadef}
D_{\ga_{x,y}}(x,y;E) &:=& \left|\frac{\det\left(-\frac{\partial^2 
R_{\ga_{x,y}}}{\partial x_k\partial y_l}\right)}{\frac{\partial^2}
{\partial t^2}R_{\ga_{x,y}}}\right| =\left|\det\left(\begin{array}{cc} 
\frac{\partial^2 S_{\ga_{x,y}}}{\partial x_k\partial y_l} & 
\frac{\partial^2 S_{\ga_{x,y}}}{\partial x_k\partial E} \\ 
\frac{\partial^2 S_{\ga_{x,y}}}{\partial y_l\partial E}
& \frac{\partial^2 S_{\ga_{x,y}}}{\partial E^2}
\end{array}\right)\right|\ ,\\ \label{muga}
\mu_{\ga_{x,y}} &:=& \left\{ \begin{array}{ccl} \nu_{\ga_{x,y}} &,&
\frac{\partial^2}{\partial t^2}R_{\ga_{x,y}} >0 \\
\nu_{\ga_{x,y}}+1 &,&
\frac{\partial^2}{\partial t^2}R_{\ga_{x,y}} <0 \end{array}
\right. \ .
\eeqa
Finally, $S_{\ga_{x,y}}(x,y;E):=R_{\ga_{x,y}}(x,y;T_{\ga_{x,y}})+E
T_{\ga_{x,y}}$ can also be expressed as the integral of $p\,dx$
along the trajectory $\ga_{x,y}$.

In a second step, the trace of $G_{po}^\vp$ will be calculated
by integrating (\ref{Greensemicl}) over the diagonal $x=y$ with
respect to $x$. Once again the method of stationary phase will
be applied, and to this end the condition of stationarity for
the phase, 
\beq
\label{staphaseS}
\left[\nabla_x S_{\ga_{x,y}}(x,y;E)+\nabla_y S_{\ga_{x,y}}(x,y;E)
\right]_{x=y}=p-q\stackrel{!}{=}0\ ,
\eeq
yields all points $x$ on classical periodic orbits with energy
$E$ as solutions. These points are obviously not isolated, as it
would be required by the simplest version of the method of stationary
phase, since already a single periodic orbit $\ga$ constitutes a
one dimensional connected manifold of stationary points.

However, in case the set of stationary points $x$ devides into a
series $M_j$, $j=1,2,3,\dots$, of connected smooth manifolds of
dimensions $m_j\leq d$ in configuration space, one can introduce
local coordinates in suitable neighbourhoods of the $M_j$'s such
that $x\mapsto (u,v)$, where $u=(u_1,\dots,u_{m_j})$ parametrizes
$M_j$. The method of stationary phase is then applied to the
integral over the transversal coordinates $v=(v_1,\dots,v_{d-
m_j})$. As a result, one obtains that
\beq
\label{Msubjcontrib}
\mbox{Tr}\,G_{po}^\vp =\sum_j\hbar^{\frac{1-m_j}{2}}\,A_{M_j}\,
e^{\frac{i}{\hbar}S_{M_j}}\,\left\{1+O(\hbar)\right\}\ ,
\eeq
where $A_{M_j}$ is an appropriate amplitude which is independent
of $\hbar$, and $S_{M_j}$ denotes the constant value of the action
$S(x,x;E)$ on $M_j$. If $\ga$ is an isolated periodic orbit, the 
corresponding manifold $M_j$ is one dimensional, so that it yields
a contribution of $O(\hbar^0)$ to $\mbox{Tr}\,G^\vp_{po}$. A
$k$-parameter family of periodic orbits leads to $m_j =k+1$ and
hence contributes $O(\hbar^{-\frac{k}{2}})$. In particular, in a 
classically integrable system each invariant torus is of dimension 
$d$ so that its corresponding contribution is $O(\hbar^{\frac{1-d}
{2}})$.

When the classical dynamics is such that all periodic orbits are
isolated, and thus all connected manifolds $M_j$ of stationary 
points are one dimensional, the amplitudes $A_{M_j}$ can be calculated
explicitly. The result being due to Gutzwiller can, e.g., be
found in \cite{Gutz2,Gutz} and reads 
\beq
\label{pocontrib}
\mbox{Tr}\,G^\vp_{po}=\frac{1}{2\pi}\sum_{\ga}\frac{T_{\ga_p}\,\hat
\vp(T_{\ga})\,e^{\frac{i}{\hbar}S_{\ga}-i\frac{\pi}{2}\tilde\mu_{
\ga}}}{|\det \left( M_{\ga}-\unmat\right)|^{\frac{1}{2}}}\,
\left\{1+O(\hbar)\right\}\ .
\eeq 
Here the sum extends over all classical periodic orbits with
energy $E$ such that their periods $T_\ga$ are contained in the
support of $\hat\vp(t)$. $T_{\ga_p}$ then denotes the primitive
period corresponding to $\ga$, i.e., the period of the primitive
periodic orbit $\ga_p$ attached to $\ga$. If $\ga$ is a $k$-fold 
traversal of $\ga_p$ we write $\ga=\ga_p^k$ and obviously find
that $T_\ga=kT_{\ga_p}$. Notice that $k$ can be both positive
and negative, corresponding to traversals of the primitive orbit
in both directions. The quantity $M_\ga$ appearing in 
(\ref{pocontrib}) denotes the monodromy matrix, or stability
matrix, of the periodic orbit $\ga$. It is given as a linearization
of the Poincar\'e recurrence map on a surface of section transversal
to $\ga$ in phase space. Collecting now the contributions 
(\ref{leaddense}) and (\ref{pocontrib}) to (\ref{regtrace2}) 
yields the regularized Gutzwiller Trace Formula (GTF) 
\beqa
\label{GTF}
\sum_n\vp\left(\frac{E_n -E}{\hbar}\right)
  &=&\frac{\mbox{vol}\,(\Om_E)}{(2\pi\hbar)^{d-1}}\,\frac{\hat\vp(0)}
     {2\pi}\,\left\{1+O(\hbar)\right\}\nonumber \\
  & &+\sum_{\ga_p}\sum_{k\neq 0}\frac{\hat\vp(kT_{\ga_p})}{2\pi}\,
     \frac{T_{\ga_p}\,e^{\frac{i}{\hbar}kS_{\ga_p}-i\frac{\pi}{2}k\tilde
     \mu_{\ga_p}}}{|\det \left( M_{\ga_p}^k -\unmat\right)|^{\frac{1}
     {2}}}\,\left\{1+O(\hbar)\right\}\ ,
\eeqa
where the sum over all periodic orbits with energy $E$ has been
replaced by a sum over primitive orbits and their $k$-fold
repetitions. Exploiting the relation (\ref{tracephid}) now allows
to obtain a periodic-orbit representation for the spectral density
\cite{Gutz2,Gutz},
\beq
\label{densepo}
d(E)=\frac{\mbox{vol}\,(\Om_E)}{(2\pi\hbar)^d}\,\left\{1+O(\hbar)
\right\}+\frac{1}{\pi\hbar}\sum_{\ga_p}\sum_{k=1}^\infty\frac{T_{
\ga_p}\,\cos\left(\frac{k}{\hbar}S_{\ga_p}-\frac{\pi}{2}k\tilde
\mu_{\ga_p}\right)}{|\det \left( M_{\ga_p}^k -\unmat\right)|^{\frac{1}
{2}}}\,\left\{1+O(\hbar)\right\}\ .
\eeq
At this place a remark seems to be in order. The test function
$\hat\vp(t)$ cuts off the sum over periodic orbits in (\ref{GTF})
since it has a compact support, i.e., it vanishes outside a finite
interval $[T_1,T_2]$. In hyperbolic classical dynamical systems
the number of periodic orbits with periods not exceeding $T$ is
finite and obeys the asymptotic law
\beq
\label{Huber}
\cN(T):=\#\left\{\ga;\ 0<T_\ga\leq T\right\}\sim\frac{e^{h_{top}T}}
{h_{top}T}\ ,\ \ \ \ T\rto\infty\ .
\eeq
Here $h_{top}>0$ denotes the topological entropy of the classical
dynamics on the energy shell $\Om_E$. Therefore the sum on the 
r.h.s.\ of (\ref{GTF}) is actually of finite length. However, 
the corresponding sum in (\ref{densepo}) is infinite, and due to 
the exponential proliferation (\ref{Huber}) of the number of 
periodic orbits it indeed is divergent. But this divergence is 
to be expected because the sum approximates the spectral density 
which is a singular object. Thus, (\ref{densepo}) should be 
understood as a formal relation whose actual meaning is provided 
by (\ref{GTF}).

A glance at (\ref{densepo}) reveals that for each value $E$ at which the
spectral density shall be evaluated one has to determine the classical
perodic orbits of energy $E$, their actions, stabilities, etc. In general
this is a formidable task which renders an application of the trace
formula (\ref{densepo}) for $d(E)$ almost impossible. However, for
a certain class of dynamical systems considerable simplifications 
emerge in that periodic orbits need only be calculated at a fixed
reference energy $E_0$. A rather simple scaling relation then
determines all required quantities at arbitrary values $E$ of 
the energy. Such a mechanism to apply requires a mechanical similarity
which allows to associate uniquely a periodic orbit at each value $E$,
given one at $E_0$. Furthermore, the actions of periodic orbits have 
to be homogeneous functions of the energy, $S_\ga (\la E)=\la^\al
S_\ga (E)$ for all $\la >0$. One can now fix $E_0$ and obtain
the energy dependence of $S_\ga(E)$ as
\beq
\label{scale}
S_\ga (E)=E^\al\,E_0^{-\al}\,S_\ga(E_0)\ .
\eeq
It then proves useful to discuss the GTF in terms of the scaling 
variable $E^\al$. Examples for systems with a mechanical similarity
and homogeneous actions can be found among all the cases 1.-3.\
mentioned at the beginning of section \ref{sec1}. Hamilton functions
$H(p,x)=\frac{p^2}{2m}+V(x)$ with scaling potentials, $V(\la x)=\la^\ka
V(x)$ for all $\la>0$, yield homogeneous actions of degree $\al =
\frac{1}{2}+\frac{1}{\ka}$. Billiards and geodesic flows on Riemannian
manifolds always show the property (\ref{scale}) with $\al=\frac{1}{2}$,
since $S_\ga(E)=\int_\ga p\,dx=\sqrt{2mE}\,l_\ga$, where $l_\ga$
denotes the geometric length of the periodic orbit $\ga$. We remark 
that in quantum systems whose classical limit is scaling in the above
sense, the semiclassical limit $\hbar\rto 0$ obviously is completely
equivalent to the limit $E^\al\rto\infty$. Once $\al>0$ this in
turn is equivalent to the high-energy limit $E\rto\infty$. An error
term of the form $O(\hbar^k)$, $\hbar\rto 0$, can therefore be replaced 
by $O(E^{-\al k})$, $E^\al\rto\infty$.

As an example, let us now discuss the GTF for quantum billiards with
two degrees of freedom in some more detail. For simplicity, units
will be chosen such that $2m=1$, and by abuse of notation we write
$p:=+\sqrt{E}\geq 0$ so that $S_\ga (E)=pl_\ga$. The spectral density
shall then be expressed in terms of the momentum variable $p$,
\beq
\label{densemom}
d(E)=\sum_n\de\left(E-E_n\right)=\sum_n\de\left(p^2-p_n^2\right)=
\frac{1}{2p}\sum_n\de\left(p-p_n\right)=:\frac{1}{2p}\,\tilde d(p)\ .
\eeq
The trace formula for the spectral density $\tilde d(p)$ thus
reads
\beq
\label{tildedtrace}
\tilde d(p)=\tilde d_0(p)+\frac{1}{\pi\hbar}\sum_{\ga_p}\sum_{
k=1}^\infty\frac{l_{\ga_p}\,\cos\left(\frac{p}{\hbar}kl_{\ga_p}-
\frac{\pi}{2}k\tilde\mu_{\ga_p}\right)}{|\det \left( M_{\ga_p}^k 
-\unmat\right)|^{\frac{1}{2}}}\,\left\{1+O\left(\frac{1}{p}\right)
\right\}\ ,\ \ \ \ p\rto\infty\ ,
\eeq
where $\tilde d_0(p)$ denotes the analogue to the first term on
the r.h.s.\ of (\ref{densepo}) and provides a mean behaviour for
the spectral density. If $d=2$ the leading order asymptotics
follows from (\ref{leaddense}). In addition, the subleading term
for $\tilde d_0(p)$ is also known,
\beq
\label{dbarasym}
\tilde d_0(p)=\frac{A}{2\pi\hbar^2}\,p\mp\frac{L}{4\pi\hbar}+
O\left(\frac{1}{p}\right)\ ,\ \ \ \ p\rto\infty\ ,
\eeq
where $A$ denotes the area of the billiard domain, and $L$ the
length of its boundary. The negative and positive sign correspond 
to Dirichlet and Neumann boundary conditions for the Laplacian,
respectively.   

The sum over periodic orbits in (\ref{tildedtrace}) does not
converge, as has already been discussed for the corresponding
sum in (\ref{densepo}). Again, (\ref{tildedtrace}) shall rather 
be viewed as a distributional relation that has to be evaluated
on suitable test functions in order to yield an analogue of
(\ref{GTF}). Due to the scaling property on does, however, no
longer need an external parameter fixing the energy at which the
periodic orbit sum is to be evaluated. Let us therefore choose
a smooth test function $h(p)$, whose further properties will
follow from a subsequent discussion. Since the variable $p$
derives from the energy variable $E=p^2$, $h(p)$ should be
chosen as an even function, $h(p)=h(-p)$, so that it can also
be defined for negative $p$. An evaluation of (\ref{dbarasym})
on such a test function thus leads to a regularized trace
formula for billiards,
\beq
\label{billliardtrace}
\sum_n h(p_n)=\int_0^\infty\tilde d_0(p)\,h(p)\ dp+\frac{1}
{\hbar}\sum_{\ga_p}\sum_{k=1}^\infty\frac{l_{\ga_p}\,\cF_{\ga_p^k}
[h]\left(\frac{kl_{\ga_p}}{\hbar}\right)}{|\det \left( M_{\ga_p}^k 
-\unmat\right)|^{\frac{1}{2}}}\,\left\{1+O(\hbar)\right\}\ ,
\ \ \ \ \hbar\rto 0\ ,
\eeq
where the following definition enters,
\beq
\label{Fdef}
\cF_\ga[h](u):=\frac{1}{\pi}\int_0^\infty h(p)\,\cos\left(pu-
\frac{\pi}{2}\tilde\mu_\ga\right)\ dp\ .
\eeq
Since
\beq
\label{coseval}
\cos\left(pu-\frac{\pi}{2}\tilde\mu_\ga\right)=\left\{
\begin{array}{ccl}(-1)^{\frac{\tilde\mu_\ga}{2}}\,\cos(pu) &,& 
\tilde\mu_\ga\ \mbox{even}\ , \\ (-1)^{\frac{\tilde\mu_\ga-1}{2}}
\,\sin(pu) &,& \tilde\mu_\ga\ \mbox{odd}\ , \end{array}\right.
\eeq
in case $\tilde\mu_{\ga_p}$ is even this yields
\beq
\label{Feven}
\cF_{\ga_p^k}[h](u)=e^{-i\frac{\pi}{2}k\tilde\mu_{\ga_p}}\,
\frac{1}{2\pi}\int_{-\infty}^{+\infty}h(p)\,e^{ipu}
\ dp=:e^{-i\frac{\pi}{2}k\tilde\mu_{\ga_p}}\,g(u)\ .
\eeq
The version of the GTF that emerges in this situation hence reads
\beq
\label{traceeven}
\sum_n h(p_n)=\int_0^\infty\tilde d_0(p)\,h(p)\ dp+\frac{1}
{\hbar}\sum_{\ga_p}\sum_{k=1}^\infty\frac{l_{\ga_p}\,e^{-i\frac{\pi}
{2}k\tilde\mu_{\ga_p}}\,g\left(\frac{kl_{\ga_p}}{\hbar}\right)}
{|\det \left( M_{\ga_p}^k -\unmat\right)|^{\frac{1}{2}}}\,\left\{
1+O(\hbar)\right\}\ ,\ \ \ \ \hbar\rto 0\ ,
\eeq
and was in this form given in \cite{SieStei}.

The criteria that fix the class of test functions $h(p)$ to be 
admitted in the trace formula (\ref{traceeven}) derive from the
necessity that all terms entering (\ref{traceeven}) be finite. First
of all, the leading asymptotic behaviour (\ref{dbarasym}) of the
spectral density yields Weyl's law
\beq
\label{Weyl}
N(p):=\#\left\{n;0\leq p_n\leq p\right\}\sim\frac{A}{4\pi\hbar^2}
\,p^2\ ,\ \ \ \ p\rto\infty\ ,
\eeq
upon integrating $\tilde d(p)$ once. Since obviously $N(p_n)=n$,
(\ref{Weyl}) can be inverted to observe $p_n\sim const.\sqrt{n}$,
$n\rto\infty$. This requires the test function to obey $h(p)=O
(|p|^{-2-\de})$, $|p|\rto\infty$, for some $\de>0$, in order to
render the sum on the l.h.s.\ of (\ref{traceeven}) convergent. Due
to (\ref{dbarasym}) the same condition also ensures the convergence
of the first term on the r.h.s.\ of (\ref{traceeven}). The 
exponential proliferation (\ref{Huber}) of the number of periodic 
orbits leads one to anticipate that the Fourier transform $g(u)$ 
of the test function $h(p)$ has to be required to decrease 
exponentially for $|u|\rto\infty$, i.e., $g(u)=O(e^{-(\si+\ve)|u|})$ 
for some $\ve>0$ and some characteristic constant $\si>0$ that is 
determined by the distribution of classical periodic orbits. Due
to the definition (\ref{Feven}) of $g(u)$ as a Fourier transform
the exponential asymptotic estimate for $g(u)$ is clearly equivalent 
to demand that the test function $h(p)$ itself be holomorphic in a 
strip $|\mbox{Im}\,p|\leq\si+\ve$. 

A means to characterize the constant $\si$ that determines the 
strip of holomorphy to be demanded for the test function $h(p)$
is provided by the thermodynamic formalism, see \cite{Ruelle,Walters}.
In the latter theory one introduces the Ruelle zeta function
\beq
\label{Ruellez}
\ze_\be(s):=\prod_{\ga_p}\left(1-e^{-sl_{\ga_p}-\be u_{\ga_p}}
\right)^{-1}\ ,
\eeq
where $e^{u_{\ga_p}}>1$ denotes the modulus of one of the eigenvalues
of the monodromy matrix $M_{\ga_p}$; the other eigenvalue of
this matrix then is of modulus $e^{-u_{\ga_p}}<1$. Thus the product 
over primitive periodic orbits in (\ref{Ruellez}) converges for 
those complex $s$ such that $\mbox{Re}\,s$ is large enough; here 
$\be$ is considered as a parameter and is kept fixed. Indeed, one 
denotes the abscissa of convergence for (\ref{Ruellez}) as $P(\be)$, 
so that the condition $\mbox{Re}\,s>P(\be)$ defines the right 
half-plane where the Euler product in (\ref{Ruellez}) converges 
absolutely. Thus, $P(\be)$ has the following representation,
\beq
\label{Pofbeta}
P(\be)=\inf\left\{t\in\rz;\ \sum_{\ga_p}e^{-tl_{\ga_p}-\be u_{\ga_p}}
<\infty\right\}\ .
\eeq
In the thermodynamic formalism $P(\be)$ is known as the topological
pressure of the classical dynamics. Once the latter is hyperbolic,
$P(\be)$ is shown to be of a certain universal form; this in 
particular implies that $P(\frac{1}{2})>0$. Now, since $|\det\left(
M_{\ga_p}-\unmat\right)|^{-\frac{1}{2}}=e^{-\frac{1}{2}u_{\ga_p}}\,
[1+o(1)]$, the condition which ensures an absolutely convergent
sum over periodic orbits in (\ref{traceeven}) reads
\beq
\label{convcond}
\sum_{\ga_p}l_{\ga_p}\,e^{-\frac{1}{2}u_{\ga_p}}\,\left| g\left(
\frac{l_{\ga_p}}{\hbar}\right)\right|<\infty\ .
\eeq
Hence, (\ref{convcond}) is satisfied as long as $g(u)=O(e^{-(
P(\frac{1}{2})+\ve)\hbar |u|})$ for $|u|\rto\infty$, and thus one 
identifies $\si=\hbar P(\frac{1}{2})>0$.

In summary one concludes that the trace formula for billiards
as given in (\ref{traceeven}) contains only finite quantities, once
the test function $h(p)$ satisfies the three conditions
\begin{enumerate}
\item $h(p)=h(-p)$,
\item $h(p)$ is holomorphic in the strip $|\mbox{Im}\,p|\leq\si
+\ve$ for some $\ve>0$, where $\si$ is a positive characteristic 
constant, $\si=\hbar P(\frac{1}{2})$,
\item $h(p)=O(|p|^{-2-\de})$ for $|p|\rto\infty$, where $\de>0$
is arbitrary.
\end{enumerate}

At this point we would like to add a remark on the mathematical
status of the trace formula. Gutzwiller originally derived 
\cite{Gutz1,Gutz2} his trace formula for the spectral density 
$d(E)$, and obtained the relation (\ref{densepo}). Slightly later, 
but seemingly completely independently, similar investigations were 
performed in the mathematical community. The first rigorous proof 
of a trace formula was given in a special case by Duistermaat
and Guillemin \cite{DG}: The `quantum Hamiltonian' they 
considered was an elliptic positive pseudodifferential operator of 
degree one on a compact smooth manifold without boundary; an example 
for such an operator is the square root of minus the Laplacian on 
the manifold. The corresponding classical dynamics then is generated 
by a `Hamiltonian function' which is given by the square root
of the kinetic energy for a single particle. Since thus the 
mechanical similarity as discussed above applies, one can choose
$E_0 =1$ as a reference energy. Then the classical dynamics 
generated by the kinetic energy term and its square root, 
respectively, coincide on $\Om_{E_0=1}$. Translating now the 
result of \cite{DG} into a relation for the spectral density 
$\tilde d(p)$, one recovers (\ref{tildedtrace}) with $\hbar=1$. 
Subsequently, generalizations in several directions were achieved. 
Finally, under certain assumptions on the class of quantum 
Hamiltonians and on the regularity of the classical dynamics, 
the semiclassical trace formula as given in (\ref{GTF}) was 
proven \cite{PU,Mein}. In principle, the strategy employed in 
\cite{DG,PU,Mein} was to start with an ansatz like (\ref{Kansatz}) 
and then to give the integral on the r.h.s.\ as well as the 
expansion (\ref{ampexpand}) a precise mathematical meaning. 
Thereafter the multiple integral in (\ref{tracephiansatz}) could 
be evaluated essentially by employing the method of stationary 
phase. In our presentation, we tried to discuss the trace formula 
in a manner that somehow interpolates between Gutzwiller's original 
investigation and the rigorous mathematical treatment, which 
itself requires an extensive technical apparatus. 

\subsection{Some Applications of Trace Formulae}
\label{sec1.2}
In order to illustrate the use of semiclassical trace formulae
in quantum chaos, we are now going to discuss some of their 
applications. For simplicity the following considerations will be 
restricted to the cases 2.-3.\ of the list at the beginning of
section \ref{sec1}. Hence the relevant form of the trace formula
is given by (\ref{traceeven}). To begin with let us study the test
function $h(p):=e^{-p^2 t}$, $t>0$, which yields on the l.h.s.\ of
the trace formula the trace of the heat kernel,
\beq
\label{Heattr}
\Th_{\hat H}(t)=\mbox{Tr}\,e^{-\hat Ht}=\sum_n e^{-p_n^2 t}\ .
\eeq
For quantum billiards as well as for Laplacians on Riemannian
manifolds the asymptotic behaviour of $\Th_{\hat H}(t)$ as $t\rto 
0$, or equivalently as $\hbar\rto 0$, is well known to yield 
\cite{McKSing}
\beq
\label{Heatasym}
\Th_{\hat H}(t)=\frac{A}{4\pi\hbar^2 t}\mp\frac{L}{8\sqrt{\pi t}
\hbar}+c_0 +\sum_{n=1}^{N-1}c_n\,\hbar^n t^{\frac{n}{2}} +O\left(
\hbar^N t^{\frac{N}{2}}\right)\ ,
\eeq
with some appropriate coefficients $c_n$ that can in principle be
successively determined. The r.h.s.\ of the trace formula can be
evaluated once the Fourier transform $g(u)=\frac{1}{\sqrt{4\pi t}}
\,e^{-\frac{u^2}{4t}}$ of the test function is inserted in 
(\ref{traceeven}). Thus
\beq
\label{HeatGTF}
\Th_{\hat H}(t)=\int_0^\infty\tilde d_0(p)\,e^{-p^2 t}\ dp+
\frac{1}{\sqrt{4\pi\hbar^2 t}}\sum_{\ga_p}\sum_{k=1}^\infty
\frac{l_{\ga_p}\,e^{-i\frac{\pi}{2}k\tilde\mu_{\ga_p}}\,e^{-
\frac{k^2 l_{\ga_p}^2}{4\hbar^2 t}}}{|\det \left( M_{\ga_p}^k 
-\unmat\right)|^{\frac{1}{2}}}\,\left\{1+O(\hbar)\right\}\ .
\eeq
For $\hbar\rto 0$ each term in the sum over periodic orbits is of 
$O(\hbar^\infty)$ so that in (\ref{Heatasym}) all power-like 
contributions in $\hbar$ have to derive from $\tilde d_0 (p)$.
A term-by-term inversion then yields the semiclassical expansion
\beq
\label{dzeroex}
\tilde d_0(p)=\frac{A\,p}{2\pi\hbar^2}\mp\frac{L}{4\pi\hbar}+
O(\hbar)\ .
\eeq
As remarked earlier, the asymptotics for $\hbar\rto 0$ can be 
converted into an asymptotics for $p\rto\infty$, so that 
(\ref{dzeroex}) exactly reproduces (\ref{dbarasym}). In fact, 
the above reasoning is a proper justification for 
(\ref{dbarasym}).

Among the first applications of trace formulae in quantum chaos
one finds efforts to introduce semiclassical quantization rules 
for classically chaotic systems. These should serve as substitutes
for the semiclassical EBK-quantization scheme which only applies
to classically integrable systems. A first guess of a quantization
rule could be to use the representation (\ref{tildedtrace}) for
the spectral density $\tilde d(p)$: one has to find classical
periodic orbits and to calculate their lenghts, stabilities, and 
Maslov phases. Upon evaluating the r.h.s.\ of (\ref{tildedtrace})
with these data one would then approximate $\tilde d(p)$ and thus
be able to identify the singularities of the spectral density.
However, due to the lack of convergence of the periodic orbit
sum in (\ref{tildedtrace}) one has no good control over the
quality of the approximation that occurs by cutting off the sum
after finitely many terms. Therefore a regularization is called
for which renders the periodic orbit sum absolutely convergent.
One hence is advised to consult the smoothed trace formula 
(\ref{traceeven}) with a suitable test function. Having in mind
to approximate the spectral density, as a natural choice for an 
admissible test function a Gaussian
\beq
\label{Gausstest}
h(p')=\frac{1}{\ve\sqrt{\pi}}\left[e^{-\frac{(p'-p)^2}{\ve^2}}+
e^{-\frac{(p'+p)^2}{\ve^2}}\right]
\eeq
was suggested in \cite{ASS}, so that the l.h.s.\ of (\ref{traceeven}) 
yields a smoothed spectral density. This approaches $\tilde d(p)$ in 
the limit $\ve\rto 0$,
\beq
\label{denselimit}
\lim_{\ve\rto 0 }\sum_n h(p_n)=\sum_n\left[\de (p_n-p)+\de (p_n+p)
\right]=\tilde d(p)\ ,
\eeq
if $p>0$. The second Gaussian in (\ref{Gausstest}) is necessary
to yield an even test function, $h(-p')=h(p')$. If $p$ is not too
small, the first Gaussian yields peaks of height $\frac{1}{\ve
\sqrt{\pi}}$ at each $p=p_n$, and the second Gaussian only adds a 
negligable value. Thus, upon scanning $p$ one can detect the quantum
energies as $E_n=p_n^2$. Now the r.h.s.\ of (\ref{traceeven})
can be evaluated with the Fourier transform $g(u)=\frac{1}{\pi}\,
\cos(pu)\, e^{-\frac{\ve^2 u^2}{4}}$. Therefore the periodic orbit 
sum
\beq
\label{poGauss}
\frac{1}{\pi\hbar}\sum_{\ga_p}\sum_{k=1}^\infty\frac{l_{\ga_p}\,
e^{-i\frac{\pi}{2}k\tilde\mu_{\ga_p}}\,\cos\left(\frac{p}{\hbar}
kl_{\ga_p}\right)}{|\det \left( M_{\ga_p}^k -\unmat\right)|^{
\frac{1}{2}}}\,e^{-\frac{\ve^2}{4}k^2 l_{\ga_p}^2}
\eeq
clearly converges absolutely due to the Gaussian suppression of 
the exponential proliferation (\ref{Huber}) of the number of 
terms entering (\ref{poGauss}). Moreover, one can explicitly 
observe that the limit $\ve\rto 0$ applied to (\ref{poGauss})
recovers the periodic orbit sum in (\ref{tildedtrace}). 

Integrating the spectral density once with respect to $p$ yields
the spectral staircase (\ref{Weyl}). The trace formula with the 
test function (\ref{Gausstest}) will therefore now be integrated
in $p$ over the interval $[0,q]$. On the l.h.s.\ one then
obtains the approximation
\beq
\label{Napprox}
N_\ve (q)=\sum_{0\leq p_n<q}\left[1+O\left(\ve\,e^{-\frac{c}{\ve^2}}
\right)\right]+\sum_{p_n =q}\left[\frac{1}{2}+O\left(\ve\,e^{-\frac{c}
{\ve^2}}\right)\right]+\sum_{p_n>q}O\left(\ve\,e^{-\frac{c}{\ve^2}}
\right)
\eeq
to the spectral staircase. However, in the limit $\ve\rto 0$ the
expression (\ref{Napprox}) approaches the symmetrized staircase
function
\beq
\label{symmstair}
N_0 (q)=\lim_{\de\rto 0}\frac{1}{2}\,\left[N(q+\de)+N(q-\de)
\right]\ ,
\eeq
which is identical to $N(q)$ whenever $q\neq p_n$. Otherwise
one obvioulsy observes that $N(p_n)=n-\frac{1}{2}$. This allows
to set up the quantization rule
\beq
\label{cosquant}
\cos\left[\pi N_0(q)\right]=0\ \ \ \ \Leftrightarrow\ \ \ \ q\in
\{p_1,p_2,p_3,\dots\}\ .
\eeq
According to (\ref{poGauss}) the leading periodic orbit contribution
to (\ref{Napprox}) reads
\beq
\label{cospo}
N_{\ve,fl}(q)=\frac{1}{\pi}\sum_{\ga_p}\sum_{k=1}^\infty\frac{e^{-i
\frac{\pi}{2}k\tilde\mu_{\ga_p}}}{k}\,\frac{\sin\left(\frac{q}{\hbar}
kl_{\ga_p}\right)}{|\det \left( M_{\ga_p}^k -\unmat\right)|^{
\frac{1}{2}}}\,e^{-\frac{\ve^2}{4}k^2 l_{\ga_p}^2}\left\{1+O(\hbar)
\right\}\ ,
\eeq
which again is an absolutely convergent sum as long as $\ve>0$. In 
\cite{AMSS} it was suggested to use (\ref{cospo}) in (\ref{cosquant}) 
to obtain an approximate semiclassical quantization rule. It was 
furthermore demonstrated in several examples that this procedure 
yields good numerical approximations to the quantum energies $E_n=
p_n^2$, and that this scheme is quantitatively superior to the direct
use of a Gaussian test function (\ref{Gausstest}) in the trace 
formula.

Inspecting the periodic orbit sum (\ref{cospo}) one observes that,
viewed as a function of $q$, it constitutes a superposition of
sine-oscillations with wave lengths $\frac{2\pi\hbar}{kl_{\ga_p}}$.
Adding those terms finally yields $N_\ve (q)$. One can now ask how 
many of these terms become efficient when one is interested in
resolving all quantum energies $E_n\leq p^2$ with the quantization
rule (\ref{cosquant}), see \cite{AB} for related discussions.
At the point $p$ the mean density of zeros of the expression 
$\cos[\pi N_0 (p)]$ is given by (\ref{dbarasym}) so that their 
mean separation $\De p$ asymptotically reads $\De p\sim\frac{2\pi
\hbar^2}{Ap}$. In order to effectively resolve the zeros $p_n$ in 
the vicinity of $p$ with the help of the periodic orbit sum 
(\ref{cospo}) one hence has to add all oscillatory terms with wave 
lengths down to $\De p$, i.e., one needs all periodic orbits $\ga=
\ga_p^k$ with $\frac{2\pi\hbar}{kl_{\ga_p}}\geq\De p\sim\frac{2\pi
\hbar^2}{Ap}$. Thus, a cut-off at approximately $kl_{\ga_p}\approx
\frac{Ap}{\hbar}$ ensures that one has taken all relevant contributions 
into account. Since the exponential proliferation (\ref{Huber}) of 
the number of periodic orbits reads in terms of their geometrical 
lengths
\beq
\label{Huberl}
\tilde\cN(l)=\#\left\{\ga;\ l_\ga\leq l\right\}\sim\frac{e^{\tau l}}
{\tau l}\ ,\ \ \ \ l\rto\infty\ ,
\eeq
where $\tau$ denotes a scaled topological entropy which is 
independent of $p$, the number $N_p$ of terms to be included in
(\ref{cospo}) in order to resolve all quantum energies with $E_n
\leq p^2$ is given by
\beq
\label{Nsubp}
N_p\sim\frac{\hbar}{A\tau p}\,e^{\frac{A\tau}{\hbar}p}\ ,\ \ \ \ 
p\rto\infty\ .
\eeq
Recalling now the asymptotics (\ref{Weyl}), the number $N_n$ of
periodic orbits required to resolve the lowest $n$ quantum energies
thus derives from (\ref{Nsubp}) to behave as
\beq
\label{Nsubn}
N_n\sim\frac{1}{2\tau\sqrt{\pi An}}\,e^{2\tau\sqrt{\pi An}}\ ,
\ \ \ \ n\rto\infty\ .
\eeq
Therefore, the computational effort to obtain more and more quantum 
energies by using the semiclassical quantization rule grows
enormously. 

Similar considerations apply to all quantization
procedures that are based on semiclassical trace formulae since
the structure of the periodic orbit sums involved together with
the exponential proliferation of the number of periodic orbits
will always produce estimates of the effort to be spent which are 
similar to (\ref{Nsubn}); one will certainly not achieve an 
improvement that comes down to a power-law behaviour for $N_n$. 
However, in some specific systems, especially in classically chaotic 
scattering systems, it might happen that one can resolve a certain 
number of zeros using only a few periodic orbits. In these situations 
the specific features of the systems cause the number of required
periodic orbits to blow up only at considerably large $p_n$. In 
this context one should bear in mind that (\ref{Nsubn}) is only an 
asymptotic statement. The principal effect, resulting in the 
estimate (\ref{Nsubn}), is however inherent in semiclassical
quantization procedures based on trace formulae and cannot be 
simply overcome. At this point also the analogy between the zeros
of the Riemann zeta function, see below, and quantum energies of
classically chaotic systems breaks down since the density of 
Riemann zeros is considerably lower than (\ref{dbarasym}). This
results in a power-law estimate for $N_n$, and that has to be 
compared with (\ref{Nsubn}), see \cite{AB} for a
brief account. Basically the same problem occurred in the 
mathematical literature on the Selberg trace formula, where it
was noted that well known estimates that hold in the theory of 
the Riemann zeta function could not be carried over to the Selberg      
zeta function, see \cite{Hejhal} for an extensive discussion. 

As a consequence, various studies have revealed that it is 
possible to calculate a certain number, say some ten, quantum
energies from classical data employing semiclassical trace formulae.
However, the huge numerical effort to be spent in order to increase
the output of the number of eigenvalues prohibits to apply these
semiclassical methods on a large scale. It turns out that numerical
methods to solve the eigenvalue problem of the quantum Hamiltonian 
directly are far more efficient than semiclassical ones, if one is
interested in obtaining a large number of quantum energies.

\subsection{The Selberg Trace Formula and the Selberg Zeta Function}
\label{sec1.3}
The trace formulae discussed so far essentially yield the leading 
semiclassical approximation to the spectral density. In addition,
one has to cope with a further approximation in applications because
in practise one can only take finitely many terms of a periodic 
orbit sum into account. Thus, it is often difficult to trace back 
numerical inaccuracies to either one of the approximations applied. 
Luckily there exist cases in which an exact trace formula is at hand, 
which therefore does not include any semiclassical approximation. 
The classical systems to be considered are geodesic motions on
manifolds of constant negative curvature, and their quantizations
are provided by the quantum Hamiltonians $\hat H=-\frac{\hbar^2}
{2m}\De$, where $\De$ denotes the Laplace-Beltrami operator for
the manifold. Due to the negative curvature the classical
dynamics show a strong chaotic behaviour, and thus these models
serve as convenient playgrounds for quantum chaos. 

In the simplest case of $d=2$ degrees of freedom one has to deal
with the geometry of hyperbolic surfaces, see for example 
\cite{Rat,BalVoros}. Here we choose as a model for two dimensional 
hyperbolic geometry the unit disc
\beq
\label{UD}
\cD:=\left\{z=x+iy\in\kz;\ |z|<1\right\}
\eeq
endowed with the Poincar\'e metric $ds^2=4\,(1-x^2-y^2)^{-1}\,
(dx^2+dy^2)$ of constant negative Gaussian curvature $K=-1$. The 
classical dynamics of a single particle on $\cD$ is that of a free 
motion along geodesics. The latter are those circular arcs and 
straight lines in $\cD$ that are perpendicular to the unit circle. 
The hyperbolic distance $d(z,w)$ between two points $z\neq w$ in $\cD$ 
is measured with the Poincar\'e metric $ds^2$ and is given as the 
hyperbolic length of the unique geodesic arc connecting $z$ and $w$. 
In explicit terms it reads
\beq
\label{hypdist}
\cosh d(z,w)=1+\frac{2|z-w|^2}{(1-|z|^2)(1-|w|^2)}\ .
\eeq
From this relation one immediately concludes that any point $w$
on the unit circle is infinitely far away from any point $z$ in
the interior of $\cD$.

Due to the latter observation the Poincar\'e unit disc itself is 
an infinitely extended two dimensional hyperbolic space. In order
to yield compact and closed surfaces, and thus examples for the 
case 3.\ of the list at the beginning of section \ref{sec1}, one
has to confine oneself to bounded domains in the unit disc supplied
with appropriate boundary identifications. In as much as a flat 
torus can be constructed from a parallelogram in the euclidean
plane with opposite edges identified, one can obtain compact
surfaces with a metric of constant negative curvature by an 
analogous construction. The boundary
identifications that produce a torus out of a parallelogram are
translations on the euclidean plane. The group generated by the
two identifications of a torus then is a discrete subgroup of
the group of motions of the plane, i.e., the group of translations
and rotations. The analogous transformations on the Poincar\'e
disc are the M\"obius transformations
\beq
\label{Moebius}
z\mapsto\frac{\al z+\be}{\overline{\be}z+\overline{\al}}\ ,
\eeq
with $|\al|^2-|\be|^2=1$, which indeed map the interior of $\cD$
to itself. Introducing the matrix group $SU(1,1):=
\left\{\left(\al\,\be\atop\overline{\be}\,\overline{\al}\right);\ 
|\al|^2-|\be|^2=1\right\}$, a composition of two M\"obius 
transformations (\ref{Moebius}) corresponds to a matrix multiplication 
in $SU(1,1)$. The latter group is therefore the analogue of the
group of motions on the euclidean plane. Given now a suitable 
bounded domain $\cF\subset\cD$, with finitely many geodesic arcs
as its boundary components, the M\"obius transformations 
(\ref{Moebius}) identifying pairs of edges generate a discrete 
subgroup $\Ga\subset SU(1,1)$.

The quantum Hamiltonian $\hat H=-\frac{\hbar^2}{2m}\De$ requires
to know the Laplace-Beltrami operator $\De$ for the Poincar\'e
disc $\cD$,
\beq
\label{LaBel}
-\De=-\frac{1}{4}\,\left(1-x^2-y^2\right)\,\left(\frac{\partial^2}
{\partial x^2}+\frac{\partial^2}{\partial y^2}\right)\ .
\eeq
If this operator is defined on the smooth functions on $\cD$ it
has a continuous spectrum $[\frac{1}{4},\infty)$. Due to the 
peculiar value of the bottom of the spectrum one parametrizes the
quantum energies as $E=\frac{\hbar^2}{2m}(p^2+\frac{1}{4})$, with 
$p\geq 0$.
For the following we choose units such that $\hbar=1=2m$, and 
hence $E=p^2+\frac{1}{4}$. If one now constructs a compact and
closed surface from a suitable domain $\cF\subset\cD$ as described 
above, one imposes periodic boundary conditions on the eigenfunctions
of $\hat H=-\De$. Let $g_1,\dots,g_k$ be the M\"obius transformations
(\ref{Moebius}) identifying pairs of edges of $\cF$, then one
demands that $\psi(g_j z)=\psi(z)$, $j=1,\dots,k$. Since the 
transformations $g_1,\dots,g_k$ generate the discrete group
$\Ga\subset SU(1,1)$, the periodicity extends to all elements of
$\Ga$, $\psi(gz)=\psi(z)$ for all $g\in\Ga$. The Laplacian now 
being defined on a compact surface has a discrete spectrum
$0=E_0<E_1\leq E_2\leq\dots$. If one defines $p_n:=\sqrt{E_n-
\frac{1}{4}}$, the spectral staircase satisfies the Weyl 
asymptotics (\ref{Weyl}). 

In the 1950's Selberg \cite{Selberg} employed the above 
constructions to obtain the Selberg Trace Formula (STF)
\beq
\label{STF}
\sum_n h(p_n)=\frac{A}{2\pi}\int_0^\infty p\,\tanh(\pi p)\,h(p)\ 
dp+\sum_{\ga_p}\sum_{k=1}^\infty\frac{l_{\ga_p}\,g(kl_{\ga_p})}
{2\sinh(\frac{kl_{\ga_p}}{2})}\ ,
\eeq
see also \cite{Hejhal}. Here $A$ denotes the area of the given 
surface, measured with the Poincar\'e metric, and the sum over 
the $\ga_p$'s extends over all primitive periodic orbits of the
associated classical dynamics. Since the latter are given by the
geodesic motion, the periodic orbits are the closed geodesics on 
the compact surface. The test function $h(p)$ is required to 
fulfill    
\begin{enumerate}
\item $h(-p)=h(p)$,
\item $h(p)$ is holomorphic in the strip $|\mbox{Im}\,p|\leq\frac{1}
{2}+\ve$ for some $\ve>0$,
\item $h(p)=O(|p|^{-2-\de})$ for $|p|\rto\infty$, where $\de>0$
is arbitrary.
\end{enumerate}
Furthermore, $g(u)=\frac{1}{2\pi}\int_{-\infty}^{+\infty}h(p)\,
e^{ipu}\,dp$ is the Fourier transform of the test function $h(p)$.

A comparison of the STF (\ref{STF}) with the trace formula 
(\ref{traceeven}) reveals that both relations are almost identical.
If one were to set up the trace formula (\ref{traceeven}) for the
systems presently under study one would first of all notice that due
to the STF all higher-order corrections to the periodic orbit
sum which are not explicitly contained in (\ref{traceeven}) 
indeed vanish. One is then furthermore able to read off from the
STF a number of quantities entering the semiclassical trace formula:
\begin{enumerate}
\item The mean spectral density is completely known,
$$\tilde d_0(p)=\frac{A}{2\pi}\,p\,\tanh(\pi p)\sim\frac{A}{2\pi}
\,p+O(e^{-2\pi p})\ ,\ \ \ \ p\rto\infty\ .$$
Since the surface has no boundary, in a comparison with 
(\ref{dbarasym}) one has to choose $L=0$.
\item The eigenvalues of the monodromy matrices $M_\ga$ are given
by $e^{\pm u_\ga}=e^{\pm l_\ga}$, that is $u_\ga =l_\ga$, for all
periodic orbits $\ga$.
\item No Maslov phases occur, $e^{-i\frac{\pi}{2}\tilde\mu_\ga}
=1$.
\item The width of the strip of holomorphy demanded for $h(p)$
is explicitly known, and hence $P(\frac{1}{2})=\frac{1}{2}$.
Indeed, a further study of the classical dynamics yields that
the topological pressure is linear, $P(\be)=1-\be$.
\end{enumerate}
Of course, all of the applications discussed for the semiclassical
trace formula in section \ref{sec1.2} carry over to the STF.

All the information on the distribution of classical periodic
orbits and on the quantum energies which is contained in the STF
can be encoded in a function of the complex variable $s=\frac{1}
{2}-ip$, where $p=\sqrt{E-\frac{1}{4}}$ is the momentum variable
introduced above, but now extended to the whole complex plane,
see \cite{Selberg,Hejhal}. This Selberg zeta function
\beq
\label{Selzeta}
Z(s):=\prod_{\ga_p}\prod_{n=0}^\infty\left(1-e^{-(s+n)l_{\ga_p}}
\right)
\eeq
is defined by a product over the classical primitve periodic orbits
$\ga_p$ which converges in the right half-plane $\mbox{Re}\,s>1$.
Since the latter condition is equivalent to $\mbox{Im}\,p>\frac{1}
{2}$ the domain of convergence excludes the real momentum axis where
all quantum energies, with the exception of possibly finitely many, 
are to be found as $E_n =p_n^2+\frac{1}{4}$.

In order to obtain the analytic properties of the Selberg zeta
function one chooses the test function
\beq
\label{zetatest}
h(p)=\frac{1}{p^2+(s-\frac{1}{2})^2}-\frac{1}{p^2+(\si-\frac{1}{2}
)^2}\ ,
\eeq
with $\mbox{Re}\,s$, $\mbox{Re}\,\si>1$, and inserts this into the 
STF. The Fourier transform of $h(p)$ reads $g(u)=\frac{1}{2s-1}\,
e^{-|u|(s-\frac{1}{2})}-\frac{1}{2\si-1}\,e^{-|u|(\si-\frac{1}{2})}$
so that the sum over periodic orbits on the r.h.s.\ of (\ref{STF})
can be evaluated,
\beqa
\label{zetasum}
\frac{1}{2s-1}\sum_{\ga_p}\sum_{k=1}^\infty\frac{l_{\ga_p}}{2\sinh
(\frac{kl_{\ga_p}}{2})}\,e^{-kl_{\ga_p}(s-\frac{1}{2})}
  &=& \frac{1}{2s-1}\sum_{\ga_p}\sum_{n=0}^\infty l_{\ga_p}
      \sum_{k=1}^\infty e^{-(s+n)kl_{\ga_p}}\nonumber\\
  &=& \frac{1}{2s-1}\sum_{\ga_p}\sum_{n=0}^\infty\frac{e^{-(s+n)l_{
      \ga_p}}}{1-e^{-(s+n)l_{\ga_p}}} \\
  &=& \frac{1}{2s-1}\frac{d}{ds}\log Z(s)\ .\nonumber
\eeqa
Extracting the contribution of the lowest eigenvalue $E_0=0$ on the 
l.h.s.\ of the STF then yields
\beqa
\label{STFform2}
\sum_{n=1}^\infty \left[\frac{1}{E_n +s(s-1)}-\frac{1}{E_n +
\si(\si-1)}\right]
    &=& \frac{1}{\si(\si -1)}-\frac{1}{s(s-1)}-\frac{A}{2\pi}\left[
        \psi(s)-\psi (\si)\right] \nonumber \\
    & & +\frac{1}{2s-1}\,\frac{Z'(s)}{Z(s)}-\frac{1}{2\si -1}
        \frac{Z'(\si)}{Z(\si)}\ ,
\eeqa
where $\psi(z)=\frac{d}{dz}\log\Ga(z)$. Up to now we have restricted 
the above expressions to the domain $\mbox{Re}\,s$, $\mbox{Re}\,
\si>1$ in order to keep the integrals and sums convergent. The
relation (\ref{STFform2}), however, allows to analytically continue 
$Z(s)$ to all $s\in\kz$. To this end we perform the limit $\si\rto 1$
and arrive at
\beqa
\label{regresolSTF}
\sum_{n=1}^\infty \left[\frac{1}{E_n +s(s-1)}-\frac{1}{E_n}\right]
=-\ga_\De -\frac{1}{s(s-1)}-\frac{A}{2\pi}\,\psi (s)+\frac{1}{2s-1}
\frac{Z'(s)}{Z(s)}\ ,
\eeqa
where $\ga_\De$ denotes an appropriate constant, see \cite{Steiner}.
An integration of this expression with respect to $s$ results in
the product representation \cite{Steiner}
\beq
\label{Selzetaprod}
Z(s)=Z'(1)\,s(s-1)\,e^{s(s-1)\ga_\De}\,
\left[(2\pi)^{(1-s)}\,e^{-s(s-1)}\,
G(s)\,G(s+1)\right]^{\frac{A}{2\pi}}\,\prod_{n=1}^\infty\left[\left(
1+\frac{s(s-1)}{E_n}\right)\,e^{-\frac{s(s-1)}{E_n}}\right] 
\eeq 
for the Selberg zeta function, where Barnes' double $\Ga$-function
\beq
\label{Barnes}
G(z+1)=(2\pi)^\frac{z}{2}\,e^{-\frac{z}{2}-\frac{1+\ga}{2}z^2}\,
\prod_{n=1}^\infty\left[\left(1+\frac{z}{n}\right)^n\ e^{-z+\frac{
z^2}{2n}}\right]
\eeq
appears. The analytic continuation of the Selberg zeta function  based 
on (\ref{STFform2}) reveals that $Z(s)$ is an entire holomorphic
function with zeros at
\begin{itemize}
\item $s=\frac{1}{2}\pm ip_n$ of multiplicity $d_n$, when $E_n =
p_n^2 +\frac{1}{4}$, is an eigenvalue of $-\De$ of
multiplicity $d_n$,
\item $s=1$ of multiplicity one,
\item $s=0$ of multiplicity $\frac{A}{2\pi}+1$,
\item $s=-k$, $k\in\nz$, of multiplicity $\frac{A}{2\pi}
(k+1)$.
\end{itemize} 
We remark that by topological reasons the area of a compact surface
of constant negative curvature is such that $\frac{A}{4\pi}$ is 
an integer. The analytic structure of $Z(s)$ can also be directly
read off from the product representation (\ref{Selzetaprod}).

If one substitutes $s\mapsto 1-s$ in (\ref{Selzetaprod}) and 
subtracts the resulting equation from (\ref{Selzetaprod}) one 
obtains the functional equation
\beq
\label{functeq}
Z(1-s)=Z(s)\,\exp\left\{- A\int_0^{s-\frac{1}{2}}u\,\tan(\pi u)\,du
\right\}
\eeq 
upon integrating the difference with respect to $s$. We now introduce
a mean spectral staircase 
\beq
\label{meanstair}
\overline{N}(p):=\int_0^p\tilde d_0 (q)\ dq=\frac{A}{2\pi}\int_0^p
q\,\tanh(\pi q)\ dq\ ,
\eeq
and then evaluate the functional equation (\ref{functeq}) for $s=
\frac{1}{2}-ip$, $p\in\rz$,
\beq
\label{fcteqcrit}
Z\left(\frac{1}{2}+ip\right)=Z\left(\frac{1}{2}-ip\right)\,\exp
\left\{ -2\pi i\,\overline{N}(p)\right\}\ .
\eeq
Thus the function
\beq
\label{Selxi}
\xi (p):=Z\left(\frac{1}{2}-ip\right)\,\exp\left\{-i\pi\overline{N}
(p)\right\}=\left|Z\left(\frac{1}{2}-ip\right)\right|\,\exp\left\{
-i\pi\left[\overline{N}(p)-\frac{1}{\pi}{\rm arg}\,Z(\frac{1}{2}-ip)
\right]\right\}
\eeq
is real when $p\in\rz$ and satisfies $\xi(-p)=\xi(p)$. The zeros of
$Z(s)$ at $s_n=\frac{1}{2}\pm p_n$ appear as zeros of $\xi(p)$ at
$\pm p_n$. Thus, upon increasing $p$ starting at $p=0$, one 
successively passes through all $p_n$'s related to the quantum
energies $E_n=p_n^2+\frac{1}{4}$. Hence, the spectral staircase 
$N(p)$ counts all zeros of the function $\xi(p')$ in the interval
$0\leq p'\leq p$. When passing through $p_n$ the sign of $\xi(p)$
changes according to the multiplicity $d_n$ of the eigenvalue $E_n$,
i.e., $\xi(p)$ is multiplied by $(-1)^{d_n}$. A comparison with
the r.h.s.\ of (\ref{Selxi}) therefore shows that
\beq
\label{stairdcomp}
N(p)=\overline{N}(p)+\frac{1}{\pi}\,\mbox{arg}\,Z \left(\frac{1}{2}
+ip\right)\ ,
\eeq
see also \cite{Hejhal} for a detailed discussion of this relation.
The r.h.s.\ of (\ref{stairdcomp}) can be interpreted as a 
decomposition of the spectral staircase into a mean part and a part
$N_{fl}(p)=\frac{1}{\pi}\,\mbox{arg}\,Z(\frac{1}{2}+ip)$ describing
the fluctuations of the staircase.

At this stage we remark that since the zeros $p_n$ of the function
$\xi (p)$ are related to the quantum energies, one can introduce 
the quantization condition 
\beq
\label{xiquant}
\xi(p)\stackrel{!}{=}0\ \ \ \ \Leftrightarrow\ \ \ \ p^2+\frac{1}
{4}\in\left\{E_0,E_1,E_2,\dots\right\}\ .
\eeq
In \cite{SieStei1,Tanner} methods to evaluate for real $p$ the zeta 
function, or rather its semiclassical analogue for billards,
which use cassical periodic orbits are devised, and then zeros 
of $\mbox{Re}\,\xi(p)$ are calculated. This procedure leads
to good approximations to quantum energies and constitutes an
alternative to the quantization rules presented in section
\ref{sec1.2}.  

In view of the previous discussion one notices that the Selberg zeta
function $Z(s)$ enjoys many properties which are similar to those
of the Riemann zeta function
\beq
\label{Rzeta}
\ze(s):=\sum_{n=1}^\infty\frac{1}{n^s} =\prod_p\left( 1-p^{-s}
\right)^{-1}\ ,\ \ \ \ \mbox{Re}\,s>1\ ,
\eeq
where the product on the r.h.s.\ extends over all primes $p$, see
\cite{Titchmarsh} for a detailed discussion. $\ze (s)$ can be 
extended to a meromorphic function with one simple pole at $s=1$
and zeros on the real axis at $s=-2,-4,-6,\dots$. The infinitely
many complex zeros are located in the strip $0<\mbox{Re}\,s<1$;
according to the famous Riemann hypothesis these are of the form
$s_n=\frac{1}{2}\pm it_n$, $t_n>0$. One can hence introduce the 
counting function
\beq
\label{Rcount}
N_R(T):=\#\left\{ n;\ 0\leq t_n\leq T\right\}=\overline{N}_R (T)+
\frac{1}{\pi}\,\mbox{arg}\,\ze\left(\frac{1}{2}+iT\right)\ ,
\eeq
where $\overline{N}_R (T)=\frac{T}{2\pi}\log\frac{T}{2\pi}-\frac{T}
{2\pi}+\frac{7}{8}+O(\frac{1}{T})$ for $T\rto\infty$. The Riemann
zeta function also satisfies a functional equation,
\beq
\label{Rfuncteq}
\ze(1-s)=\frac{2^{-s}\,\pi^{1-s}}{\Ga(1-s)\,\sin(\frac{\pi s}{2})}
\,\ze (s)\ ,
\eeq
so that
\beq
\label{Rxidef}
\xi_R (t):=\ze\left(\frac{1}{2}-it\right)\,e^{-2\pi i\overline{N}_R
(t)}
\eeq
is real and even for $t\in\rz$, $\xi_R (-t)=\xi_R (t)$.

The close similarity between the Riemann zeta function $\ze(s)$ and 
the Selberg zeta function $Z(s)$, compare in particular 
(\ref{stairdcomp}) and (\ref{Rcount}), suggests to view the complex 
zeros $s_n=\frac{1}{2}\pm it_n$ as analogues of the spectral zeros 
$s_n=\frac{1}{2}\pm ip_n$ of $Z(s)$. Thus the quantities $t_n^2+
\frac{1}{4}$ might serve as `model quantum energies' and their 
distribution might imitate the distribution of the quantum energies 
$E_n=p_n^2+\frac{1}{4}$. 

\xsection{Eigenvalue Statistics and Periodic Orbit Theory}
\label{sec2}
As already mentioned in the Introduction, one of the basic
assumptions in the field of quantum chaos is that one should
be able to observe fingerprints of classical chaos in the
distribution of quantum energies. That is, given the sequence
\beq
\label{specseq}
0\leq E_1\leq E_2\leq E_3\leq\dots
\eeq
of eigenvalues of some quantum Hamiltonian $\hat H$, the
distribution of the quantum energies is expected to follow 
certain universal rules which allow to identify whether the
corresponding classical dynamics are chaotic or regular. In 
order to investigate statistical properties of a spectrum 
suitable quantities that reflect the distribution of eigenvalues
are required. Obviously, the spectral staircase
\beq
\label{Enstair}
N(E):=\#\left\{n;\ E_n\leq E\right\}=\int_0^E d(E')\ dE'
\eeq
encodes all information on spectral statistics and thus turns
out to be a fundamental object for our further purposes. Due to
the trace formula (\ref{densepo}) for the spectral density $d(E)$,
\beq
\label{NofEStairdecomp}
N(E)=\overline{N}(E)+N_{fl}(E)
\eeq
arises as a natural decomposition 
of the staircase function into a part $\overline{N}(E)$ that 
derives from the contribution (\ref{partun1}) of $t=0$ to the trace
formula, and a further contribution $N_{fl}(E)$ reflecting the 
effect of the non-trivial classical periodic orbits. When discussing
the Selberg trace formula we noticed that for certain classically
chaotic systems the contribution of the classical periodic orbits
to the spectral staircase is contained in the phase of the Selberg
zeta function via $N_{fl}(E)=\frac{1}{\pi}\,\mbox{arg}\,Z(\frac{1}
{2}+ip)$ with $E=p^2+\frac{1}{4}$, see (\ref{stairdcomp}). Since
$\overline{N}(E)$ essentially gives the volume of that part of the
classical phase space where $H(p,x)\leq E$, see (\ref{stairsemicl}),
the interesting details of the spectral statistics are contained
in $N_{fl}(E)$; the latter describes the fluctuations of the staircase 
about its mean behaviour $\overline{N}(E)$. It therefore proves 
useful to defold the spectrum, i.e., to introduce the scaled
eigenvalues $x_n:=\overline{N}(E_n)$. By abuse of notation we
now denote the counting function for the $x_n$'s by
\beq
\label{Nofx}
N(x):=\#\left\{n;\ x_n\leq x\right\}=x+N_{fl}(x)\ .
\eeq
The corresponding spectral density then reads
\beq
\label{xdense}
d(x)=\sum_n\de\left(x-x_n\right)=1+d_{fl}(x)\ ,
\eeq
so that the scaled eigenvalues $x_n$ have a unit mean separation.
Defolded spectra can thus be directly compared in their statistical
properties.

\subsection{Measures for the Distribution of Eigenvalues}
\label{sec2.1}    
The basic idea behind spectral statistics is to view the list
$x_1,x_2,x_3,\dots$ of scaled eigenvalues of a quantum Hamiltonian
as a sequence of random events. The spectral staircase function
(\ref{Enstair}) is then considered as a sample function of a 
random process. The ultimate goal therefore is to characterize
and classify all possible such random processes in terms of 
properties of the --classical and/or quantum mechanical-- physical
systems giving rise to the defolded spectra. 

At first one might wonder what can be random about a spectrum
of a fixed quantum Hamiltonian? In order to answer this question
one should recall the mathematical concept of probability. In 
general, the starting point is a probability space, i.e., a
measure space with a normalized measure. Any measurable real valued
function on a probability space is then considered as a random 
variable. Now, $N(x)$ obviously is an integer valued and piecewise
continuous function on the positive real line $\rz_+ =(0,\infty)$.
In order to turn the latter into a probability space one needs to
define a normalized measure. Since $\rz_+$ is unbounded one cannot
simply use Lebesgue measure because it is not finite. Instead,
one restricts attention to a bounded region on the positive real
line, say to an interval $I_x :=[x-\De x,x+\De x]$ of width $2\De x$
that is centered at $x$. In order to obtain information on the
asymptotic distribution of eigenvalues one then has to consider the 
limit $x\rto\infty$. As this is equivalent to the semiclassical
limit $\hbar\rto 0$ the asymptotic analysis of spectral statistics 
allows for an application of semiclassical methods as, e.g., the use
of trace formulae.

In defolded spectra the mean separation of eigenvalues is one so 
that in the mean the intervals $I_x$ contain $2\De x$ quantum 
energies. The latter number should approach infinity in order to 
yield reasonable limit distributions as $x\rto\infty$. We therefore
choose $\De x=c\,x^\al$, with some suitable constants $c>0$ and 
$\al$. If not stated otherwise, $\al$ will be fixed to obey $\frac{1}
{2}<\al <1$. In view of the limit $x\rto\infty$ we intend to perform,
the unboundedness of the spectral staircase suggests an alternative
choice for the random process to be studied as a means to characterize
spectral statistics. One customarily introduces the number
\beq
\label{nsubL}
n_L (y):= N(y+L)-N(y)\ ,\ \ \ \ \ y\in I_x\ ,
\eeq
of eigenvalues in intervals $(y,y+L]$ of length $L>0$. On the interval 
$I_x$, from which $y$ has to be chosen randomly, we make the simplest 
choice for a probability measure, $d\mu_x (y):=\frac{1}{2\De x}\,
\chi_{I_x}(y)\,dy$, where $\chi_{I_x}(y)$ denotes the characteristic 
function of the interval $I_x$. Given an integrable function $f(y)$
on $I_x$ its mean value with respect to $d\mu_x (y)$ is thus
given by
\beq
\label{fmean}
\ez_x [f]:=\int f(y)\ d\mu_x (y)=\frac{1}{2\De x}\int_{x-\De x}^{x
+\De x}f(y)\,dy\ .
\eeq
One can choose as a probability measure on $I_x$ any normalized
measure of the form $w(y)\,dy$, where $w(y)\geq 0$ is continuous
and vanishes outside $I_x$. The limit distributions then induced
by the random variable $n_L(y)$ on $I_x$, for $x\rto\infty$,
do not depend on the choice of $w(y)$, see \cite{Bleher} for
a discussion on this point. We therefore stick to our favourite
choice $d\mu_x(y)$. 

At this point we introduce some notation that will be kept throughout
the rest of this presentation: Above we indicate the interval $I_x$
by a subscript $x$ which we attach to the random variables and their
distributions. When passing to the limit $x\rto\infty$, the 
resulting distributions take care of the asymptotic properties
of a given spectrum. All quantities corresponding to this limit
will then carry no subscript. The same notation will be used in case
a random process is stationary, i.e., no dependence on $x$ occurs
anyway. 
   
Now, for each value of the parameter $L$ the function $n_L(y)$
is a random variable on the interval $I_x$, and thus upon varying
$L$ it defines a continuous parameter random process. The ultimate
aim then is to investigate the resulting limit random process as
$x\rto\infty$ as far as possible. On the way, however, we will
notice that already at finite $x$ one can obtain interesting 
information. As long as $L$ is finite, $n_L(y)$ provides information
on correlations among approximately $L$ eigenvalues and therefore
yields the local spectral statistics. In order to reach the global
scale one has to perform the limit $L\rto\infty$. Starting with
$n_L(y)$ on $I_x$ one is therefore confronted with a competition
of the two limits $x\rto\infty$ and $L\rto\infty$. Phenomena 
occurring in this context are the central topic of the third part 
of these lectures. 

For the further analysis it proves useful to introduce the 
probability $E_x(k;L)$ to observe a value of $k$ for the random 
variable $n_L(y)$ when $y$ is uniformly distributed on the interval
$I_x$,
\beq
\label{ExkLdef}
E_x(k;L):=\ez_x\left[\de_{n_L,k}\right]=\frac{1}{2\De x}\int_{x-
\De x}^{x+\De x}\de_{n_L(y),k}\ dy\ ,
\eeq
where $\de_{k,l}$ denotes Kronecker's symbol. According to 
(\ref{xdense}) and (\ref{fmean}) the mean value of $n_L(y)$ reads
\beq
\label{nsubLmean}
\ez_x[n_L]=\frac{1}{2\De x}\int_{x-\De x}^{x+\De x}\left[N(y+L)
-N(y)\right]\ dy=L+\frac{1}{2\De x}\int_{x-\De x}^{x+\De x}\left[
N_{fl}(y+L)-N_{fl}(y)\right]\ dy\ .
\eeq
A simple result emerges in the limit $x\rto\infty$ because $N_{fl}
(y)$ describes the fluctuations of the spectral staircase about its
mean behaviour $\overline{N}(y)=y$. Thus in particular
\beq
\label{Nflmean}
\lim_{x\rto\infty}\frac{1}{2\De x}\int_{x-\De x}^{x+\De x}N_{fl}
(y)\ dy=0\ ,
\eeq
so that the mean value $\ez_x[n_L]$ asymptotically approaches $L$.
See \cite{PhD} for a discussion of (\ref{Nflmean}) which makes use
of the trace formula. Having in mind the limit $x\rto\infty$ as our 
actual goal, we introduce the $m$-th moments of the shifted random 
variable $n_L(y)-L$,
\beq
\label{Simdef}
\Si^m_x (L):=\ez_x\left[(n_L -L)^m\right]=\frac{1}{2\De x}\int_{x-
\De x}^{x+\De x}\left[N(y+L)-N(y)-L\right]^m\,dy\ .
\eeq
Due to the above remark $\Si^1_x(L)$ vanishes as $x\rto\infty$
so that the lowest non-trivial moment is given by the so-called
number variance
\beq
\label{Numvardef}
\Si^2_x(L)=\frac{1}{2\De x}\int_{x-\De x}^{x+\De x}\left[N(y+L)-N(y)
-L\right]^2\,dy =\sum_{k=0}^\infty (k-L)^2\,E_x (k;L)\ .
\eeq
One can easily determine the behaviour of the number variance for
small $L$: Since $I_x$ contains finitely many scaled eigenvalues
there exists a positive minimal separation $s_n:=x_{n+1}-x_n$ of
neighbouring levels. Once $L$ is smaller than the minimal $s_n$
every interval $(y,y+L]$ with $y\in I_x$ either contains one or
no $x_n$, and thus $E_x (k;L)=0$ for $k\geq 2$. One then employs
the normalization of the probabilities, $E_x (0;L)+E_x(1;L)=1$,
to observe
\beq
\label{Si2limit}
\Si^2_x(L)=\sum_{k=0}^1 (k-L)^2\,E_x(k;L)=(1-2L)\,\ez_x [n_L]
+L^2\ .
\eeq
The relation (\ref{nsubLmean}) then implies that
\beq
\label{Si2limit1} 
\Si^2_x(L)\sim L\ ,\ \ \ \ \ L\rto 0\ .
\eeq
The behaviour of $\Si^2_x(L)$ for small $L$ does not provide
detailed information on the distribution of eigenvalues since for
too small $L$ the number variance merely reflects the fact that
$N(x)$ is a staircase function. 

The opposite limit $L\rto\infty$ cannot so easily be dealt with, 
but on the other hand reflects specific properties of the underlying 
physical system. From (\ref{Numvardef}) we obtain
\beqa
\label{Numvarform}
\Si^2_x(L)&=&\frac{1}{2\De x}\int_{x-\De x}^{x+\De x}\left[N(y+L)-
             N(y)-L\right]^2\,dy\nonumber\\
          &=&\ez_{x+L}\left[N_{fl}^2\right]+\ez_{x}\left[N_{fl}^2
             \right]-2\ez_{x}\left[N_{fl}(y+L)N_{fl}(y)\right]\ .
\eeqa
Certainly, the limit $L\rto\infty$ also requires to perform $x\rto
\infty$, since otherwise the intervals $(y,y+L]$ for $y\in I_x$
hardly overlap with $I_x$; by the same reason $L$ should not
grow faster than $x$. We therefore choose $L=L(x)=a\,x^\ga$ with 
some positive constants $a$ and $\ga\leq 1$. Now two effects on 
the behaviour of $\Si^2_x(L)$ in this limit are to be observed on
the r.h.s.\ of (\ref{Numvarform}). On the one hand, a large exponent
$\ga$ will decouple $N_{fl}(y+L)$ and $N_{fl}(y)$ to a certain
extent so that the contribution of the correlation function on the
r.h.s.\ of (\ref{Numvarform}) becomes smaller in absolute value.
On the other hand, a small exponent $\ga$ will turn the two 
expectation values of $N_{fl}^2$ almost equal. However, in the 
asymptotic regime for $x\rto\infty$, $\ga$ need not really be small 
because once $\ez_x [N_{fl}^2]$ increases for $x\rto\infty$, then 
$\ez_{x+L}[N_{fl}^2]\sim\ez_{x}[N_{fl}^2]$ for any choice $L=a\,x^\ga$ 
with $\ga<1$. It indeed turns out that in typical cases the 
expectation values of $N_{fl}^2(y)$ on $I_x$ behave either like
\beq
\label{expectasym}
\ez_{x}\left[N_{fl}^2\right]\sim b_1\,\log x\ \ \ \ \ \mbox{or}
\ \ \ \ \ \ez_{x}\left[N_{fl}^2\right]\sim b_2\,x^\rho\ ,
\eeq
with some $\rho>0$. We hence expect that as long as $\ga$ is larger
than some critical exponent $\ga_c$ the number variance is asymptotically 
given by the dominant and $L$-independent contribution $2\,\ez_{x}
[N_{fl}^2]$, with $L$-dependent modifications provided by the correlation
function $\ez_{x}[N_{fl}(y+L)N_{fl}(y)]$. Theoretical \cite{BerryN}
as well as numerical \cite{BerryN,AurStei} studies indeed confirm
this qualitative picture. In section \ref{sec2.3} we will return to a 
more detailed discussion of this point.

The characteristic function of a probability distribution is
defined as the Fourier transform of the probability measure. In the 
case of the random variable $n_L(y)$ it is given by
\beq
\label{chardef}
J_{n_L,x}(\xi):=\ez_x\left[e^{i\xi n_L}\right]=\frac{1}{2\De x}
\int_{x-\De x}^{x+\De x}e^{i\xi n_L(y)}\,dy\ .
\eeq
The probabilities $E_x(k;L)$, see (\ref{ExkLdef}), can be recovered 
from the characteristic function because
\beqa
\label{probrecov}
E_x(k;L)&=&\frac{1}{2\De x}\int_{x-\De x}^{x+\De x}\de_{n_L(y),k}\,dy
           =\frac{1}{2\De x}\int_{x-\De x}^{x+\De x}\frac{1}{2\pi}
           \int_{-\pi}^{+\pi}e^{i\xi[n_L(y)-k]}\,d\xi\ dy\nonumber \\
        &=&\frac{1}{2\pi}\int_{-\pi}^{+\pi}e^{-i\xi k}\,J_{n_L,x}(\xi)
           \,d\xi\ .
\eeqa
Hence the knowledge of the characteristic function is equivalent 
to knowing the probabilities $E_x(k;L)$ themselves. Characteristic 
functions can be used to test the statistical indepedence of random
variables since the characteristic function of the sum of independent
random variables factorizes into the product of the respective
characteristic functions of the individual random variables. In our
case, $n_L(y)=L+N_{fl}(y+L)-N_{fl}(y)$ essentially is the sum of 
the two random variables $N_{fl}(y+L)$ and $-N_{fl}(y)$, with $y\in
I_x$. If one wants to test their statistical independence one 
therefore has to study whether or not the characteristic function
(\ref{chardef}) factorizes. Certainly, a complete statistical indepence
can only be expected for the limit distributions as $x\rto\infty$,
since only these take infinitely many eigenvalues into account. 
As argued above, a decoupling of the distributions of $N_{fl}(y+L)$ 
and $N_{fl}(y)$ will then require to perform also $L\rto\infty$,
with $L=a\,x^\ga$ and a positive exponent $\ga$ which is large 
enough. One way to determine the minimal such $\ga$, which then
yields $\ga_c$, could be to test the factorization of the characteristic 
function $J_{n_L,x}(\xi)$. At the end of section \ref{sec2.3} we will 
return to this topic in some more detail.

\subsection{Random Matrix Theory}
\label{sec2.2}
So far we discussed the distribution of eigenvalues of individual
quantum Hamiltonians, without specifying any result about, e.g.,
the probabilities $E_x(k;L)$. However, we already addressed the
expectation to observe fingerprints of the classical dynamics
on the corresponding distribution of quantum energies. Ideally,
this could mean a universal behaviour of the probabilities
$E_x(k;L)$, as $x\rto\infty$ of course, depending only on the
type of the classical dynamics. For example, one could expect
that all classically chaotic systems lead to one and the same
limit distribution for their respective random variables $n_L$;
classically integrable systems then should produce a distinctively
different distribution. This expectation being very strong could
be weakened in replacing `all systems' by `almost all systems'
in order to account for `special' or `non-generic' cases where
exceptions might occur. Of course, one then has to specify what
`almost all' means in this context. To this end it is
necessary to consider a family of quantum Hamiltonians $\hat 
H_\la$, where $\la$ ranges over some parameter space $\La$. Since
now one is dealing with families $\{E_1(\la),E_2(\la),E_3(\la),
\dots\}$ of quantum spectra all spectral quantities depend on the
parameter $\la$, e.g., $n_{L,\la}(y)$, $E_{x,\la}(k;L)$. Provided
the parameter space $\La$ is endowed with a suitable measure
$d\nu(\la)$, a weak form of universality in spectral statistics 
would be reflected in the fact that the limit probabilities
$E_\la(k;L)$, arising from $E_{x,\la}(k;L)$ as $x\rto\infty$,
are given by $E_\La (k;L)$ for almost all $\la\in\La$ with
respect to the measure $d\nu(\la)$. If the latter were normalized,
$\int_\La d\nu(\la)=1$, i.e., it were a probability measure, then 
one would obtain
\beq
\label{EKLmean}
E_\La (k;L)=\int_\La E_\La (k;L)\,d\nu(\la)=\int_\La 
E_\la(k;L)\,d\nu(\la)\ .
\eeq
Thus, even in case it is not known whether $E_\la(k;L)=E_\La
(k;L)$ for almost all $\la\in\La$ it would be interesting to 
calculate the mean value on the r.h.s.\ of (\ref{EKLmean}) which
describes the spectral statistics of the quantum Hamiltonians
$\hat H_\la$ on average. In this perspective the limit of 
$E_{x,\la}(k;L)$ as $x\rto\infty$ may even not exist for a set
of $\la$ which is of zero measure in $\La$. 

Examples for families of classically chaotic systems to which
the above programme might be applied are provided by geodesic motions
on surfaces of constant negative curvature as discussed in section
\ref{sec1.3}. The compact such surfaces with the topology of a sphere
with $g$ handles ($g\geq 2$) form a $(6g-6)$-parameter family. For
$g=2$ Aurich and Steiner \cite{AurStei1} calculated numerically
eigenvalues of quantum Hamiltonians on $30$ surfaces which they picked
out randomly from the corresponding $6$-parameter family, and computed
averages of the probabilities $E_{x,\la}(k;L)$ over their sample
of $30$ values for the parameter $\la$. Apart from the fact that $x$
necessarily had to be kept finite and thus the saturation effects to 
be discussed in section \ref{sec2.3} showed up, it was observed that 
the `ensemble averages' agreed with the respective probabilities in the 
Gaussian orthogonal ensemble (GOE) of random matrix theory. We remark 
that in this example it turns out to be essential not to expect the 
GOE result for all values of $\la$ since certain exceptions, which go 
under the notion of arithmetic quantum chaos \cite{Bogo,Steil,Luo}, 
are known to exist. However, with respect to any reasonable measure 
$d\nu(\la)$ these will form a set of zero measure. 

In a purely mathematical model the programme has recently been 
completely carried out by Katz and Sarnak \cite{KS}. As model 
operators they consider matrices from the classical compact
Lie groups $O(N)$, $SO(N)$, $U(N)$, $SU(N)$, and $USP(N)$. Then 
the respective group itself can serve as the parameter space
$\La$, and the most natural choice for $d\nu(\la)$ is provided
by the Haar measure on the group. Katz and Sarnak prove that in the 
limit $N\rto\infty$ almost all matrices from any one of the classical
groups show the spectral statistics of the Gaussian unitary
ensemble (GUE) of random matrix theory. 

In random matrix theory, see \cite{Mehta} and Bohigas' contribution
to these proceedings, one attempts to construct from certain 
assumptions the probabilities $E_\La (k;L)$ which are expected 
for `almost all' quantum Hamiltonians of a given type. In this 
context the first step consists of cutting off the dimension of
the Hilbert space on which a Hamiltonian $\hat H$ is defined at a 
finite value $N$. Then, upon choosing some orthonormal basis (onb) 
in the Hilbert space, $\hat H$ is represented by a hermitian 
$N\times N$-matrix $\hz$. In case the physical system from which 
$\hz$ derives is time-reversal invariant one can always choose the 
onb in such a way that $\hz$ is real symmetric. For simplicity we
will in the following always focus our attention to this case. 
One then still has the freedom to choose another onb without 
changing $\hat H$ and its spectral properties; only the matrix 
representation is affected. The base changes in the $N$-dimensional 
Hilbert space are provided by orthogonal transformations $O\in O(N)$. 
The matrix $\hz$ is thus conjugated to $O^{-1}\hz O$. As a model 
space of truncated Hamiltonians one therefore
considers the set of all real symmetric matrices $\hz$, with an
identification of $\hz$ and $O^{-1}\hz O$ whenever $O\in O(N)$.
Altogether a real symmetric $N\times N$-matrix has $\frac{1}{2}
N(N+1)$ independent entries. The symmetry group $O(N)$ is of
dimension $\frac{1}{2}N(N-1)$ so that the set of classes of 
equivalent real symmetric matrices has dimension $N$. A possible
choice for coordinates of the space of truncated Hamiltonians is
obtained after diagonalizing each $\hz$: There exists an $O\in O(N)$
such that $O^{-1}\hz O$ is diagonal. Therefore each $\hz$ is 
equivalent to a diagonal matrix, with its eigenvalues $E_1,\dots,
E_N$ as entries on the diagonal. The eigenvalues then yield $N$ 
parameters to describe $\hz$, and the additional $\frac{1}{2}N(N-1)$
coordinates required to characterize $\hz$ completely are provided
by the independent entries of $O$.   

The basic assumption of random matrix theory now concerns the 
measure $d\nu_N (\hz)$ on the space of real symmetric $N\times 
N$-matrices. It is based on the principle of minimal knowledge 
and consists of two parts: (i) $d\nu_N (\hz)=P_N (\hz)
\,d\hz$, where $d\hz=\prod_{k\leq l}d\hz_{kl}$, and $P_N (\hz)=P_N
(O^{-1}\hz O)$, so that the measure does not depend on a particular
choice of an onb. (ii) The independent entries $\hz_{kl}$, $k\leq 
l$, are statistically independent. By these requirements the measure
$d\nu_N (\hz)$ is already fixed to read \cite{Mehta}
\beq
\label{measureN}
d\nu_N (\hz)=const.\  \exp\left\{-a\,\mbox{tr}\,\hz^2+b\,\mbox{tr}
\,\hz\right\}\,d\hz\ ,
\eeq
with some real constants $a>0$ and $b$. In order the obtain a more
explicit expression one changes variables from $\hz_{kl}$, $k\leq 
l$, to $E_1,\dots,E_N$ and $\frac{1}{2}N(N-1)$ further variables
which parametrize the orthogonal matrix $O$ that diagonalizes $\hz$.
One finally observes \cite{Mehta}
\beq
\label{GOEN}
d\nu_N (\hz)=d\rho(O)\,\prod_{k<l}\left|E_k-E_l\right|\,\prod_{j=1}^N
e^{-aE_j^2}\,dE_j\ ,
\eeq
where $d\rho(O)$ denotes some measure that only depends on $O$. 
Upon integrating out $d\rho(O)$ one has thus obtained a measure on
the model space of truncated Hamiltonians, where the latter are being 
parametrized by their eigenvalues. Notice that the eigenvalues are 
not statistically independent since the measure does not factorize. 
The term $\prod_{k<l}|E_k-E_l|$ rather introduces a correlation of the 
eigenvalues that makes small separations $E_{k+1}-E_k$ of neighbouring 
eigenvalues less probable. This effect is known as level repulsion.

The limit measure resulting from (\ref{GOEN}) as $N\rto\infty$ is 
commonly called Gaussian orthogonal ensemble (GOE). Given the latter 
one can then calculate all statistical properties of eigenvalues of 
random symmetric matrices. For example, the number variance reads
\beqa
\label{GOEnumvar}
\Si^2_{GOE}(L)&=&\frac{2}{\pi^2}\left[\log(2\pi L)+\ga+1-\cos(2\pi L)
                 -\mbox{Ci}(2\pi L)\right]\nonumber\\
              & &+2L\left[1-\frac{2}{\pi}\mbox{Si}(2\pi L)\right]
                 +\frac{1}{\pi^2}\mbox{Si}(\pi L)^2-\frac{1}{\pi}
                 \mbox{Si}(\pi L)\\
              &=&\frac{2}{\pi^2}\left[\log(2\pi L)+\ga+1-\frac{\pi^2}
                 {8}\right]+O\left(\frac{1}{L}\right)\ ,\nonumber
\eeqa
where
\beq
\label{intsin}
\mbox{Si}(x):=\int_0^x\frac{\sin y}{y}\,dy\ \ \ \ \mbox{and}\ \ \ \ 
\mbox{Ci}(x):=\ga+\log x+\int_0^x\frac{\cos y -1}{y}\,dy\ .
\eeq
Here $\ga=0.577215\dots$ denotes Euler's constant. For the GOE one 
does not know, however, an explicit closed expression for the 
probabilities $E_{GOE}(k;L)$ to find $k$ defolded eigenvalues in 
spectral intervals of length $L$; but these quantities can be 
calculated numerically to a high precision. That way Aurich and 
Steiner \cite{AurStei1} found
\beq
\label{EkLasym}
E_{GOE}(k;L)\sim\frac{1}{\sqrt{2\pi\Si^2_{GOE}(L)}}\,e^{-\frac{(k-L)^2}
{2\Si^2_{GOE}(L)}}\ ,\ \ \ \ L\rto\infty\ .
\eeq
Later Costin and Lebowitz \cite{CL} made precise the way the limit of 
large $L$ has to be understood. Since both the mean value $\ez_{GOE}
[n_L]=L$ and the variance (\ref{GOEnumvar}) of the random variable 
$n_L$, when considered in the GOE, diverge as $L\rto\infty$ one has 
to employ a renormalization before one performs the limit. Costin and
Lebowitz proved that after renormalization the distribution of the 
random variable obtained from $n_L$ weakly approaches a standard 
Gaussian, see section \ref{sec3.1} for a further discussion.

The construction of the GOE relies on a number of assumptions that
cannot be derived from first principles when one is discussing
the distribution of eigenvalues of actual quantum Hamiltonians.
Historically, random matrix theory was introduced to describe
the statistics of energy levels and of resonances of large atomic
nuclei. In view of the huge number of degrees of freedom involved 
the principle of minimal knowledge seemed to be reasonable in this 
context. It indeed turned out to result in a correct description
of experimental data, see \cite{Porter} for a collection of the
original contributions. Later Bohigas, Giannoni, and Schmit
\cite{BGS} conjectured that the eigenvalue statistics of quantum
Hamiltonians with even a low number ($\geq 2$) of degrees of
freedom universally follow the predictions of random matrix theory, 
if only the corresponding classical dynamics were chaotic. Again, 
the conjecture found confirmation by numerical calculations of
eigenvalues and their statistics in many examples.  

For classically integrable systems the situation appears to be 
completely different. Since at least a second classical integral
of motion exists one has a quantum mechanical observable at hand that 
commutes with the Hamiltonian. For each value of the quantum number 
that corresponds to the second conserved quantity one obtains a
subspectrum of the quantum energy spectrum. It now seems that the 
entire energy spectrum behaves like an independent superposition 
of these, typically infinitely many, subspectra. Provided the
superposition of subspectra were indeed uncorrelated, the complete
spectrum would have the local statistics of a Poissonian random
process, i.e.,
\beq
\label{PoissonEkl}
E_{Poisson}(k;L)=\frac{L^k}{k!}\,e^{-L}\ .
\eeq
In particular, the density $P(s)$ of the distribution of nearest
neighbour level spacings $s_n:=x_{n+1}-x_n$ of defolded eigenvalues
would be exponential, $P_{Poisson}(s)=e^{-s}$. Based on a semiclassical
analysis the Poissonian behaviour of the local eigenvalue statistics
was predicted by Berry and Tabor \cite{BerryTabor} and meanwhile has 
been verified by numerical calculations in many examples. See e.g.\ 
\cite{ChengLeb} for a careful numerical analysis.

But again, exceptions are known to exist. In order to give an 
example suppose that $e_1,e_2\in\rz^2$ are linearly independent 
such that $\cL=\gz e_1+\gz e_2$ is a lattice in $\rz^2$. Then $\rz^2/
\cL$ is a two dimensional flat torus. The classical geodesic motion 
on $\rz^2/\cL$ is integrable, and the spectrum of $\hat H=-
\frac{\hbar^2}{2m}\De$ on $L^2(\rz^2/\cL)$ is explicitly known; it
comprises of the eigenvalues $E_{k,l}=\frac{\hbar^2}{2m}4\pi^2
(kf_1+lf_2)^2$. Here $k$ and $l$ run through $\gz$, and $\{f_1,
f_2\}$ is a basis for the dual lattice $\cL^*$ of $\cL$, i.e., 
$e_i\cdot f_j=\de_{ij}$. Hence
\beq
\label{Quadform}
E_{k,l}=\frac{\hbar^2}{2m}\,Q(k,l)=\frac{\hbar^2}{2m}\left[A\,k^2+
B\,kl+C\,l^2\right]\ ,
\eeq
where $A,B,C$ are suitable real constants which determine $Q(k,l)$
to be a positive definite binary quadratic form. If $\frac{B}{A}$
and $\frac{C}{A}$ are rational numbers, $Q(k,l)$ is a rational 
quadratic form. In this case the value of $Q(k,l)$ itself is a
rational number for all $k,l\in\gz$; it can in particular be
an integer, $Q(k,l)=n$, $n\in\nz$. If we denote the number of 
pairs $(k,l)\in\gz^2$ that represent $n$ as $n=Q(k,l)$ by 
$r_Q(n)$, the multiplicity of the eigenvalue $E=\frac{\hbar^2}
{2m}\,n$ is $r_Q(n)$. Now, for rational forms it is known that   
$r_Q(n)$ is unbounded as $n\rto\infty$. One can then conclude that
the distribution of level spacings becomes singular at $s=0$;
it is actually given by the `density' $P(s)=\de(s)$, see for
example \cite{Sarnak}.

The coefficients $A,B,C$ in (\ref{Quadform}) provide parameters
for the set $\La$ of all quantum Hamiltonians $\hat H_\la$ that
derive from Laplacians on flat tori. The positivity of the 
eigenvalues only requires the discriminant $D_Q=B^2-4AC$ to be
negative. The set of Hamiltonians in $\La$ corresponding to rational 
quadratic forms then clearly is of zero Lebesgue measure, so that
the tori leading to the singular distribution of level spacings
as described above are non-generic in a well defined sense. It
is widely believed, however not known, that for a set of full
Lebesgue measure in $\La$ the level spacings distribution has an
exponential density, $P(s)=e^{-s}$. As an important step towards
a proof of Poissonian local eigenvalue statistics for Laplacians
on flat tori Sarnak recently proved \cite{Sarnak} that the spectral 
statistics, when measured with the pair correlation function, is 
Poissonian on a set of full Lebesgue measure in $\La$. Moreover,
if $B=0$ and $\frac{C}{A}$ satisfies a certain Diophantine condition,
Bleher and Lebowitz \cite{BleLeb} proved that the number variance $\Si^2
(L)$ is Poissonian. In view of the above discussion a further result of 
Sarnak is rather amazing. He could show that there exists a set 
$\tilde\La\subset\La$ of tori, which is generic in a topological 
sense, and on which the level spacings distribution has no exponential 
density; in fact $\int_0^{1/3}P_\la(s)\,ds$ does even not exist for 
$\la\in\tilde\La$. The set $\tilde\La$ is generic in the sense that 
it is of second Baire category, i.e., it is a countable intersection 
of open dense sets. The lesson to learn from this example is that 
universality in spectral statistics can only hold in a weak form, as 
for example expressed around (\ref{EKLmean}). One even has to be very
careful in the way one is declaring particular quantum Hamiltonians 
as `generic'.

\subsection{A Semiclassical Analysis of Spectral Statistics}
\label{sec2.3}
It seems that the only tools available to investigate the distribution
of eigenvalues of a single quantum Hamiltonian $\hat H$ are provided
by semiclassical trace formulae. These relate the spectrum of 
$\hat H$ to classical quantities: periodic orbits and their stabilities
and Maslov phases. One therefore might anticipate that the asymptotic 
distribution of eigenvalues is completely determined by the corresponding
classical dynamics. The first to perform such a semiclassical analysis, 
to classically integrable systems, were Berry and Tabor 
\cite{BerryTabor}. Later, Berry extended this analysis to classically 
chaotic systems as well, see \cite{A400,BerryN} and his contribution to
\cite{LesHouches}. Here we only want to consider quantum systems with 
a strongly chaotic classical limit, which shall in particular mean that 
all classical periodic orbits be unstable and isolated in phase space. 
To be definite, we further restrict our attention to cases where the 
trace formula (\ref{traceeven}) applies, and moreover, the spectral 
density has an asymptotic expansion as in (\ref{dbarasym}). Thus, 
the classical dynamics are given by either billiards or geodesic 
motions on surfaces. The case of the Selberg trace formula (\ref{STF}) 
is therefore included.

The starting point for our discussion is the representation
\beq
\label{nLrep}
n_L(y)=N(y+L)-N(y)=\int_y^{y+L}d(u)\ du=L+\int_y^{y+L}d_{fl}(u)\ du
\eeq
of the random variable $n_L(y)$, $y\in I_x$, in terms of $d_{fl}(u)$,
compare (\ref{xdense}). Now, the fluctuating part of the spectral
density, when expressed in terms of the momentum variable $p$, can be
given in terms of a periodic orbit sum as in (\ref{poGauss}). The 
number variance, being the second moment of $n_L(y)-L$, can hence
be represented as
\beqa
\label{Si2rep}
\Si_x^2(L)&=&\frac{1}{2\De x}\int_{x-\De x}^{x+\De x}\left[n_L(y)-L
             \right]^2\,dy\nonumber\\
          &=&\frac{1}{2\De x}\int_{x-\De x}^{x+\De x}\int_y^{y+L}
             \int_y^{y+L}d_{fl}(u_1)\,d_{fl}(u_2)\ du_1\,du_2\ dy.
\eeqa
To proceed further, one customarily introduces the two-level form factor
\beq
\label{Kdef}
K_x(\tau):=\int_{-\infty}^{+\infty}e^{2\pi it\tau}\frac{1}{2\De x}
\int_{x-\De x}^{x+\De x}d_{fl}(y)\,d_{fl}(y+t)\ dy\ dt .
\eeq
By Fourier inversion and a change of variables one obtains from
(\ref{Si2rep}) and (\ref{Kdef}) that
\beq
\label{Si2K}
\Si^2_x(L)=\int_{-\infty}^{+\infty}d\tau\int_0^L d\rho_1\int_0^L
d\rho_2\,e^{-2\pi i\tau(\rho_1-\rho_2)}\,K_{x+\rho_1}(\tau)\ .
\eeq
As we are finally interested in the limit of $x\rto\infty$, with 
$0\leq\rho_1\leq L\ll x$, we approximate $K_{x+\rho_1}(\tau)\sim
K_x(\tau)$ and observe
\beq
\label{Si2Kfin}
\Si_x^2(L)\sim\frac{2}{\pi^2}\int_0^\infty\frac{\sin^2(\pi L\tau)}
{\tau^2}\,K_x(\tau)\,d\tau\ .
\eeq
After having performed $x\rto\infty$ the above asymptotic relation
indeed turns into an identity. When differentiating the latter with
respect to $L$ and applying a sine-Fourier inversion afterwards,
one obtains \cite{AurStei}
\beq
\label{Kinv}
K(\tau)=\left.\frac{d}{dL}\Si^2(L)\right|_{L=0}-\frac{1}{2\pi\tau}
\int_0^\infty\sin(2\pi L\tau)\,\frac{d^3}{dL^3}\Si^2(L)\ dL\ ,
\eeq
from which one concludes that
\beq
\label{Kasym}
K(\tau)=1+O\left(\frac{1}{\tau}\right)\ ,\ \ \ \ \tau\rto\infty\ ,
\eeq
since $\Si^2(L)\sim L$ for $L\rto 0$, see (\ref{Si2limit1}).

We are now going to express the form factor in terms of sums over
classical periodic orbits. To this end $d_{fl}(x)$ in (\ref{Kdef})
has to be represented by the periodic orbit sum (\ref{poGauss}).
However, since the form factor (\ref{Kdef}) is given in terms of
the defolded spectral variable $x=\overline{N}(E)$, we first have
to rewrite it as a function of the energy $E=p^2$. From the relation
(\ref{dbarasym}) we conclude that $\overline{N}(E)\sim\frac{A}{4\pi
\hbar^2}E$ as $E\rto\infty$ or $\hbar\rto 0$, and thus $\overline{d}
(E)\sim\frac{A}{4\pi\hbar^2}=:\overline{d}$. We therefore observe that
\beq
\label{dflucrel}
d_{fl}(x)=\frac{d}{dx}\,N_{fl}(x)=\frac{dE}{dx}\,\frac{dN_{fl}(E)}
{dE}=\frac{d_{fl}(E)}{\overline{d}(E)}\sim\overline{d}^{-1}\,d_{fl}
(E)\ .
\eeq
The two-level form factor, when expressed by $E$ instead of $x$, hence 
reads 
\beq
\label{KofE}
K(\tau;E)\sim\overline{d}^{-1}\int_{-\infty}^{+\infty}e^{2\pi
i\tau\overline{d}\la}\,\frac{1}{2\De E}\int_{E-\De E}^{E+\De E}
d_{fl}\left(E'-\frac{\la}{2}\right)\,d_{fl}\left(E'+\frac{\la}{2}
\right)\ dE'\ d\la\ ,
\eeq
where we now correlate $d_{fl}$ at the symmetrically placed points 
$E'\pm\frac{\la}{2}$ instead of $E'$ and $E'+\la$.

We are now in a position to employ the semiclassical periodic orbit
sum (\ref{poGauss}) for $\tilde d_{fl}(p)=2p\,d_{fl}(E)$, $p=\sqrt{E}$,
in (\ref{KofE}). The result, which was first obtained by Berry 
\cite{A400}, then reads
\beq
\label{Kpot}
K(\tau;E)\sim\frac{4p^2}{l_H^2}\left<\sum_{l_n,l_m}\sum_{r,s\neq 0}
A_{n,r}\,A_{m,s}\ e^{\frac{i}{\hbar}p'(rl_n-sl_m)}\ \de\left(\tau-
\frac{rl_n+sl_m}{2l_H}\right)\right>\ ,
\eeq
where
\beq
\label{Ampl}
A_{n,r}=\frac{g_nl_n}{2p}\,\frac{e^{-i\frac{\pi}{2}r\tilde\mu_n}}{|\det(
M_n^r-\unmat)|^{\frac{1}{2}}}\sim\frac{g_nl_n}{2p}\,e^{-i\frac{\pi}{2}
r\tilde\mu_n}\,e^{-\frac{r}{2}u_n}\ ,\ \ \ \ l_n\rto\infty\ ,
\eeq
is an amplitude attached to the $r$-fold repetition of the primitive
length $l_n$, which is of multiplicity $g_n$. The inner double sum in
(\ref{Kpot}) extends over all non-zero integers $r,s$, both positive 
and negative. The braces $<\dots>$ appearing in (\ref{Kpot}) denote 
the average in $E'$ over the interval $[E-\De E,E+\De E]$. Furthermore, 
the quantity $l_H:=4\pi p\hbar\overline{d}$ is introduced. Apart from
the factor of $4\pi$, which is subject to convention, $l_H$ is a
combination of quantities which has the dimension of a length and 
can be used as a `semiclassical parameter' in that the semiclassical 
limit corresponds to $l_H\rto\infty$. (The subscript refers to the fact 
that $l_H$ is sometimes called `Heisenberg length'.) Notice that
in the discussion following (\ref{cospo}) $l_H$ has been identified
as the length where the periodic orbit sum (\ref{cospo}) can be 
truncated without leaving out its most important contributions.
Following this philosophy, we therefore also truncate the periodic
orbit sums in (\ref{Kpot}) at $|rl_n|,|sl_m|\leq l_H$. Then obviously
the resulting expression can at most represent the form factor in
the range $|\tau|\leq 1$; for larger values of $|\tau|$ the truncated
periodic orbit sum yields no contributions. But already in the interval
$\frac{1}{2}<\tau<1$ some terms are missing so that a reliable
statement can in fact only be made for $|\tau|\leq\frac{1}{2}$. Since 
now we are dealing with finite sums the semiclassical representation 
of $K(\tau;E)$ is no longer plagued with convergence problems. The 
price to pay for this convenient fact is the restriction to $|\tau|
\leq\frac{1}{2}$, and the inaccuracies which derive from the omission 
of the -infinite- tails of the periodic orbit sums. However, now the 
non-diagonal terms on the r.h.s.\ of (\ref{Kpot}), with $rl_n\neq 
sl_m$, are suppressed by the average $<\dots>$ over $E'$, or 
equivalently over $p'$. In the limit $E,\De E\rto\infty$ the 
non-diagonal terms actually vanish. We have therefore arrived at 
the semiclassical representation
\beq
\label{Ksc1}
K(\tau;E)\sim K_{diag}(\tau;E):=
\frac{4p^2}{l_H^2}\sum_{l_n}\sum_{r\geq 1\atop rl_n
\leq l_H}A_{n,r}^2\,\de\left(\tau-\frac{rl_n}{l_H}\right)
\eeq
for $0\leq\tau\leq\frac{1}{2}$ and $E\rto\infty$. The subscript
refers to the fact that $K_{diag}(\tau;E)$ is solely defined by
the diagonal terms of the periodic orbit sums. Due to the exponential
proliferation of the number of periodic orbits $\ga$ with lengths
$l_\ga\leq l$, see (\ref{Huberl}), the sum over $r\geq 1$ is dominated 
by its contribution from $r=1$. If we moreover substitute for the
amplitude $A_{n,1}$ its leading asymptotic behaviour (\ref{Ampl})
we obtain
\beq
\label{Ksc2}
K(\tau;E)\sim\frac{1}{l_H^2}\sum_{l_n\leq l_H}g_n^2\,l_n^2\,e^{-u_n}
\,\de\left(\tau-\frac{l_n}{l_H}\right)\ .
\eeq 
The sum over the length spectrum in (\ref{Ksc2}) will now be rewritten
as a sum over periodic orbits, with the effect that one factor of $g_n$
is being absorbed by the sum. We furthermore now assume that the
multiplicities $g_n$ of lengths $l_n$ asymptotically approach a 
constant value of $\overline{g}$. Thus
\beq
\label{Ksc3}
K(\tau;E)\sim\frac{\overline{g}}{l_H^2}\sum_{\ga,\,l_\ga\leq l_H}
l_\ga^2\,e^{-u_\ga}\,\de\left(\tau-\frac{l_\ga}{l_H}\right)\ .
\eeq
We then integrate $K(\tau';E)$ with respect to $\tau'\in [0,\tau]$,
$\tau\leq\frac{1}{2}$,
\beq
\label{Kint}
\int_0^\tau K(\tau',E)\,d\tau'\sim\frac{\overline{g}}{l_H^2}\sum_{\ga,
\,l_\ga\leq l_H\tau}l_\ga^2\,e^{-u_\ga}\ .
\eeq
Periodic orbit sums like the one appearing in (\ref{Kint}) frequently
occur in the thermodynamic formalism for hyperbolic dynamical systems,
see for example \cite{PP}. The asymptotic behaviour as $l_H\rto\infty$ 
of the r.h.s.\ of (\ref{Kint}) can be obtained from relations known
in the thermodynamic formalism, which relate the various dynamical 
entropies and exploit properties of the topological pressure $P(\be)$, 
see (\ref{Pofbeta}). For hyperbolic systems with compact phase space
one can in particular show that $P(1)=0$, and from this result one 
can conclude that (\ref{Kint}) yields $\frac{\overline{g}}{2}\tau^2$, 
so that after differentiation with respect to $\tau$,
\beq
\label{Ksc4}
K(\tau;E)\sim\overline{g}\,\tau\ ,\ \ \ \ \tau\leq\frac{1}{2}\ ,
\ \ \ \ E\rto\infty\ .
\eeq
The linear behaviour of the form factor was first obtained by 
Berry \cite{A400}. He concluded the result (\ref{Ksc4}) from 
(\ref{Ksc3}) with the help of a sum rule due to Hannay and Ozorio De
Almeida \cite{HOdA}.

Above we derived the general behaviour (\ref{Kasym}) of the form factor
as $\tau\rto\infty$. Since it must approach one, the linear increase
(\ref{Ksc4}) cannot extend beyond $\tau\approx\frac{1}{2}$. We recall 
that the restriction to $\tau\leq\frac{1}{2}$ resulted from the 
truncation of the periodic orbit sums at $l_H$. It now becomes obvious 
that for large $\tau$ the r.h.s.\ of (\ref{Kpot}) has to behave in a 
way that cannot be controlled by as simple semiclassical considerations 
as for small $\tau$. A further observation that can be made with the 
relation (\ref{Ksc4}) is that the behaviour of $K(\tau;E)$ depends 
on the average asymptotic multiplicities of lengths of periodic 
orbits. In `generic' classical dynamical systems with time-reversal 
invariance one obtains that $\overline{g}=2$ because a periodic orbit 
and its time-reversed image share the same length. Here `generic' means, 
among other things, that the classical dynamics have been completely 
desymmetrized, i.e., all discrete symmetries have been removed; a 
multiplicity of $g_n\geq 3$ is then of an accidental kind. And, moreover, 
only if an orbit of length $l_n$ is self-retracing, one observes $g_n=1$. 
Typically, the asymptotic fraction of self-retracing orbits vanishes 
so that indeed `generically' $\overline{g}=2$. For small $l_n$, 
however, effects of multiplicities $g_n=1$ remain. Hence, the relation 
(\ref{Ksc4}) can only apply once one is deep enough in the 
semiclassical regime, i.e., for sufficiently large $l_H$. When the 
classical dynamics shows no time-reversal invariance, one typically 
finds $g_n=1$ so that then $\overline{g}=1$. Thus quantum systems with 
`generic' time-reversal invariant classical limits will yield a number 
variance that differs from those quantum systems whose classical limits 
are not time-reversal invariant. In random matrix theory a lack of
time-reversal invariance forces one to deal with complex hermitian
matrices and unitary base canges. Then the measure replacing the
GOE is the Gaussian unitary ensemble (GUE). 

At this point we can also hint at the reason for the exceptional 
behaviour of the spectral statistics in arithmetic quantum chaos, 
see \cite{Bogo,Steil,Luo,Schur}. There the classical 
systems are completely chaotic, but the multiplicities of lengths
of periodic orbits are exceptionally large; indeed one can show
that $g_n\sim c\,\frac{e^{l_n/2}}{l_n}$ for $l_n\rto\infty$, 
\cite{Nonlin,Stefan}. Clearly, the above discussion of the form 
factor has to be modified in an essential way. As a result, one 
can show \cite{PhD} that $K(\tau;E)$ increases exponentially instead 
of linearly.  

We are now going to analyse the properties of the number variance
by means of the relation (\ref{Si2Kfin}). The form factor will be
modelled by a simplified version that, however, captures its main
features as they have been worked out above. For small $\tau$,
close to zero, the form factor will be represented by its diagonal
approximation (\ref{Ksc1}), which we deliberately will sometimes
use in the form (\ref{Ksc2}) or (\ref{Ksc3}). For larger values of 
$\tau$, however still below $\frac{1}{2}$, $K(\tau;E)$ can be well 
approximated by its linear overall increase (\ref{Ksc4}) with 
$\overline{g}=2$. Then, at $\tau=\frac{1}{2}$, both the approximations 
employed to arrive at (\ref{Ksc4}) break down and the form factor 
reaches its asymptotic value of one, see (\ref{Kasym}). For simplicity, 
and because no further information on the behaviour of the form factor
beyond $\tau=\frac{1}{2}$ is available, we merely set $K(\tau;E)
=1$ in our model. This therefore reads in explicit terms
\beq
\label{Kmodel}
K_M(\tau;E):=\left\{\begin{array}{ccl}K_{diag}(\tau;E)&,&0\leq\tau
\leq\tau^* \\ 2\tau&,&\tau^*<\tau\leq\frac{1}{2} \\ 1&,&\tau\geq
\frac{1}{2}\end{array}\right.\ ,
\eeq
see also \cite{A400}. At this point we remark that recently Bogomolny
and Keating \cite{BogoKea} managed to take non-diagonal contributions to
(\ref{Kpot}) into account and hence provided a further justification for
(\ref{Kmodel}). The value of $\tau^*$, beyond which we replace $K_{diag}
(\tau;E)$ by its semiclassical asymptotics (\ref{Ksc4}), seems to be 
somewhat arbitrary. Its actual role will become clearer in the course 
of our subsequent discussion.

As a first illustration of the meaning of $\tau^*$ we remark that 
(\ref{Ksc4}) results from the estimate 
\beq
\label{sumrule}
\sum_{\ga,l_\ga\leq l}l_\ga^2\,e^{-u_\ga}\sim\frac{1}{2}\,l^2\ ,
\ \ \ \ l\rto\infty\ ,
\eeq
applied to (\ref{Kint}) with $l=l_H\tau$. As long as the cut-off $l$
is not connected to $l_H$ no quantum mechanical or semiclassical 
quantities appear; (\ref{sumrule}) thus is a purely classical
asymptotics. One can hence introduce a classical length scale $l_{cl}$
such that the r.h.s.\ of (\ref{sumrule}) approximates the l.h.s.\ 
to a given precision. We then define $\tau^*:=l_{cl}/l_H$. Thus,
for $\tau\geq\tau^*$ one obtains that $l_H\tau\geq l_{cl}$ which
allows to use the asymptotics (\ref{sumrule}). The arbitrariness
in the choice of $\tau^*$ in (\ref{Kmodel}) now appears as the 
-restricted- freedom to fix a value for $l_{cl}$. Certainly, $l_{cl}$
has to be considerably larger than the length $l_1$ of the shortest
periodic orbit. Moreover, since $\tau^* <\frac{1}{2}$ and indeed the
semiclassical limit corresponds to $l_H\rto\infty$, we have three
well separated length scales $l_1\ll l_{cl}\ll l_H$.

The model number variance now follows from (\ref{Kmodel}) through
\beq
\label{Simodel}
\Si^2_M (L;E):=\frac{2}{\pi^2}\int_0^\infty\frac{\sin^2(\pi L\tau)}
{\tau^2}\,K_M(\tau;E)\,d\tau\ ,
\eeq   
and consists of three parts which are due to the three $\tau$-ranges
as they appear in (\ref{Kmodel}). The first part derives from the
integration over the interval $0\leq\tau\leq\tau^*$ and reads
\beq
\label{SiM1}
\Si^2_{M,1}(L;E)\sim\frac{2}{\pi^2}\sum_{l_n}\sum_{k\geq 1\atop
kl_n\leq l_{cl}}\frac{g_n^2}{k^2}\,e^{-ku_n}\,\sin^2\left(\frac{kl_n}
{l_H}\pi L\right)\ .
\eeq
It depends on $\tau^*$ throught the cut-off $l_{cl}$. The second part,
which emerges from the integration over $\tau^*\leq\tau\leq\frac{1}
{2}$, is given by
\beq
\label{SiM2}
\Si^2_{M,2}(L;E)=\frac{2}{\pi^2}\,\left[-\log(2\tau^*)-\mbox{Ci}
(\pi L)+\mbox{Ci}(2\pi L\tau^*)\right]\ ,
\eeq
recall (\ref{intsin}) for the definition of $\mbox{Ci}(x)$. Finally, 
the remaining part reads
\beq
\label{SiM3}
\Si^2_{M,3}(L;E)=\frac{2}{\pi^2}\,\left[1-\cos(\pi L)-\pi L\,
\mbox{Si}(\pi L)+\frac{\pi^2 L}{2}\right]\ .
\eeq
An immediate observation that can be made with (\ref{SiM1})--(\ref{SiM3})
is that $\Si^2_M(L;E)\sim L$ for $L\rto 0$, which is in accordance
with the general statement (\ref{Si2limit1}).

Having explicit expressions at hand for the number variance, we are
now in a position to analyse its behaviour for large $L$. To this end
we first notice that $\Si^2_M(L;E)$ can be split into a sum of
two contributions, one of which is explicitly independent of $\tau^*$ 
or $l_{cl}$, respectively, and reads
\beqa
\label{Sial}
\Si^2_{M,\al}(L;E)&:=&\frac{2}{\pi^2}\left[1-\cos(\pi L)-\pi L\,
                      \mbox{Si}(\pi L)+\frac{\pi^2 L}{2}+\log(2\pi L)
                      -\mbox{Ci}(\pi L)\right]\nonumber\\
                   &=&\frac{2}{\pi^2}\,\log(2\pi L)+\frac{2}{\pi^2}
                      +O\left(\frac{1}{L}\right)\ ,\ \ \ \ L\rto\infty\ .
\eeqa
The second part
\beq
\label{Sibe}
\Si^2_{M,\be}(L;E):=\frac{2}{\pi^2}\left\{\left[\sum_{l_n}\sum_{
k\geq 1\atop kl_n\leq l_{cl}}\frac{g_n^2}{k^2}\,e^{-ku_n}\,\sin^2\left(
\frac{kl_n}{l_H}\pi L\right)\right]-\log(4\pi L\tau^*)+\mbox{Ci}(2\pi
L\tau^*)\right\}
\eeq
contains $\tau^*$ and $l_{cl}$. We will see below, however, that
the leading order asymptotics as $L\rto\infty$ is independent of
the parameter $l_{cl}$. Inspecting (\ref{Sibe}) one observes that 
$L$ and $\tau^*$ appear in the combination $L\tau^*$. Now, $\tau^*=
l_{cl}/l_H\rto 0$ in the semiclassical limit $l_H\rto\infty$, which
forces us to analyse the limit $L\rto\infty$ in the two cases
$L\tau^*\rto 0$ and $L\tau^*\rto\infty$. For convenience we now
introduce the dimensionless quantity $L_{max}:=\frac{1}{\tau^*}$
so that in the following we discuss (i) $L/L_{max}\rto 0$ and (ii)
$L/L_{max}\rto\infty$.  

In the first case the argument of $\sin^2$ can be estimated as
$\frac{kl_n}{l_H}\pi L\leq\pi\frac{L}{L_{max}}\ll 1$ so that the 
periodic orbit sum in (\ref{Sibe}) yields a contribution of
$O((L/L_{max})^2)$. Altogether, one obtains for (\ref{Sial}) and 
(\ref{Sibe}) in the limit $L\rto\infty$, $l_H\rto\infty$, $L/L_{max}
\rto 0$,
\beq
\label{Llim1}
\Si^2_M(L;E)=\frac{2}{\pi^2}\left[\log(2\pi L)+1+\ga-\log 2\right]
+O\left(\left(\frac{L}{L_{max}}\right)^2\right)+O\left(\frac{1}{L}
\right)\ .
\eeq
Our first observation with (\ref{Llim1}) is that, as announced before,
all terms that have been evaluated explicitly do not depend on 
$l_{cl}$. Furthermore, a comparison with the number variance 
(\ref{GOEnumvar}) in the GOE reveals that the leading terms coincide.
Concerning the next-to-leading order, however, the constant in 
(\ref{Llim1}) contains $-\log 2\approx -0.693$ instead of 
$-\frac{\pi^2}{8}\approx -1.234$ as in the GOE.
Thus, our model (\ref{Kmodel}) reproduces the number variance
obtained in random matrix theory to leading order. The next-to-leading
terms, however, differ slightly. This effect is most probably due
to the crude approximation (\ref{Kmodel}) to the actual form factor
in the domain $\tau\geq\frac{1}{2}$. One might expect that the true
form factor will fully reproduce the GOE result.

Remarkable deviations from random matrix theory show up in the 
second case $L/L_{max}\rto\infty$. Here the arguments of the 
$\sin^2$-terms are no longer necessarily small; hence the periodic 
orbit sum contributes oscillatory terms. We are interested in the 
overall behaviour of the number variance as $L\rto\infty$, $l_H\rto
\infty$, $L/L_{max}\rto\infty$ so that we now determine the mean value 
$\Si^2_\infty(E)$ about which the oscillations of the number variance
take place. This can be achieved by replacing each $\sin^2$-term 
by its average value $\frac{1}{2}$. Thus the periodic orbit sum no 
longer contains $L$; it merely yields a constant $C(l_{cl})$ that 
is solely determined by the value of the cut-off length $l_{cl}$.
Collecting then the asymptotics of (\ref{Sial}) and (\ref{Sibe})
in the limit $L\rto\infty$, $l_H\rto\infty$, $L/L_{max}\rto\infty$
yields
\beq
\label{Llim2}
\Si^2_\infty(E)=\frac{2}{\pi^2}\left[\log L_{max}+1-\log 2+C(l_{cl})
\right]+O\left(\frac{L_{max}}{L}\right)\ .
\eeq
It now seems that the result (\ref{Llim2}) depends on $l_{cl}$.
However, the following estimate shows that to leading order this
dependence disappears. To this end take only the leading contribution 
of the primitive periodic orbits into account,
\beq
\label{poestim}
C(l_{cl})\approx\frac{1}{\pi^2}\sum_{l_n\leq l_{cl}}g_n^2\,e^{-u_n}
\approx\frac{\overline{g}}{\pi^2}\sum_{\ga,l_\ga\leq l_{cl}}
e^{-u_\ga}\ .
\eeq
Then, in analogy to (\ref{sumrule}) the thermodynamic formalism
allows to estimate the r.h.s.\ of (\ref{poestim}) for large $l_{cl}$.
Together with $\overline{g}=2$ one thus observes that $C(l_{cl})
\sim\frac{2}{\pi^2}\log l_{cl}$, $l_{cl}\rto\infty$, with the effect
that finally
\beq
\label{Siinftyest}
\Si^2_\infty(E)\approx\frac{2}{\pi^2}\left[\log l_H +1-\log 2
\right]\ ,
\eeq
where now $l_{cl}$ no longer appears. However, if one wants to
calculate a numerical value for $\Si^2_\infty(E)$ the approximations
employed might not be satisfactory so that the value of the
constant on the r.h.s.\ of (\ref{Llim2}) cannot really be fixed
by our considerations. The relation (\ref{Llim2}) then only
shows that this constant depends on the distribution of the short 
periodic orbits (with lenghts $\leq l_{cl}$). If in addition we 
introduce the explicit form $l_H=4\pi\hbar\overline{d}\sqrt{E}$
for the Heisenberg length, we can obtain from (\ref{Llim2}) the 
following high energy asymptotics,
\beq
\label{Siinftyasym}
\Si^2_\infty(E)\sim\frac{1}{\pi^2}\,\log E\ ,\ \ \ \ E\rto\infty\ .
\eeq 

If we now compare (\ref{Llim1}) and (\ref{Llim2}) we realize that
the number variance, as a function of $L$, grows logarithmically
up to approximately $L\approx L_{max}$. Beyond this scale $\Si^2
(L;E)$ ceases to increase, but rather oscillations
caused by classical periodic orbits set in which take place about
the saturation value $\Si^2_\infty(E)$. Both the transition scale
$L_{max}=l_H/l_{cl}$ and the saturation value depend on $E$: The
larger $E$, the larger are both $L_{max}$ and $\Si^2_\infty(E)$. 
Since the number variance is `generically' well described by
random matrix theory in the domain $L\ll L_{max}$, this is sometimes
called the universality regime. On the other hand, the oscillations
of the number variance about $\Si^2_\infty(E)$ in the domain
$L\gg L_{max}$ suggest to call this the saturation regime.  
This general picture has been confirmed numerically in the case
of the zeros of the Riemann zeta function \cite{BerryN} and
for some two dimensional classically chaotic systems, e.g.\ in  
\cite{AurStei,Baecker}. 

As a final remark on the number variance let us compare the results
of the semiclassical analysis with the general observations made
in section \ref{sec2.1}. To this end we first have to recall the
connection between the variables $E$ and $x$: since $x=\overline{N}
(E)\sim\frac{A}{4\pi\hbar^2}E=\overline{d}E$, one observes that
$l_H=4\pi\hbar\overline{d}\sqrt{E}\sim4\pi\hbar\sqrt{\overline{d}x}
=\sqrt{4\pi A}\sqrt{x}$. While discussing the relation 
(\ref{Numvarform}) we noticed that the limit $L\rto\infty$ has to
be considered in conjunction with $x\rto\infty$. Due to the above
the latter is equivalent to the semiclassical limit $l_H\rto\infty$.
We also related the two limits in choosing $L=a\,x^\ga$ with some
exponent $0<\ga<1$. Since $L_{max}=l_H/l_{cl}=const.\,\sqrt{x}$,
the two cases $L/L_{max}\rto 0$ and $L/L_{max}\rto\infty$, which
we were forced to distinguish in the semiclassical analysis, 
therefore correspond to $0<\ga<\frac{1}{2}$ and $\frac{1}{2}<\ga<1$, 
respectively. We hence now have a means to characterize what we 
before alluded to as a large exponent $\ga$. The latter should 
result in a decoupling of the two random variables $N_{fl}(y)$
and $N_{fl}(y+L)$ on $I_x$, when $x\rto\infty$ and $L=a\,x^\ga$.
Now, choosing $\ga$ larger than the critical exponent $\ga_c=
\frac{1}{2}$ corresponds to $L/L_{max}\rto\infty$ and therefore to 
the saturation regime of the number variance. Here $\Si^2(L;E)$ 
oscillates about the $L$-independent saturation value $\Si^2_\infty
(E)$. Recalling the relation (\ref{Numvarform}), we hence conclude 
that the oscillations can only be caused by the correlation function 
on the r.h.s.\ of (\ref{Numvarform}), whereas the $L$-independent 
contribution $\Si^2_\infty(E)$ comes from the first two terms on the 
r.h.s. Moreover, (\ref{Siinftyasym}) implies that
\beq
\label{varasym}
\ez_x\left[N_{fl}^2\right]\sim\frac{1}{2\pi^2}\log x\ ,\ \ \ \ 
x\rto\infty\ ,
\eeq
compare also (\ref{expectasym}). So far the whole analysis concerned
two dimensional systems. An immediate generalization to the $d$
dimensional case reveals that $L_{max}\sim const.\,x^{1-\frac{1}{d}}$,
so that the critical exponent $\ga_c$ above which saturation takes 
place is in general $\ga_c=1-\frac{1}{d}$. Thus, in larger dimension 
the saturation regime for the limit $L\rto\infty$ becomes 
smaller. However, it could only disappear in infinite dimensional
systems.

The above analysis therefore leads us to conjecture that the two random 
variables $N_{fl}(y)$ and $N_{fl}(y+L)$ on $I_x$ become statistically 
independent in the limit $x\rto\infty$ when $L=a\,x^\ga$ with $\ga>
\ga_c$. Once $\ga<\ga_c$, strong correlations remain that force the 
number variance to follow the predictions of the GOE. For certain two 
dimensional classically integrable systems Bleher and Lebowitz 
\cite{BL1} proved that indeed the saturation behaviour, with $\ez_x
[N_{fl}^2]\sim const.\,\sqrt{x}$, occurs as well as the statistical 
independence for $\ga>\ga_c =\frac{1}{2}$. The difference in $\ez_x
[N_{fl}^2]$ between classically integrable and chaotic systems 
(\ref{varasym}) stems from the occurrence of periodic orbits as 
one-parameter families in the former case. Under certain regularity 
assumptions, Sarnak \cite{Schur} showed that the existence of 
one-parameter families of closed geodesics in the unit tangent bundle 
over a surface implies $\Si^2(L;E)\gg\sqrt{E}$ for $L\sim E\rto\infty$. 
To this end he used the trace formula in order to represent $N_{fl}(E)$, 
and in particular the structure (\ref{Msubjcontrib}) implied by 
$k$-parameter families which yield exponents $m_j=k+1$. In the example 
of the stadium billiard, where the bouncing ball orbits occur as a 
one-parameter family that can be dealt with in full detail, Sieber 
et al.\ \cite{BB} determined the saturation value of the number 
variance as $\Si^2_\infty(E)=const.\,\sqrt{E}$, and thus confirmed 
the effect of non-isolated periodic orbits even in classically 
chaotic systems. 

\xsection{From Local to Global Eigenvalue Statistics}
\label{sec3}
The principal measure for the distribution of the eigenvalues
of some quantum Hamiltonian that was discussed in section
\ref{sec2} was the probability $E(k;L)$ to observe $k$ defolded
eigenvalues in an interval of length $L$. As long as $L$ is finite
one thus measures correlations on a scale $L$, i.e., among 
approximately $L$ eigenvalues. One is therefore dealing with
local eigenvalue statistics. If one wants to include correlations
among infinitely many levels, one is forced to consider the limit
$L\rto\infty$. At several stages in the previous discussion we
already considered asymptotics for large $L$, and hence the transition 
to the global scale. It is the goal of the present section to
investigate this limit in some more detail. We will in particular 
emphasize the distinction as to whether one passes to the global 
scale within the universality regime $L/L_{max}\rto 0$, or in the 
saturation regime $L/L_{max}\rto\infty$. It will then be noticed 
that the latter approach offers an interesting opportunity to 
identify fingerprints of classical chaos in the distribution of
eigenvalues.

\subsection{The Renormalized Random Variable}
\label{sec3.1}
Our discussion of the local eigenvalue statistics in section 
\ref{sec2} was based on the random variable $n_L(y)=N(y+L)-N(y)$
for $y\in I_x=[x-\De x,x+\De x]$. We noticed that its expectation 
value reads $\ez_x[n_L]\sim L$, $x\rto\infty$, see (\ref{nsubLmean}). 
Moreover, its variance is essentially given by the number variance 
(\ref{Numvardef}), which grows logarithmically for classically chaotic
and linearly for classically integrable systems as $L\rto\infty$ once 
one confines oneself to the universality regime $L\ll L_{max}=const.
\,\sqrt{x}$, see (\ref{Llim1}). In the saturation regime $L\gg L_{max}$ 
it oscillates about $\Si^2_\infty$. In any case, both the mean value 
and the variance of the random variable $n_L(y)$ diverge as $L\rto
\infty$, $x\rto\infty$. One is therefore advised to renormalize the 
random variable so as to yield finite first and second moments in 
order to have a chance to obtain a finite limit distribution. The 
most simple choice
\beq
\label{etadef}
\eta_L(y):=\frac{n_L(y)-L}{\sqrt{\Si^2_x(L)}}\ ,\ \ \ \ y\in I_x\ ,
\eeq
certainly ensures that the lowest two moments are finite, since
\beq
\label{eta1}
\ez_x\left[\eta_L\right]=\frac{1}{2\De x}\int_{x-\De x}^{x+\De x}
\eta_L (y)\,dy=\frac{\ez_x[n_L]-L}{\sqrt{\Si^2_x(L)}}\rto 0\ ,
\eeq
for $x\rto\infty$, and
\beq
\label{eta2}
\ez_x\left[\eta_L^2\right]=\frac{\ez_x\left[\left(n_L-L\right)^2
\right]}{\Si^2_x(L)}=1\ .
\eeq
In general, the $m$-th moments of the renormalized random variable
$\eta_L(y)$ can be expressed in terms of the moments $\Si_x^m(L)$, 
see (\ref{Simdef}), of $n_L(y)-L$,
\beq
\label{Mmdef}
M_x^m(L):=\ez_x\left[\eta_L^m\right]=\frac{1}{\left[\Si^2_x(L)
\right]^{\frac{m}{2}}}\,\ez_x\left[\left(n_L-L\right)^m\right]=
\frac{\Si_x^m(L)}{\left[\Si^2_x(L)\right]^{\frac{m}{2}}}\ .
\eeq
Due to the normalization of $\eta_L(y)$, as expressed by (\ref{eta1})
and (\ref{eta2}), one is hence led to expect a limit distribution for
$L\rto\infty$ to exist.

The characteristic function of $\eta_L(y)$ can now be expressed in 
terms of the characteristic function of $n_L(y)$,
\beq
\label{etachar}
J_{\eta_L,x}(\xi)=\ez_x\left[e^{i\xi\eta_L}\right]=e^{-i\xi
\frac{L}{\sqrt{\Si^2_x(L)}}}\,J_{n_L,x}\left(\frac{\xi}{\sqrt{
\Si^2_x(L)}}\right)\ ,
\eeq
compare (\ref{chardef}). This quantity allows to characterize the
distribution of the random variable $\eta_L(y)$ through
\beqa
\label{etadischar}
\nu_{\eta_L,x}(d\eta)&=&\frac{1}{2\De x}\int_{x-\De x}^{x+\De x}\de
                        (\eta_L(y)-\eta)\,dy\,d\eta\nonumber\\
                     &=&\frac{1}{2\De x}\int_{x-\De x}^{x+\De x}
                        \frac{1}{2\pi}\int_{-\infty}^{+\infty}
                        e^{i[\eta_L(y)-\eta]\xi}\,d\xi\,dy\,d\eta\\
                     &=&\frac{1}{2\pi}\int_{-\infty}^{+\infty}
                        e^{-i\eta\xi}\,J_{\eta_L,x}(\xi)\,d\xi\,
                        d\eta\nonumber\ .
\eeqa
Going back to (\ref{probrecov}), we realize that
\beqa
\label{EkLrenorm}
\sqrt{\Si^2_x(L)}\,E_x\left(\eta\sqrt{\Si^2_x(L)}+L;L\right)\,d\eta
    &=&\frac{\sqrt{\Si^2_x(L)}}{2\pi}\int_{-\pi}^{+\pi}e^{-i\rho
       (\eta\sqrt{\Si^2_x(L)}+L)}\,J_{n_L,x}(\rho)\,d\rho\,d\eta
       \nonumber\\
    &=&\frac{1}{2\pi}\int_{-\pi\sqrt{\Si^2_x(L)}}^{+\pi\sqrt{\Si^2_x
       (L)}}e^{-i\xi\eta}\,J_{\eta_L,x}(\xi)\,d\xi\,d\eta\ ,
\eeqa
where we changed variables from $\rho$ to $\xi:=\rho\sqrt{\Si^2_x(L)}$
and employed (\ref{etachar}). Hence a comparison of (\ref{etadischar}) 
and (\ref{EkLrenorm}) yields
\beq
\label{etadis}
\nu_{\eta_L,x}(d\eta)\sim\sqrt{\Si^2_x(L)}\,E_x\left(\eta\sqrt{\Si^2_x
(L)}+L;L\right)\,d\eta\ ,
\eeq
whenever the number variance approaches infinity, i.e., when $L\rto
\infty$, $x\rto\infty$. Since we concluded (\ref{etadis}) from 
the relation (\ref{etachar}) of the characteristic functions, the
asymptotic equivalence of both sides of (\ref{etadis}) is to be
understood in the sense of weak limits. This means that for any 
bounded continuous function $g(\eta)$
\beq
\label{weaklim}
\lim_{x\rto\infty\atop L\rto\infty}\int_{-\infty}^{+\infty}g(\eta)\,
\nu_{\eta_L,x}(d\eta)=\lim_{x\rto\infty\atop L\rto\infty}\sqrt{\Si^2_x
(L)}\,\int_{-\infty}^{+\infty}g(\eta)\,E_x\left(\eta\sqrt{\Si^2_x
(L)}+L;L\right)\,d\eta\ .
\eeq

As a first example, let us discuss a Poissonian random process $n_L(y)$.
Since this is stationary there appears no dependence of its 
distribution on $x$. The probability to find $k$ events (i.e., 
eigenvalues) in an interval of length $L$ is then given by 
\beq
\label{EkLPoisson1}
E_{Poisson}(k;L)=\frac{L^k}{k!}\,e^{-L}\ .
\eeq
Now, it is well known and easy to see that (\ref{EkLPoisson1})
approaches a Gaussian for $L\rto\infty$,
\beq
\label{Poissonlim}
E_{Poisson}(k;L)\sim\frac{1}{\sqrt{2\pi L}}\,e^{-\frac{(k-L)^2}
{2L}}=\frac{1}{\sqrt{2\pi\Si^2(L)}}\,e^{-\frac{(k-\ez[n_L])^2}
{2\Si^2(L)}}\ ,
\eeq
since $\ez_{Poisson}[n_L]=L$ and $\Si^2_{Poisson}(L)=L$. Hence,
according to (\ref{etadis}) the distribution of $\eta_L(y)$ approaches
\beq
\label{etaPoisdis}
\nu_{\eta_L}(d\eta)\rto\frac{1}{\sqrt{2\pi}}\,e^{-\frac{1}{2}\eta^2}
\,d\eta\ ,\ \ \ \ L\rto\infty\ .
\eeq
In random matrix theory one obtains a similar asymptotic result.
Here the random process $n_L(y)$ is also stationary, after one has
performed the limit of matrix dimension $N\rto\infty$, and hence
no $x$-dependence occurs either. In section \ref{sec2.2} we 
already reported on the numerical finding (\ref{EkLasym}). Due to
the relation (\ref{etadis}) one therefore would conclude that 
(\ref{etaPoisdis}) also holds for the GOE. Indeed, Costin and 
Lebowitz \cite{CL} calculated the cumulants of the distribution
of $\eta_L(y)$ in the GOE, GUE, and GSE (Gaussian symplectic 
ensemble) and proved that $\nu_{\eta_L}(d\eta)$ weakly converges
to a Gaussian with zero mean and unit variance. 

At this point we recall that the stationarity of the random process 
in the Poissonian as well as in the random matrix case makes the 
limit $L\rto\infty$ unique, i.e., no saturation effects and related
phenomena occur. As we saw in section \ref{sec2.3} the situation
is different for spectra of individual quantum Hamiltonians. Now,
provided one accepts that the local eigenvalue distributions of 
`generic' quantum Hamiltonians are described by Poisson statistics 
in the classically integrable case and by random matrix theory  
in the chaotic case, and that this description extends to hold
when passing to the global scale within the universality regime, 
the above findings suggest that
\beq
\label{nulimactual}
\lim\int_{-\infty}^{+\infty}g(\eta)\,\nu_{\eta_L,x}(d\eta)=
\frac{1}{\sqrt{2\pi}}\int_{-\infty}^{+\infty}g(\eta)\,e^{-\frac{1}
{2}\eta^2}\,d\eta\ , 
\eeq
for $L\rto\infty$, $x\rto\infty$, $L/L_{max}\rto 0$, and any bounded
continuous function $g(\eta)$. Thus, a central limit theorem would 
hold true for the global distribution of eigenvalues in the universality
regime, irrespective of the type of classical dynamics. 

As can be anticipated from the discussion in section \ref{sec2.3}, 
the situation changes drastically, if one passes to the global scale 
within the saturation regime $L/L_{max}\rto\infty$. In \ref{sec2.3}
we interpreted the saturation phenomenon in terms of the relation
(\ref{Numvarform}) for the number variance and concluded that all
evidence seems to be in favour of an asymptotic statistical
independence of the two random variables $N_{fl}(y+L)$ and $N_{fl}
(y)$, $y\in I_x$, as $L/L_{max}\rto\infty$. We remark that instead
of choosing the arguments $y+L$ and $y$, with $y\in I_x$, one can
obviously also consider $N_{fl}(y)$ with $y\in I_x$ and $y\in 
I_{x+L}$, respectively. In analogy to the discussion above, one
should also renormalize these random variables and hence introduce
\beq
\label{Wxofy}
W (y):=\frac{N_{fl}(y)}{\sqrt{\ez_x[N_{fl}^2]}}\ ,
\eeq
for $y\in I_x$ and $y\in I_{x+L}$, respectively. The first and second 
moments of this random variable are clearly analogous to (\ref{eta1}) 
and (\ref{eta2}),
\beq
\label{Wmom}
\ez_x[W]\rto 0\ ,\ \ \ \ x\rto\infty\ ,\ \ \ \ \mbox{and}\ \ \ \ 
\ez_x[W^2]=1\ .
\eeq
In view of the saturation of the number variance,
\beq
\label{varsatu}
\Si_x^2(L)\sim\Si^2_\infty\sim 2\,\ez_x[N_{fl}^2]\sim 2\,\ez_{x+L}
[N_{fl}^2]
\eeq
for $x\rto\infty$, $L\rto\infty$, $L/L_{max}\rto\infty$, the random
variable
\beq
\label{etaW}
\eta_L(y)\sim\frac{N_{fl}(y+L)-N_{fl}(y)}{\sqrt{\Si^2_\infty}}
\eeq
appears as the difference of the asymptotically independent random
variables $W(y)$, $y\in I_x$, and $W(y)$, $y\in I_{x+L}$. Thus, instead 
of the limit distribution of $\eta_L(y)$ as $L/L_{max}\rto\infty$, one 
can equivalently consider the limit distribution of $W(y)$ as 
$x\rto\infty$. Provided the statistical independence alluded to
above indeed holds, both limit distributions coincide.

The quantity $W(y)$ has a nice and simple interpretation in that 
it describes the -normalized- fluctuations of the spectral staircase
$N(E)$ about its mean behaviour $\overline{N}(E)$. Furthermore, it
can easily be calculated from numerical data, and its distribution
is readily computed. It thus appears that the global eigenvalue 
distribution in the saturation regime is most conveniently measured
in terms of the limit distribution of $W(y)$ for large $x$.    

\subsection{Classically Integrable Systems}
\label{sec3.2}
In contrast to the case of classically chaotic systems, spectra of
quantum Hamiltonians that arise as quantizations of integrable
classical systems allow for a fairly explicit treatment. Due to
the semiclassical EBK-quantization scheme the behaviour of the
spectral staircase $N(E)$ can be related to a lattice point problem.
Given an integrable classical system with $d$ degrees of freedom,
one introduces action-angle variables $(I,\th)$. The classical 
Hamiltonian function then depends only on the actions, $H(I)=H(
I_1,\dots,I_d)$. According to the EBK-procedure, semiclassical
approximations to the quantum energies can be obtained once one
quantizes the actions as $I_k=(n_k+\frac{\al_k}{4})\hbar$. Here
$n_k\in\gz$, or $n_k\in\nz$, yields a quantum number, and $\al_k
\in\{0,1,2,3\}$ is an appropriate Maslov index. Then
\beq
\label{EBKenergy}
E^{EBK}_{n_1,\dots,n_d}:=H\left(\left(n_1+\frac{\al_1}{4}\right)
\,\hbar,\dots,\left(n_d+\frac{\al_d}{4}\right)\,\hbar\right) 
\eeq
defines semiclassical energies which approximate the eigenvalues
of the quantum Hamiltonian $\hat H$ for small $\hbar$. The EBK-spectral
staircase
\beq
\label{EBKstair}
N^{EBK}(E):=\#\left\{(n_1,\dots ,n_d);\ E^{EBK}_{n_1,\dots,n_d}\leq 
E\right\}
\eeq
counts the number of lattice points $(n_1,\dots ,n_d)$ inside the 
domain that is bordered by the $(d-1)$-dimensional manifold defined
by $H(I_1,\dots,I_d)=E$. As ususal in lattice point problems, the
leading behaviour of $N^{EBK}(E)$ in the limit $E\rto\infty$ is 
provided by the volume of the manifold which is given by the condition 
$H(I_1,\dots,I_d)\leq E$. If one then takes the complete phase space 
into account by introducing an additional integration over the angle 
variables one observes that
\beq
\label{EBKlead} 
N^{EBK}(E)\sim\frac{1}{(2\pi\hbar)^d}\int\int\Th\left(E-H(I)\right)
\ dI\,d\th\ .
\eeq
The r.h.s.\ is identical to the r.h.s.\ of (\ref{stairsemicl}) because
the transformation $(I,\th)\mapsto (p,x)$ is canonical and hence 
preserves phase space volumes. Therefore (\ref{EBKstair})
yields the actual spectral staircase at least to leading order as
$\hbar\rto 0$. However, as discussed at length in section \ref{sec2} 
the fine structure in the eigenvalue distribution, which is of interest
in quantum chaos, is encoded in the remainder term to the leading
asymptotics (\ref{stairsemicl}) of the spectral staircase. In
order to analyse spectral statistics in terms of the lattice point
problem (\ref{EBKstair}) one therefore has to ensure that 
the limit distributions arising from $N(E)$ and $N^{EBK}(E)$,
respectively, coincide. For the free motion of a particle on a 
surface of revolution that obeys a certain non-degeneracy condition 
Bleher \cite{Blrev} proved this to be the case. The same fact was 
proven by Kosygin et al.\ \cite{KMS} for the free motion on a torus 
with Liouville metric, which is also integrable.

As a first and very simple example of a classically integrable
system let us discuss a one dimensional billiard, i.e., the free
motion on the interval $[0,l]$ with elastic reflections from the
boundary points at $0$ and $l$. In units where $\hbar=1=2m$ the
eigenvalues of the quantum Hamiltonian $\hat H=-\frac{d^2}{dx^2}$ 
with Dirichlet boundary conditions read $E_n=\frac{\pi^2}{l^2}n^2$,
$n\in\nz$. If we denote the integer part of a real number $r$ by
$[r]$, the spectral staircase is given by $N(E)=[\frac{l}{\pi}
\sqrt{E}]\sim\frac{l}{\pi}\sqrt{E}$, $E\rto\infty$. As the defolded 
energy variable we thus introduce $x:=\frac{l}{\pi}\sqrt{E}$. Then 
the fluctuations of the spectral staircase read
\beq
\label{onedimstair}
N_{fl}(x)=\frac{1}{2}-\{x\}=\frac{1}{\pi}\sum_{n=1}^\infty\frac{1}
{n}\,\sin(2\pi nx)\ ,
\eeq
where $\{x\}=x-[x]$ denotes the fractional part of $x$. The $\frac{1}
{2}$ appearing in the middle term of (\ref{onedimstair}) is chosen 
such that $N_{fl}(x)$ fluctuates about zero. Clearly, $|N_{fl}(x)|\leq
\frac{1}{2}$, and the r.h.s.\ of (\ref{onedimstair}) constitutes 
a representation of the one-periodic and piecewise continuous function 
$N_{fl}(x)$ by its Fourier series. The variance of $N_{fl}(x)$ can 
easily be determined to yield $\ez_x[N_{fl}^2]\sim\frac{1}{12}$ for 
$x\rto\infty$, so that $W(y)=\sqrt{12}N_{fl}(y)$. The characteristic 
function for the distribution of $W(y)$ is given by
\beq
\label{JW}
J_{W,x}(\xi)=\frac{1}{2\De x}\int_{x-\De x}^{x+\De x}e^{i\xi W(y)}
\,dy\ .
\eeq
In the limit $x\rto\infty$ we obtain
\beq
\label{Jinfty}
J_W(\xi)=\int_0^1e^{i\xi\sqrt{12}(\frac{1}{2}-y)}\,dy=\frac{1}{\sqrt{3}
\xi}\,\sin\left(\sqrt{3}\xi\right)=\frac{1}{2\sqrt{3}}\int_{-\sqrt{3}}^{
+\sqrt{3}}e^{i\xi w}\,dw\ .
\eeq
Since the density $p(w)$ of the limit distribution of $W(y)$ is
related to the characteristic function by Fourier inversion, we
observe from (\ref{Jinfty}) that
\beq
\label{pofw}
p(w)=\left\{\begin{array}{ccl}\frac{1}{2\sqrt{3}}&,&-\sqrt{3}\leq
w\leq +\sqrt{3} \\ 0&,&|w|>\sqrt{3}\end{array}\right.\ .
\eeq
Thus, $W(y)$ is asymptotically uniformly distributed on a finite
interval. The fact that $p(w)=0$ for $|w|>\sqrt{3}$ results from
the bound $|W(y)|=\sqrt{12}|N_{fl}(y)|\leq\sqrt{3}$; this clearly 
determines the support of the probability density $p(w)$. The result 
(\ref{pofw}) could also be naively anticipated from the explicit form 
and the periodicity of $N_{fl}(x)$.

A more prototypical example, however, of a classically integrable 
system is provided by the free motion of a particle on a two 
dimensional torus. As discussed at the end of section \ref{sec2.2}, 
the quantum energies for this system are exactly represented by an 
expression that is proportional to the positive definite binary 
quadratic form (\ref{Quadform}). The spectral staircase $N(E)$ is 
hence exactly given by a lattice point problem; more precisely, 
$N(E)$ is the number of points of the lattice $\gz^2$ inside an 
ellipse that is defined by the quadratic form $Q$. In this example 
we again choose units such that $\hbar=1=2m$.

Let us now discuss the torus $\rz^2/\cL$ with $\cL=\gz^2$ in some
more detail. Here $N(E)$ is the number of lattice points inside a 
circle of radius $\frac{\sqrt{E}}{2\pi}$, i.e., we are dealing with
the circle problem. Employing the Poisson summation formula, one can
readily demonstrate that for $E\rto\infty$
\beqa
\label{Torustrace}
N(E)&=&\frac{E}{4\pi}+\frac{E^{\frac{1}{4}}}{\sqrt{2\pi^3}}
       \sum_{(n_1,n_2)\neq (0,0)}\frac{1}{(n_1^2+n_2^2)^{\frac{3}{4}}}
       \,\cos\left(\sqrt{n_1^2+n_2^2}\,\sqrt{E}-\frac{3\pi}{4}\right)
       +O\left(E^{-\frac{1}{4}}\right)\ ,\nonumber\\
    &=&\frac{E}{4\pi}+\frac{E^{\frac{1}{4}}}{\sqrt{2\pi^3}}
       \sum_{m\in\nz\atop m\,{\rm square\,free}}\frac{1}{m^{\frac{3}
       {4}}}\sum_{k=1}^\infty\frac{1}{k^\frac{3}{2}}\,\cos\left(k
       \sqrt{m}\,\sqrt{E}-\frac{3\pi}{4}\right)+O\left(E^{-\frac{1}{4}}
       \right)\ .
\eeqa
Since $\sum_{n_1,n_2}(n_1^2+n_2^2)^{-\frac{3}{4}}$ diverges, one 
cannot estimate the sum on the r.h.s.\ of (\ref{Torustrace}) in a 
simple way as a function of $E$; to obtain improved bounds for this 
sum still constitutes an active area of research in analytic number 
theory. A novel perspective on the circle problem, which inspired
much of the work on the global distribution of quantum energies, 
was introduced by Heath-Brown \cite{HB}. Instead of estimates of
the remainder term in the circle problem, he considered its 
distribution. As we will soon realize, this corresponds to the 
limit distribution of $W(y)$. According to (\ref{Torustrace}), where
the first term on the r.h.s.\ defines $\overline{N}(E)$, the 
defolded spectral variable reads $x=\frac{E}{4\pi}$. Thus
\beq
\label{Nfltorus}
N_{fl}(x)=\left(\frac{x}{\pi}\right)^{\frac{1}{4}}\,\Phi(x)+O
\left(x^{-\frac{1}{4}}\right)\ ,\ \ \ \ x\rto\infty\ ,
\eeq
with
\beq
\label{Phidef}
\Phi(x)=\sum_{m\in\nz\atop m\,{\rm square\,free}}f_m\left(
\sqrt{\frac{x}{\pi}}\,\sqrt{m}\right)\ ,
\eeq
where
\beq
\label{fmdef}
f_m (t):=\frac{1}{\pi m^{\frac{3}{4}}}\sum_{k=1}^\infty\frac{1}
{k^{\frac{3}{2}}}\,\cos\left(2\pi kt-\frac{3\pi}{4}\right)\ .
\eeq
It is furthermore known, see for example \cite{HB}, that $\ez_x[
N_{fl}^2]\sim const.\,\sqrt{x}$, $x\rto\infty$, so that $W(y)=
const.\,\Phi(y)$. Apart from the normalization of the variance,
the limit distributions of $W(y)$ and $\Phi(y)$ therefore coincide.
Notice that the representation (\ref{Torustrace}) of the spectral
staircase can be interpreted in terms of a semiclassical trace 
formula. To this end one observes that $\sqrt{n_1^2+n_2^2}$, with 
$(n_1,n_2)\in\gz^2\backslash\{(0,0)\}$, yields all lengths of 
periodic orbits on the torus. In the same spirit, the sum in 
(\ref{Phidef}) can be identified as extending over all lengths 
$\sqrt{m}$ of primitive periodic orbits; their repetitions are then 
accounted for by the sum over $k$ in (\ref{fmdef}). 

The functions $f_m(t)$ in (\ref{fmdef}) are clearly continuous and
periodic with period 1. For such functions Heath-Brown proved 
\cite{HB} that whenever $\ga_1,\dots,\ga_k$ are real numbers linearly 
independent over $\qz$, then $f_{m_1}(\ga_1 t),\dots,f_{m_k}(\ga_k t)$, 
with $0\leq t\leq T$, become statistically independent in the limit 
$T\rto\infty$. If one now chooses $\ga_m=\sqrt{m}$, $m$ square free, 
any set $\{\ga_{m_1}\neq\dots\neq\ga_{m_k}\}$ is indeed linearly 
independent implying that the random variables $f_m(t\sqrt{m})$, 
$0\leq t\leq T$, as they appear in (\ref{Phidef}), asymptotically 
become independent for $T\rto\infty$. A further analysis reveals 
\cite{Bleher,BleDuke} that under fairly general assumptions about a 
set of continuous real valued periodic functions $b_1(t),b_2(t),\dots$ 
with
\beq
\label{bcond}
\int_0^1b_n(t)\,dt=0\ \ \ \ \mbox{and}\ \ \ \ \sum_{n=1}^\infty
\int_0^1 b_n(t)^2\,dt<\infty\ ,
\eeq
the series $F(t)=\sum_n b_n(\ga_n t)$, $0\leq t\leq T$, has a 
distribution that weakly converges to the distribution of the random
series $\sum_n b_n(\th_n)$, where the $\th_n$'s are independent
random variables uniformly distributed on the interval $[0,1]$. 
The latter fact arises because the periodicity of the functions 
$b_n(t)$ results in $\ga_n t$ to be taken mod 1. As $t$ varies over
the interval $[0,T]$, the values of $\ga_n t$ mod 1 become uniformly
distributed on $[0,1]$ as $T\rto\infty$.

We remark that for the following results to hold the second condition 
in (\ref{bcond}) becomes essential; it roughly means that the maximum 
of $b_n(t)$ must decrease sufficiently fast as $n\rto\infty$. Then a 
sufficient condition for the existence of the limit distribution is 
for example given by \cite{HB,Bleher}
\beq
\label{almostper}
\lim_{N\rto\infty}\limsup_{T\rto\infty}\frac{1}{T}\int_0^T\min\left\{
1,\left|F(t)-\sum_{n\leq N}b_n(\ga_n t)\right|\right\}\,dt=0\ .
\eeq
This means that $F(t)$ can be represented, in the sense indicated in
(\ref{almostper}), by the infinite sum $\sum_n b_n(\ga_n t)$. 
Under further assumptions on the behaviour of the functions $b_n(t)$
it was moreover proven \cite{Bleher} that the limit distribution has 
a density $p(x)$ which behaves asymptotically as $e^{-c|x|^\rho}$
for $x\rto\pm\infty$. For the circle problem it was verified 
\cite{Bleher} that the functions $f_m(t)$ meet all requirements,
and the exponent was estimated from above and below by $\rho_\pm =
4\pm\ve$ for any $\ve>0$, respectively; thus the limit distribution 
is not Gaussian.

We are interested in the limit distribution of $W(y)$ for the
eigenvalues of the quantum Hamiltonian on the torus $\rz^2/\gz^2$,
and hence in the limit distribution of $\Phi(y)$, see (\ref{Nfltorus}).
By the change of variables $x=\pi t^2$ we can relate this problem
to the distribution of the random series $\sum_m f_m(\th_m)$
and then apply the results of \cite{HB,Bleher}. Therefore
\beq
\label{Wtorus}
\lim_{x\rto\infty}\frac{1}{2\De x}\int_{x-\De x}^{x+\De x}g\left(
W(y)\right)\,\rho\left(\frac{y-x}{2\De x}+\frac{1}{2}\right)\,dy=
\int_{-\infty}^{+\infty}g(w)\,p(w)\,dw
\eeq
for any piecewise continuous bounded function $g(w)$ and any
probability density $\rho(\tau)$, $\tau\in[0,1]$. Here $p(w)$ is 
the non-Gaussian probability density mentioned above. The freedom
to choose a weight $\rho(\tau)$ allows to change variables from, say, 
$y$ to $t$ and thus to compensate for the Jacobian then arising. 

The existence of a limit distribution (\ref{Wtorus}) has by now been 
established for a large class of two dimensional classically 
integrable systems, see \cite{BleMin} for a review. This includes
the free motion on more general flat tori, with and without 
Aharonov-Bohm type magnetic fluxes \cite{Bleher,Ble70}, on surfaces 
of revolution \cite{Blrev}, and on tori with Liouville metrics 
\cite{KMS,BKS}. All these cases lead to limit distributions with 
densities $p(w)$ that approximately decay as $e^{-c|w|^4}$ for $w\rto
\pm\infty$. 

In a further class of examples for two dimensional classically 
integrable systems the limit distributions of $W(y)$ were explicitly
determined by Schubert \cite{Roman}. He considered the geodesic
motion on Zoll surfaces, i.e., surfaces with the topology of a sphere
which are endowed with a metric all of whose geodesics are closed,
and all lengths of periodic orbits are multiples of a fundamental
length. The simplest example for a Zoll surface is the sphere 
with its usual round metric; however, non-trivial deformations 
thereof exist. Schubert showed that $N_{fl}(x)=\sqrt{x}\Th(\sqrt{x})$,
where $\Th(t)$ is an almost periodic function of Besicovitch-class
$B^2$, see \cite{BleDuke} for explanations. The trace formula then
allows to represent $\Th(t)$ by an $L^2$-convergent Fourier series
that is, apart from a possibly non-trivial Maslov phase, identical to
(\ref{onedimstair}). Therefore, $W(y)=\Th(\sqrt{y})$ for Zoll surfaces
has the same limit distribution (\ref{pofw}) as a one dimensional 
billiard. This seemingly surprising coincidence is caused by the fact 
that in both cases the classical side of the trace formula is given 
by a sum over multiples of one fundamental length. The extra dimension 
of a Zoll surface thus is only reflected in the overall growth of 
$N_{fl}(x)$ by $\sqrt{x}$. Notice also the difference to the case of 
tori where the overall growth of $N_{fl}(x)$ is given by $x^{\frac{1}
{4}}$.

As a final remark we observe that the decay of the propability density
$p(w)$ as $w\rto\pm\infty$ is connected to the type of convergence
of the periodic orbit sum representing $N_{fl}(x)$. For the one
dimensional billiard as well as for Zoll surfaces the series converges
in $L^2$-norm, which is considerably strong. In contrast, one
obtains only the weaker form of convergence (\ref{almostper}) for
flat tori. We hence conclude that the weaker the convergence is, the
weaker the decay of $p(w)$. This observation will be important
for classically chaotic systems. We also notice that in all examples
of classically integrable systems studied so far the distribution of
$W(y)$ has a limit as $x\rto\infty$ that is non-Gaussian; thus in the
integrable case no central limit theorem for the global distribution of
eigenvalues in the saturation regime is valid.

\subsection{Classically Chaotic Systems}
\label{sec3.3}
In analogy to the situation for classically integrable systems that has
been described above, the global distribution of eigenvalues of quantum
Hamiltonians with chaotic classical limits will now be studied by means 
of the limit distributions of $W(y)$, $y\in I_x$, as $x\rto\infty$.
Above we realized that these distributions can be obtained from periodic
orbit sums for $N_{fl}(x)$. In all of the integrable cases discussed
we noticed that $N_{fl}(x)=x^\al\,\phi(x)$, where $\al$ is some
appropriate power that describes the overall growth of the fluctuations
$N_{fl}(x)$ of the spectral staircase as $x\rto\infty$, and $\phi(x)$
is a sum over the (primitive) lengths of classical periodic orbits and 
their multiples; one in particular finds that $\ez_x[N_{fl}^2]\sim
const.\,x^{2\al}$, $x\rto\infty$. The periodic orbit sum for $\phi(x)$
was found to share some controllable convergence properties; it either
was a Fourier series or, more typically, an almost periodic function
whose representation by a trigonometric series converges in some 
Besicovitch norm. Now focussing on the chaotic case, we realize that
a representation of $N_{fl}(x)$ in terms of a sum over (primitive) 
periodic orbits, see for example (\ref{cospo}), results in a considerably 
worse convergence. Due to the exponential proliferation (\ref{Huberl}) 
of the number of periodic orbits, a Gaussian smoothing was required in  
(\ref{cospo}) in order to achieve a finite sum. Moreover, no power of 
$x$ or $\log x$ drops out explicitly in front of the periodic orbit
term that could directly indicate the overall increase of $N_{fl}(x)$
as $x\rto\infty$. However, due to (\ref{varasym}) $\ez_x[N_{fl}^2]$
is not bounded but rather grows logarithmically. Hence the overall
increase of $N_{fl}(x)$ is implicitly contained in the periodic
orbit term itself and has to be uncovered by careful estimates of
the sum. This remark should indicate the non-trivial problems that go
along with a quantitative analysis of the distribution of $W(y)$ for 
classically chaotic systems.

A strong hint at what has to be expected in the chaotic case is provided 
by a comparison with the distribution of $W(y)$ for the non-trivial
zeros of the Riemann zeta function. At the end of section \ref{sec1.3}
we pointed at the close analogy between the Selberg zeta function with
its spectral zeros on the one hand, and the Riemann zeta function with
its non-trivial zeros on the other hand. The asymptotics (\ref{Rcount})
of the counting function for the Riemann zeros now suggests to introduce
the defolded variable $x=\frac{t}{2\pi}\log\frac{t}{2\pi}$. A
`semiclassical' analysis in the spirit of section \ref{sec2.3} then
reveals that $l_H=const.\,\log x$ has to be chosen so that also
$L_{max}=const.\,\log x$. According to (\ref{Siinftyest}) one therefore 
concludes that $\ez_x[N_{fl}^2]\sim\frac{1}{2\pi^2}\log\log x$, $x\rto
\infty$. In order to arrive at this result one has to take into account 
that lengths of periodic orbits correspond to logarithms of primes. 
Since the latter obviously have multiplicity one, the semiclassical 
analysis requires to choose $\overline{g}=1$, as in the case of 
classically chaotic systems without time-reversal invariance. The 
latter point is in accordance with the conjectured -and numerically 
confirmed- GUE behaviour of the local statistics of the Riemann zeros
\cite{Odlyzko}. Already in 1946 Selberg proved \cite{Sel1}
\beq
\label{Sel46}
\ez_x\left[N_{fl}^{2k}\right]=\frac{(2k)!}{k!}\,\left(\frac{1}{2\pi^2}
\log\log x\right)^k+O\left((\log\log x)^{k-\frac{1}{2}}\right)\ ,
\eeq
so that the even moments of 
\beq
\label{WdefR}
W(y)=\frac{N_{fl}(y)}{\sqrt{\frac{1}{2\pi^2}\log\log x}}\ ,\ \ \ \ 
y\in I_x\ ,
\eeq
converge to the even moments of a standard Gaussian as $x\rto\infty$.
It was later noticed that an application of Selberg's method to the 
odd moments of $W(y)$ shows that these asymptotically vanish, see for
example \cite{Mont}. The distribution of $W(y)$, $y\in I_x$, therefore 
weakly converges to a Gaussian of zero mean and unit variance as $x\rto
\infty$. Selberg \cite{Sel2} extended this result to the corresponding 
limit distributions for a whole class of number theoretical zeta 
functions with functional equations. Since the zeta functions covered 
by Selberg's theorem are required to possess representations as ordinary 
Dirichlet series, the case of the Selberg zeta function $Z(s)$ is not 
included. But based on the analogy between the zeros of the Riemann 
zeta function and quantum energies of a classically chaotic system as 
well as on numerical evidence, in \cite{ABS} the following conjecture 
was established: If the classical dynamics are chaotic, with only 
isolated and unstable periodic orbits, and a scaling behaviour as in 
(\ref{scale}) holds, the limit distribution of $W(y)$ is a standard 
Gaussian. In analogy to (\ref{Wtorus}) it hence is conjectured that   
\beq
\label{Wchaotic}
\lim_{x\rto\infty}\frac{1}{2\De x}\int_{x-\De x}^{x+\De x}g\left(
W(y)\right)\,\rho\left(\frac{y-x}{2\De x}+\frac{1}{2}\right)\,dy=
\frac{1}{\sqrt{2\pi}}\int_{-\infty}^{+\infty}g(w)\,e^{-\frac{1}{2}w^2}
\,dw
\eeq
for any bounded continuous function $g(w)$ and any probability
density $\rho(\tau)$ on $[0,1]$. The restriction of the conjecture to
chaotic systems with scaling (\ref{scale}) stems from the analogy to
the distribution of the zeros of zeta functions. Exactly in the scaling
case one can directly encode the periodic orbit contribution to the 
spectral staircase in the complex phase of a semiclassical zeta function 
which is a function of the scaling variable $E^\al$, compare 
(\ref{scale}). Nevertheless, we expect the conjecture to hold also in 
more general situations, with the essential requirement that all 
periodic orbits be isolated and unstable. We also remark that the
conjecture includes the case of arithmetic quantum chaos, where
the local distribution of eigenvalues strongly deviates from the
`generic' situation for classically chaotic systems in that it rather 
shows a Poissonian behaviour than a GOE one \cite{Bogo,Steil}. 
Concerning the global distribution of eigenvalues the differences 
between arithmetic and `generic' systems only show up in the 
normalization of $W(y)$ because in the arithmetic case one obtains 
$\ez_x[N_{fl}^2]\sim\frac{\sqrt{\overline{d}}}{\pi}\frac{\sqrt{x}}
{\log x}$, $x\rto\infty$, see \cite{PhD}. A preliminary version of 
the conjecture (\ref{Wchaotic}), which concerns $W(y)$ evaluated 
only at the eigenvalues, is contained in \cite{SteinerF}.

In \cite{BogoSch} the distribution of $N_{fl}(y)$ has been computed 
numerically for two chaotic billiards in the unit disc (\ref{UD}) 
with the metric of constant negative curvature, one of them showing 
arithmetic quantum chaos, as well as for one integrable billiard. The 
normalization leading to the random variable $W(y)$ was not taken 
into account. But since in numerical calculations the spectral 
interval $I_x$ from which $y$ is taken has to be kept finite, the 
non-existence of a limit as $x\rto\infty$ does not become apparent 
in numerical data. The numerical results of \cite{BogoSch} show 
distributions that are somewhat closer to Gaussians in the two 
classically chaotic examples than in the classically integrable case.
A detailed numerical analysis of the distribution of $W(y)$ is
presented in \cite{EkL}. There six chaotic systems are compared with
one pseudo-integrable and five integrable systems, and a clear evidence
for Gaussian limit distributions to arise only in the classically 
chaotic cases is obtained. 
   
The influence of periodic orbits on the limit distribution of $W(y)$
entered our above discussion only indirectly through the representation
of $N_{fl}(x)$ by the complex phase of the Selberg zeta function or 
its semiclassical analogue. If one desires to uncover the role played
by unstable periodic orbits in a more direct fashion one has to express 
$N_{fl}(x)$ in terms of the Gutzwiller -or Selberg- trace formula. 
For this purpose let us again concentrate on the situation covered by 
the trace formula (\ref{traceeven}), or more specifically by the Selberg 
trace formula (\ref{STF}). Then $N_{fl}(p)$, where $p$ denotes the 
appropriate momentum variable, can be represented by $N_{\ve,fl}(p)$ 
for $\ve\rto 0$, see (\ref{cospo}). Alike in the semiclassical analysis
of section \ref{sec2.3} we now rewrite the sum over primitive periodic
orbits $\ga_p$ as a sum over their respective lengths $l_1<l_2<l_3<
\dots$, thereby introducing the multiplicities $g_n$ of $l_n$, $n=1,2,3,
\dots$. That way $N_{fl}(p)$ turns out to be a sum over the functions
\beq
\label{defan}
a_n(t):=\frac{g_n}{\pi}\sum_{k=1}^\infty\frac{1}{k}\frac{e^{-i
\frac{\pi}{2}k\tilde\mu_n}}{|\det (M_n^k -\unmat)|^{\frac{1}
{2}}}\,\sin(2\pi kt)
\eeq
evaluated at $t=\frac{p}{2\pi\hbar}l_n$. From (\ref{defan}) we obtain
that $a_n(t)$ is a one-periodic continuous function with zero mean, 
\beq
\label{anmean}
\int_0^1 a_n(t)\,dt=0\ .  
\eeq
Moreover,
\beq
\label{anvar}
\si_n^2:=\int_0^1 |a_n(t)|^2\,dt=\frac{g_n^2}{2\pi^2}\sum_{k=1}^\infty
\frac{1}{k^2}\,\frac{1}{|\det (M_n^k -\unmat)|}\sim\frac{g_n^2}{2\pi^2}
\sum_{k=1}^\infty\frac{1}{k^2}\,e^{-ku_n}\ ,\ \ \ \ n\rto\infty\ .
\eeq
If we now employ the periodic functions $a_n(t)$ in order to express 
$N_{fl}(p)$, we have to consider $\frac{p}{2\pi\hbar}l_n$ mod 1. In 
case the lengths $l_n$ of primitive periodic orbits are linearly 
independent over $\qz$, $\frac{p}{2\pi\hbar}l_n$ mod 1 becomes 
uniformly distributed on the interval $[0,1]$ as $p\rto\infty$ or 
$\hbar\rto 0$. Furthermore, due to the result of \cite{HB} already
alluded to above for integrable systems, the linear independence of
$l_{n_1},\dots,l_{n_k}$ implies 
\beq
\label{independence}
\lim_{p\rto\infty}\frac{1}{2\De p}\int_{p-\De p}^{p+\De p}a_{n_1}
\left(\frac{p'}{2\pi\hbar}l_{n_1}\right)\dots a_{n_k}\left(\frac{p'}
{2\pi\hbar}l_{n_k}\right)\ \frac{dp'}{2\pi\hbar}=\prod_{j=1}^k\int_0^1
a_{n_j}(t)\,dt\ ,
\eeq
where the interval $[p-\De p,p+\De p]$ corresponds to the spectral 
interval $I_x$ upon changing variables from the defolded energy variable 
$y$ to the momentum variable $p'$. Thus $N_{fl}(p')$ appears as a sum of 
the asymptotically independent random variables $a_n(\frac{p'}{2\pi\hbar}
l_n)$.  

When comparing the present situation with the classically integrable 
case one first of all notices that the required linear independence of
the primitive lengths is not known to hold in a single classically
chaotic example. Secondly, and more importantly, the sum
\beq
\label{varsum}
\sum_{n=1}^\infty\si_n^2=\sum_{n=1}^\infty\int_0^1 |a_n(t)|^2\,dt
\eeq
can be shown to diverge. To this end we remark that obviously the 
leading behaviour of $\si_n^2$ as $n\rto\infty$ is determined by
the contribution from $k=1$ in (\ref{anvar}). If we furthermore
recall the discussion of the r.h.s.\ of (\ref{poestim}), we obtain
that in case $\overline{g}=2$
\beq
\label{divest}
\frac{1}{2\pi^2}\sum_{l_n\leq l}g_n^2\,e^{-u_n}\approx\frac{\overline{g}}
{2\pi^2}\sum_{\ga_p,l_{\ga_p}\leq l}e^{-u_{\ga_p}}\sim\frac{1}{\pi^2}
\,\log l\ ,\ \ \ \ l\rto\infty\ .
\eeq
Thus, if we consider the random variables $a_n(\th_n)$, where the 
$\th_n$'s are independent random variables uniformly distributed on the
interval $[0,1]$, their variances $\si_n^2$ sum up to
\beq
\label{BNdef}
B_N:=\sum_{n=1}^N\si_n^2\sim\frac{1}{\pi^2}\,\log l_N\sim\frac{1}
{\pi^2}\,\log\left[\frac{1}{\tau}\,\log(2N\log 2N)\right]\sim\frac{1}
{\pi^2}\,\log\log N\ ,\ \ \ \ N\rto\infty\ .
\eeq
The asymptotic relation $l_N\sim\frac{1}{\tau}\,\log(2N\log 2N)$, 
$N\rto\infty$, leading to (\ref{BNdef}) immediately derives from 
(\ref{Huberl}). For the arithmetic systems with their exceptionally
large multiplicities of lengths \cite{Nonlin} $B_N$ diverges more 
strongly than in the `generic' case covered by (\ref{BNdef}). In the 
classically integrable situation we noticed that both conditions 
(\ref{bcond}) ensured the existence of a limit distribution and allowed 
for conclusion on its properties. In the present situation the second 
condition is violated, which indicates bad convergence properties of 
the sum $\sum_n a_n(t)$. However, it is suggested in \cite{BleMin} that 
due to the divergence of (\ref{varsum}) one should employ the 
Lindeberg-Feller version of the central limit theorem, see for example
\cite{Petrov}. This states that the distribution of the random variable
\beq
\label{defrv}
\xi_N:=\frac{\sum_{n=1}^N a_n(\th_n)}{\sqrt{B_N}}
\eeq
converges as $N\rto\infty$ to a Gaussian with zero mean and unit 
variance. 

We now again adopt the point of view, already employed in section 
\ref{sec2.3}, that the periodic orbit sum for $N_{fl}(p)$ is dominated
by the terms with lengths $l_n\leq l_H=4\pi p\hbar\overline{d}$. If
we then denote the index corresponding to $l_H$ by $N_H$, $l_{N_H}=
l_H$, we first of all notice that $N_{fl}(p)$ is essentially given
by 
\beq
\label{NHcut}
\sum_{n=1}^{N_H}a_n\left(\frac{p}{2\pi\hbar}\,l_n\right)\ .
\eeq
We now also cut off the sum of the variances $\si_n^2$ that refer to 
the individual terms in (\ref{NHcut}) at $N_H$, and hence obtain
\beq
\label{varcut}
B_{N_H}=\sum_{n=1}^{N_H}\si_n^2\sim\frac{1}{\pi^2}\,\log l_H\sim
\frac{1}{2}\Si^2_\infty(E)\sim\ez_x\left[N_{fl}^2\right]\ ,
\eeq
see (\ref{BNdef}), (\ref{Siinftyest}), and (\ref{varasym}). The 
relation (\ref{varcut}) is remarkable since it states that in the
semiclassical limit the true variance of $N_{fl}(p)$, which appears
on the very r.h.s., is asymptotic to the sum of the variances 
$\si_n^2$ when this is cut off at $N_H$. Therefore the two assumptions:
statistical independence of the $a_n$'s and truncation of the periodic
orbit sum at $l_H$, merge into a coherent picture. We conclude that
thus $\xi_{N_H}$ essentially yields $W(y)$,
\beq
\label{XiWcor}
W(y)=\frac{N_{fl}(y)}{\sqrt{\ez_x [N_{fl}^2 ]}}\sim\xi_{N_H}=
\frac{\sum_{n=1}^{N_H}a_n(\frac{p}{2\pi\hbar}\,l_n)}{\sqrt{B_{N_H}}}\ ,
\eeq
and that the limit distributions coincide. Here $y\in I_x$, and
$x$ corresponds to $l_H$ via $l_H=4\pi\hbar\sqrt{\overline{d}x}$.
Notice that the semiclassical limit $l_H\rto\infty$ has to be 
performed in (\ref{XiWcor}) as $x\rto\infty$ on the l.h.s., and as 
$N_H\rto\infty$ on the r.h.s. We therefore expect the limit 
distribution of $\xi_N$ as $N\rto\infty$, see (\ref{defrv}), to 
reproduce the limit distribution of $W(y)$, $y\in I_x$, as $x\rto
\infty$. At this point we want to stress that the above considerations
do not constitute a proof of (\ref{Wchaotic}), since the assumptions
made at various stages remain to be verified. However, we believe that
all evidence is in favour of the conjecture.  

The above discussion of the limit distribution of $W(y)$ in the
classically chaotic case was restricted to systems with only isolated
and unstable periodic orbits. Only for those the zeta function analogy
works and the periodic orbit expression for $N_{fl}(x)$ has the 
form required by (\ref{defan}). In order to illustrate the necessity
of the condition on the periodic orbits let us discuss the stadium 
billiard in some detail. The classical dynamics are known to show
the K-property \cite{Bun}, and hence the stadium billiard qualifies
by all means as a chaotic system in the usual sense. However, not
all periodic orbits are isolated and unstable because of the presence
of the bouncing ball orbits that form a one-parameter family of periodic
orbits. In addition to the contribution of the isolated and unstable
orbits the Gutzwiller trace formula for the stadium billiard therefore
contains a different term caused by the bouncing ball orbits. Thus 
the spectral staircase expressed in the momentum variable reads
\beq
\label{stairstad}
N(p)=\overline{N}(p)+N_{fl,u}(p)+N_{bb}(p)\ ,
\eeq
where $N_{fl,u}(p)$ is the ususal contribution of the unstable periodic
orbits. In \cite{BB} an explicit expression for the contribution of 
the bouncing ball orbits is derived,
\beq
\label{Nbb}
N_{bb}(p)=\frac{b}{2\sqrt{\pi^3 a}}\,p^{\frac{1}{2}}\,\sum_{n=1}^\infty
\frac{1}{n^{\frac{3}{2}}}\,\cos\left(2anp-\frac{3\pi}{4}\right)
+O\left(p^{-\frac{1}{2}}\right)\ .
\eeq
Here $b$ denotes the length of the two parallel edges of the billiard
domain, and $a$ is the radius of the two half-circles. In fact, 
(\ref{Nbb}) corresponds to the desymmetrized stadium billiard whose
domain consists of a quarter of the stadium domain, see \cite{BB}
for details. In terms of the defolded energy variable $x=\overline{d}
\,p^2$ the variances of the two periodic orbit contributions in
(\ref{stairstad}) read
\beq
\label{var2po}
\ez_x\left[N_{fl,u}^2\right]\sim\frac{1}{2\pi^2}\log x\ \ \ \ \mbox{and}
\ \ \ \ \ez_x\left[N_{bb}^2\right]\sim\frac{b^2\ze(3)}{8\pi^3a\sqrt{
\overline{d}}}\,x^{\frac{1}{2}}\ ,\ \ \ \ x\rto\infty\ .
\eeq
The first term is just (\ref{varasym}), whereas the second term
was determined in \cite{BB}. Hence
\beq
\label{totvar}
\ez_x\left[(N_{fl,u}+N_{bb})^2\right]\sim\frac{b^2\ze(3)}{8\pi^3a\sqrt{
\overline{d}}}\,x^{\frac{1}{2}}\ ,\ \ \ \ x\rto\infty\ ,
\eeq
so that $W(x)$ is asymptotically completely determined by the bouncing 
ball contribution
\beq
\label{Wbb}
W_{bb}(x):=\frac{N_{bb}(x)}{\sqrt{\ez_x [N_{bb}^2]}}\sim\sqrt{\frac{2}
{\ze(3)}}\sum_{n=1}^\infty\frac{1}{n^{\frac{3}{2}}}\,\cos\left(2an
\sqrt{\frac{x}{\overline{d}}}-\frac{3\pi}{4}\right)\ .
\eeq
Thus $W_{bb}(x)$, considered as a function of $\sqrt{x}$, is asymptotic
to a continuous periodic function with Fourier series given by 
(\ref{Wbb}). Since $\sum_n n^{-\frac{3}{2}}<\infty$ we can moreover 
estimate 
\beq
\label{West}
\left|W_{bb}(x)\right|\leq\sqrt{\frac{2}{\ze(3)}}\sum_{n=1}^\infty
\frac{1}{n^{\frac{3}{2}}}=\sqrt{\frac{2}{\ze(3)}}\,\ze\left(\frac{3}
{2}\right)\ .
\eeq
This finite bound implies that the density $p(w)$ of the limit
distribution of $W(y)$ for the complete desymmetrized stadium billiard
vanishes outside the finite interval $[-c,+c]$, where $c$ denotes
the constant on the r.h.s.\ of (\ref{West}). The limit distribution
is therefore clearly not Gaussian, and this violation of (\ref{Wchaotic})
is solely caused by the bouncing ball orbits. We remark that this
observation is not restricted to the stadium billiard, but will
occur in all systems with families of periodic orbits. These always
yield variances of their contributions to the spectral staircase
that exceed the logarithmic contribution (\ref{varasym}) of the
unstable orbits. By dividing through the rms fluctuations of the
spectral staircase in the definition of $W(y)$, the influence of the
otherwise dominating unstable orbits hence vanishes asymptotically. 

\subsection{Summary of the Global Statistics}
\label{sec3.4}
After having discussed various aspects of the global distribution
of eigenvalues for classically integrable as well as chaotic systems
in some detail we now want to summarize the emerging perception
of the global statistics. But before we again stress some points
that we already remarked before at several places.
\begin{enumerate}
\item Statements about local as well as global eigenvalue statistics
always refer to `generic' systems in a given class. A precise definition
of what this means has to be given in each particular case. A typical
example would be some statement concerning eigenvalue statistics 
on tori, see (\ref{Quadform}), where one would have to specify the
set of parameters $A,B,C$ to which the statement should apply. It
seems to be almost impossible, however, to give a general 
characterization of a `generic' classically integrable (or chaotic)
system in a unified manner.
\item Concerning the local statistics of eigenvalues it is assumed 
that it can `generically' be described by the appropriate random
matrix ensembles for classically chaotic systems \cite{BGS}, and
by Poissonian statistics for classically integrable systems
\cite{BerryTabor}. 
\item We furthermore assume that the description via random matrix
ensembles or Poisson statistics extends to hold in the limit 
$L\rto\infty$ once this is performed in the universality regime
$L/L_{max}\rto 0$. We base this assumption on the following observation:
The local statistics of eigenvalues emerges as the limit of $x\rto
\infty$ of, say, $E_x(k;L)$ for finite $L$. Passing to $L\rto\infty$
within the universality regime appears to leave the analogy to, say,
the GOE in the number variance unchanged, see (\ref{Llim1}).
\end{enumerate}

In the course of our discussion of eigenvalue distributions it became
clear that the random variable $\eta_L(y)$, $y\in I_x$, appears to 
be most suited for a unified treatment of spectral statistics on
all scales. After the limit $x\rto\infty$ has been performed, the 
distribution of $\eta_L$ with finite $L$ yields the local statistics.
According to (\ref{etadis}) and the assumption stated in 2., then
\beq
\label{locstat} 
\nu_{\eta_L}(d\eta)=\sqrt{\Si^2(L)}\ E\left(\eta\sqrt{\Si^2(L)}+L;
L\right)\,d\eta\ ,
\eeq
where $E(k;L)$ is either given by random matrix theory or by Poisson
statistics.

The transition to the global scale now proceeds under the assumption
in 3. We hence conclude that the distribution of $\eta_L(y)$, $y\in 
I_x$, converges for $x\rto\infty$, $L\rto\infty$, $L/L_{max}\rto 0$
to the same limit distribution as the respective distributions of
$\eta_L$ in the random matrix ensembles or for a Poisson random process
when $L\rto\infty$. Since the latter limit distributions are
standard Gaussians, see \cite{BL1} and (\ref{Poissonlim}), we
therefore expect the central limit theorem (\ref{nulimactual}) to 
hold for all `generic' systems, independent of their respective
classical limits. It thus appears that the global spectral statistics
in the universality regime are not sensitive enough a measure of
eigenvalue correlations in order to yield information about specific
features of a quantum system.

Finally we have to consider the limit $x\rto\infty$, $L\rto\infty$,
$L/L_{max}\rto\infty$, i.e., the transition to the global scale in
the saturation regime. In section \ref{sec3.1} we argued that the
limit distributions of $\eta_L(y)$, $y\in I_x$, then coincide with 
the limit distributions of $W(y)$, $y\in I_x$, as $x\rto\infty$.
This conclusion was drawn from the apparent asymptotic independence
of the random variables $N_{fl}(y)$, $y\in I_x$, and $N_{fl}(y)$, 
$y\in I_{x+L}$, as $L/L_{max}\rto\infty$, which was discussed at
the end of section \ref{sec2.3}. The principal observation then 
made in sections \ref{sec3.2} and \ref{sec3.3} was that the limit
distributions of $W(y)$ allow to distinguish classically integrable
systems from classically chaotic ones with only isolated and unstable
periodic orbits. Only the latter yield Gaussian limit distributions, 
whereas in all other cases the distributions of $W(y)$ are found to 
converge to a non-Gaussian limit.

We are now in a position to summarize the above findings in  
the following generalization of the conjecture (\ref{Wchaotic})
introduced in \cite{ABS}. Considering the limit distribution 
of the random variable $\eta_L(y)$, $y\in I_x$, we claim that
\beq
\label{conj}
\lim_{x\rto\infty\atop L\rto\infty}\int_{-\infty}^{+\infty}g(\eta)
\,\nu_{\eta_L,x}(d\eta)=\int_{-\infty}^{+\infty}g(\eta)\,P(\eta)\,
d\eta\ ,
\eeq
for any bounded continuous function $g(\eta)$. The limit distribution
$P(\eta)\,d\eta$ shall have a density $P(\eta)$ that depends on
the way the double limit is performed, as well as on the type of the
classical dynamics.
\begin{enumerate}
\item $L/L_{max}\rto 0$: For `generic' systems, be they classically
integrable or chaotic, one obtains 
\beq
\label{Univers}
P(\eta)=\frac{1}{\sqrt{2\pi}}\,e^{-\frac{1}{2}\eta^2}\ .
\eeq
\item $L/L_{max}\rto\infty$: The limit density $P(\eta)$ depends on
the type of the corresponding classical system. When the latter is 
integrable (or chaotic with families of periodic orbits) $P(\eta)$ 
is non-Gaussian. Most probably it will decay faster than a Gaussian,
\beq
\label{Pdecay}
\lim_{\eta\rto\pm\infty}P(\eta)\,e^{\frac{1}{2}\eta^2}=0\ .
\eeq
In certain cases it can even have a compact support. If the classical 
dynamics are chaotic, with only isolated and unstable periodic orbits, 
then a standard Gaussian as in (\ref{Univers}) is to be observed.
\end{enumerate}

In the field of quantum chaos so far mainly the local distribution
of eigenvalues was investigated; in particular the distribution of
nearest neighbour level spacings played an eminent role. The 
results as described by the random matrix/Poissonian conjecture, 
however, seem to be somewhat counter-intuitive. A Poissonian random 
process, which describes classically integrable, i.e.\ regular, systems,
is characterized by the absence of any correlations. On the other
hand, the behaviour of classically chaotic systems shows strong
correlations. In this respect the situation concerning the global
distribution of eigenvalues in the saturation regime seems to
be more confirm to intuition. The Gaussian limit distribution
emerging for $W(y)$ in the chaotic case is the most random of 
all possible probability distributions that have a density and
are of a fixed common variance. If we introduce the spectral
entropy
\beq
\label{entropy}
\cE [p]:=-\int_{-\infty}^{+\infty}p(w)\,\log p(w)\,dw\ ,
\eeq
which provides a quantitative measure of a mean unlikelihood for
$W(y)$ to have a certain value, the distribution with maximal
entropy characterizes the most random spectral fluctuations. It is
well known that under the constraint of a fixed variance, $\cE [p]$
is maximized by a Gaussian of zero mean. Therefore, quantum spectra
that have a Gaussian limit distribution for $W(y)$ or $\eta_L(y)$
possess, on a global scale, the most random fluctuations. This seems
to be a convenient characterization of quantum chaos.     

\section*{Acknowledgement}
I thank the organizers of the 3rd International Summer School/Conference 
{\it Let's face chaos through nonlinear dynamics} at the University
of Maribor for their warm hospitality and for the opportunity to present 
these lectures.


\begin{thebibliography}{99}
{\small

\bibitem{CFS} I.P.~Cornfeld, S.V.~Fomin, Ya.G.~Sinai: {\it Ergodic 
Theory}. Springer-Verlag, New York-Heidelberg-Berlin 1982.

\bibitem{Gutz} M.C.~Gutzwiller: {\it Chaos in Classical and Quantum
Mechanics}. Springer-Verlag, New York-Berlin-Heidelberg 1990.

\bibitem{LesHouches} M.-J.~Giannoni, A.~Voros, J.~Zinn-Justin (eds.):
{\it Chaos and quantum physics}. Les Houches 1989, Session LII.
North Holland, Amsterdam-London-New York-Tokyo 1991.

\bibitem{Arnold} V.I.~Arnold: {\it Mathematical Methods of Classical
Mechanics}. Second edition, Springer-Verlag, New York-Berlin-Heidelberg 
1989.

\bibitem{Robert} D.~Robert: {\it Autour de l'Approximation 
Semi-Classique}. Birkh\"auser, Boston 1987.

\bibitem{Sjoe} A.~Grigis, J.~Sj\"ostrand: {\it Microlocal Analysis
for Differential Operators}. London Mathematical Society Lecture Note
Series Vol.\ 196, Cambridge University Press, Cambridge 1994.

\bibitem{Pauli} W.~Pauli: {\it Ausgew\"ahlte Kapitel aus der
Feldquantisierung}. Boringheri, Turin 1952.

\bibitem{Gutz1} M.C.~Gutzwiller: J.\ Math.\ Phys.\ {\bf 8} (1967) 1979.

\bibitem{Gutz2} M.C.~Gutzwiller: J.\ Math.\ Phys.\ {\bf 12} (1971) 343.

\bibitem{SieStei} M.~Sieber, F.~Steiner: Phys.\ Lett.\ {\bf A144} 
(1990) 159.

\bibitem{Ruelle} D.~Ruelle: {\it Thermodynamic Formalism}. Encyclopedia
of Mathematics and its Applications Vol.\ 5, Addison-Wesley,
Reading, Massachusetts 1978.

\bibitem{Walters} P.~Walters: {\it An Introduction to Ergodic Theory}.
Springer-Verlag, New York-Heidelberg-Berlin 1982.

\bibitem{DG} J.J.~Duistermaat, V.~Guillemin: Invent.\ Math.\ {\bf 29} 
(1975) 39.

\bibitem{PU} T.~Paul, A.~Uribe: C.R.\ Acad.\ Sci.\ Paris S\'erie I
{\bf 313} (1991) 217.

\bibitem{Mein} E.~Meinrenken: Rep.\ Math.\ Phys.\ {\bf 31} (1992)
279.

\bibitem{McKSing} H.P.~McKean, I.M.~Singer: J.\ Diff.\ Geom.\ {\bf 1}
(1967) 43.

\bibitem{ASS} R.~Aurich, M.~Sieber, F.~Steiner: Phys.\ Rev.\ Lett.\ 
{\bf 61} (1988) 483.

\bibitem{AMSS} R.~Aurich, C.\ Matthies, M.~Sieber, F.~Steiner: Phys.\ 
Rev.\ Lett.\ {\bf 68} (1992) 1629.

\bibitem{AB} R.~Aurich, J.~Bolte: Mod.\ Phys.\ Lett.\ {\bf B6} (1992)
1691.

\bibitem{Hejhal} D.A.~Hejhal: {\it The Selberg Trace Formula for
$PSL(2,\rz)$ Vol.\ I}. Lecture Notes in Mathematics 548, Springer-Verlag, 
Berlin-Heidelberg-New York 1976.

\bibitem{Rat} J.G.~Ratcliffe: {\it Foundations of Hyperbolic Manifolds}.
Springer-Verlag, New York-Berlin-Heidelberg 1994.

\bibitem{BalVoros} N.L.~Balazs, A.~Voros: Phys.\ Rep.\ {\bf 143} (1986)
109.

\bibitem{Selberg} A.~Selberg: J.\ Indian Math.\ Soc.\ {\bf 20} (1956)
47.

\bibitem{Steiner} F.~Steiner: Phys.\ Lett.\ {\bf 188B} (1987) 447.

\bibitem{SieStei1} M.~Sieber, F.~Steiner: Phys.\ Rev.\ Lett.\ {\bf 67}
(1991) 1941.
 
\bibitem{Tanner} G.~Tanner, P.~Scherer, E.B.~Bogomolny, B.~Eckhardt,
D.~Wintgen: Phys.\ Rev.\ Lett.\ {\bf 67} (1991) 2410.

\bibitem{Titchmarsh} E.C.~Titchmarsh: {\it The Theory of the Riemann
Zeta-function}. Second edition revised by D.R.~Heath-Brown, Clarendon
Press, Oxford 1986. 

\bibitem{Bleher} P.M.~Bleher, Z.~Cheng, F.J.~Dyson, J.L.~Lebowitz:
Commun.\ Math.\ Phys.\ {\bf 154} (1993) 433.

\bibitem{PhD} J.~Bolte: Int.\ J.\ Mod.\ Phys.\ {\bf B7} (1993) 4451.

\bibitem{BerryN} M.V.~Berry: Nonlinearity {\bf 1} (1988) 399.

\bibitem{AurStei} R.~Aurich, F.~Steiner: Physica {\bf D82} (1995) 266.

\bibitem{AurStei1} R.~Aurich, F.~Steiner: Physica {\bf D43} (1990) 155.

\bibitem{Bogo} E.B.~Bogomolny, B.~Georgeot, M.-J.~Giannoni, C.~Schmit:
Phys.\ Rev.\ Lett.\ {\bf 69} (1992) 1477. 

\bibitem{Steil} J.\ Bolte, G.\ Steil, F.\ Steiner: Phys.\ Rev.\ Lett.\ 
{\bf 69} (1992) 2188.

\bibitem{Luo} W.~Luo, P.~Sarnak: Commun.\ Math.\ Phys.\ {\bf 161}
(1994) 419. 

\bibitem{Schur} P.~Sarnak: Israel Mathematical Conference Proceedings
{\bf 8} (1995) 183.

\bibitem{KS} N.M.~Katz, P.~Sarnak: {\it The spacing distribution
between zeros of zeta functions}. Preprint, Princeton University 1996.

\bibitem{Mehta} M.L.~Mehta: {\it Random Matrices.} Revised and enlarged
second edition. Academic Press, San Diego 1991.

\bibitem{CL} O.~Costin, J.L.~Lebowitz: Phys.\ Rev.\ Lett.\ {\bf 75}
(1995) 69.

\bibitem{Porter} C.E.~Porter (ed.): {\it Statistical Theories of Spectra:
Fluctuations}. Academic Press, New York and London 1965.

\bibitem{BGS} O.~Bohigas, M.-J.~Giannoni, C.~Schmit: Phys.\ Rev. Lett.\ 
{\bf 52} (1984) 1.

\bibitem{BerryTabor} M.V.~Berry, M.~Tabor: Proc.\ R.\ Soc.\ Lond.\ {\bf 
A356} (1977) 375.

\bibitem{ChengLeb} Z.~Cheng, J.L.~Lebowitz: Phys.\ Rev.\  {\bf A44}
(1991) R3399.

\bibitem{Sarnak} P.~Sarnak: {\it Values at integers of binary quadratic
forms}. Preprint, Princeton University 1996.

\bibitem{BleLeb} P.M.~Bleher, J.L.~Lebowitz: Ann.\ Inst.\ H.\ Poincar\'e
{\bf 31} (1995) 27.

\bibitem{A400} M.V.~Berry: Proc.\ R.\ Soc.\ Lond.\ {\bf A400} (1985) 229.

\bibitem{PP} W.~Parry, M.~Pollicott: {\it Zeta functions and the periodic
orbit structure of hyperbolic dynamics}. Ast\'erisque {\bf 187-188},
Soci\'et\'e Math\'ematique de France 1990.

\bibitem{HOdA} J.H.~Hannay, A.M.~Ozorio De Almeida: J.\ Phys.\ A:\ 
Math.\ Gen.\ {\bf 17} (1984) 3429.

\bibitem{Nonlin} J.~Bolte: Nonlinearity {\bf 6} (1993) 935.

\bibitem{Stefan} S.~Johansson: {\it Traces in Arithmetic Fuchsian 
Groups}. Preprint Dept.\ Math.\ Chalmers Univ.\ of Techn.\  and 
G\"oteborg Univ.\ 1996. 

\bibitem{BogoKea} E.B.~Bogomolny, J.P.~Keating: Phys.\ Rev.\ Lett.\ 
{\bf 77} (1996) 1472.

\bibitem{Baecker} A.~B\"acker, P.~Stifter, F.~Steiner: Phys.\ Rev.\ 
{\bf E 52} (1995) 2463-2472.

\bibitem{BL1} P.M.~Bleher, J.L.~Lebowitz: J.\ Stat.\ Phys.\ {\bf 74}
(1994) 167.

\bibitem{BB} M.~Sieber, U.~Smilansky, S.C.~Creagh, R.G.~Littlejohn: 
J.\ Phys.\ A:\ Math.\ Gen.\ {\bf 26} (1993) 6217.

\bibitem{Blrev} P.M.~Bleher: Duke Math.\ J.\ {\bf 74} (1994) 45.

\bibitem{KMS} D.V.~Kosygin, A.A.~Minasov, Ya.G.~Sinai: Uspekhi Mat.\ 
Nauk {\bf 48} (1993) 3.

\bibitem{HB} D.R.~Heath-Brown: Acta Arithmetica {\bf 60} (1992) 389.

\bibitem{BleDuke} P.M.~Bleher: Duke Math.\ J.\ {\bf 67} (1992) 461.

\bibitem{BleMin} P.M.~Bleher: {\it Trace Formula for Quantum Integrable
Systems, Lattice-Point Problem, and Small Divisors}. Preprint
Indiana Univ.-Purdue Univ.\ 96-11, Indianapolis 1996.

\bibitem{Ble70} P.M.~Bleher: Duke Math.\ J.\ {\bf 70} (1993) 655.

\bibitem{BKS} P.M.~Bleher, D.V.~Kosygin, Ya.G.~Sinai: Commun.\ Math.\ 
Phys.\ {\bf 170} (1995) 375.

\bibitem{Roman} R.~Schubert: {\it The Trace Formula and the Distribution
of Eigenvalues of Schr\"odinger Operators on Manifolds all of whose
Geodesics are Closed}. DESY-report, DESY 95-090, 1995.

\bibitem{Odlyzko} A.M.~Odlyzko: Math.\ Comp. {\bf 48} (1987) 273.

\bibitem{Sel1} A.~Selberg: Archiv for Mathematik og Naturvidenskab
{\bf B48} (1946) 89.

\bibitem{Mont} H.L.~Montgomery, in: K.E.~Aubert, E.~Bombieri,
D.~Goldfeld (eds.): {\it Number Theory, Trace Formulas and Discrete 
Groups}. Academic Press, New York 1989.

\bibitem{Sel2} A.~Selberg: {\it Collected Papers, Vol.~II}. 
Springer-Verlag, Berlin-Heidelberg-New York 1991.

\bibitem{ABS} R.~Aurich, J.~Bolte, F.~Steiner: Phys.\ Rev.\ Lett.\ 
{\bf 73} (1994) 1356.

\bibitem{SteinerF} F.~Steiner, in: R. Ansorge (ed.): {\it Schlaglichter
der Forschung. Zum 75. Jahrestag der Universit\"at Hamburg}. Dietrich
Reimer Verlag, Berlin-Hamburg 1994.

\bibitem{BogoSch} E.B.~Bogomolny, C.~Schmit: Nonlinearity {\bf 6}
(1993) 523.

\bibitem{EkL} R.~Aurich, A.~B\"acker, F.~Steiner: {\it Mode fluctuations 
as fingerprints of chaotic and non-chaotic systems}. Preprint 
ULM-TP/96-2, 1996. To appear in Int.\ J.\ Mod.\ Phys.\ B.

\bibitem{Petrov} V.V.~Petrov: {\it Limit Theorems of Probability Theory}.
Clarendon Press, Oxford 1995. 

\bibitem{Bun} L.A.~Bunimovich: Commun.\ Math.\ Phys.\ {\bf 65}
(1979) 295.

}
\end{thebibliography}
\end{document}